\documentclass[twocolumn]{revtex4-1}
\usepackage{graphicx}
\usepackage{dcolumn}
\usepackage{bm}
\usepackage{amsmath}
\usepackage{amssymb}
\usepackage{subfigure}

\newcommand{\smmu}{\mu_{\mathrm{s}}}
\newcommand{\smJij}{J_{ij}}

\newcommand{\smJijT}{\mathbf{J}_{ij}^{\mathrm{M}}}
\newcommand{\smKu}{k_{\mathrm{u}}}
\newcommand{\smKc}{k_{\mathrm{c}}}

\newcommand{\smH}{\mathbf{H}}
\newcommand{\sms}{\mathbf{S}}
\newcommand{\mmc}{\mathbf{m}_{\mathrm{mc}}}
\newcommand{\vampire}{\textsc{vampire} }
\newcommand{\Tc}{T_{\mathrm{c}} }
\newcommand{\kB}{k_{\mathrm{B}}}
\newcommand{\muB}{\mu_{\mathrm{B}}}
\newcommand{\MKu}{K_{\mathrm{u}}}

\newcommand{\MMs}{M_{\mathrm{s}}}

\usepackage{setspace}
\usepackage[usenames,dvipsnames]{color}
\usepackage{url}

\DeclareMathOperator{\sech}{sech}

\begin{document}

\title{Atomistic spin model simulations of magnetic nanomaterials}
\author{R. F. L. Evans}
\email{richard.evans@york.ac.uk}
\homepage{\url{http://www-users.york.ac.uk/~rfle500/}}
\author{W. J. Fan}
\author{P. Chureemart}
\author{T. A. Ostler}
\author{M. O. A. Ellis}
\author{R. W. Chantrell}
\affiliation{Department of Physics, The University of York, York, YO10 5DD, UK}
\date{\today}

\begin{abstract}
Atomistic modelling of magnetic materials provides unprecedented detail about the underlying physical processes that govern their macroscopic properties, and allows the simulation of complex effects such as surface anisotropy, ultrafast laser-induced spin dynamics, exchange bias, and microstructural effects. Here we present the key methods used in atomistic spin models which are then applied to a range of magnetic problems.  We detail the parallelisation strategies used which enable the routine simulation of extended systems with full atomistic resolution.
\end{abstract}


\maketitle

\section{Introduction}
Atomistic models of magnetic materials , where the atoms treated as possessing a local magnetic moment, originated with Ising in 1925 as the first model of the phase transition in a ferromagnet\cite{Ising:1925em}. The Ising model has spin-up and spin-down only states, and is amenable to analytical treatment, at least in two dimensions. Although it is still extensively used in the study of phase transitions, it is limited in applicability to magnetic materials and cannot be used for dynamic simulations. A natural extension of the Ising model is to allow the atomic spin to vary freely in 3D space\cite{watson1969,binder1969} which yields the classical Heisenberg model, where quantum mechanical effects on the atomic spins are neglected\cite{watson1969}. Monte Carlo simulations of the classical Heisenberg model allowed the study of phase transitions, surface and finite size effects in simple magnetic systems. The study of dynamic phenomena however was intrinsically limited due to the use of Monte Carlo methods until the development of dynamic\cite{Kodama:1999wk,Mitsumata:2003hd} and stochastic Landau-Lifshitz-Gilbert atomistic spin models\cite{Nowakspinmodels,Boerner2005,Skubic:2008gs}.

Today atomistic simulation of magnetic materials has become an essential tool in understanding the processes governing the complex behaviour of magnetic nanomaterials, including ultrafast laser-induced magnetisation dynamics \cite{Ostler:2011jf,Ostler:2012hx,Radu:2011krb}, exchange bias in core-shell nanoparticles\cite{Iglesias:2005cl,Evans:2011tb,Evans:2009kf} and multilayers\cite{Mitsumata:2003hd,Ali:2003vi}, surface anisotropy in magnetic nanoparticles\cite{Garanin:2003ho,Yanes:2007gu}, microstructural effects\cite{Ma:2012el,Evans:2006cf,Evans:2007tw}, spin valves\cite{Ho:eb} and spin torque\cite{Chureemart:2011cr}, temperature effects and properties\cite{Asselin:2010iv,Barker:2010hfa,Yanes:2010we,Bauer2011} and magnetic recording media\cite{Fal:2013ba,Victora:2013ki}. A significant capability of the atomistic spin model is to bridge the gap between \textit{ab initio} electronic structure calculations and micromagnetics by using a multiscale model\cite{Mryasov:2005kj,Szunyogh:2011db,Szunyogh:2009kf,FeRh2011}. Such a model is able to calculate effective parameters for larger scale micromagnetic simulations\cite{Kazantseva:2008ja}, such as anisotropies, and exchange constants\cite{Atxitia:2010fu}. The atomistic model is also able to interface directly with micromagnetic simulations to tackle extended systems by calculating interface properties atomistically while treating the bulk of the material with a micromagnetic discretisation\cite{Jourdan:2008ij,GarciaSanchez:2006tt}. Despite the broad applicability and importance of atomistic models, no easy-to-use and open-source software packages are presently available to researchers, unlike the mesoscopic micromagnetic approaches where several packages are currently available\cite{oommf,Scholz:2003ty,Fischbacher:ii}. 

Today most magnetic modelling is performed using numerical micromagnetics in finite-difference\cite{oommf} and finite-element\cite{Scholz:2003ty,Fischbacher:ii} flavours. The theoretical basis of micromagnetics is that the atomic dipoles which make up the magnetic material can be approximated as a continuous vector field where, due to the exchange interaction, the atomic dipoles in a small finite volume are perfectly aligned. Micromagnetics has proven to be an essential tool in understanding a range of complex magnetic effects\cite{Mistral:2008js,Dobin:2006hv,Schrefl:1997tl} but due to the rapid pace of technological development in magnetic materials the continuum approximation at its heart precludes its application to many problems of interest at the beginning of the 21$^{\textrm{st}}$ century, such as heat assisted magnetic recording\cite{Kryder:2008kt}, ultrafast laser-induced demagnetisation\cite{Beaurepaire:1996es,Stanciu:2007co}, exchange bias in spin valves\cite{OGrady:2010gn}, surface and interface anisotropy\cite{Ikeda:2010iz,Jamet:2001em} and high anisotropy materials for ultrahigh density recording media such as FePt\cite{Klemmer:2002wq}. The common theme to these problems is a sub-nanometre spatial variation in the magnetisation caused by high temperatures, atomic level ordering (anti- and ferri- magnets), or atomic surface and interface effects. To tackle these problems requires a formalism to take account of the detailed atomic structure to express its impact on the macroscopic behaviour of a nano particle, grain or complete device.

Some but not all of these problems can adequately be tackled by next-generation micromagnetic approaches utilising the Landau-Lifshitz-Bloch equation \cite{Garanin:1997uf,Evans:2012ex,Atxitia:2012bk}, which is based on a physically robust treatment of the coupling of a macrospin to a heat bath, allowing the study of high temperature processes\cite{McDaniel:2012ep}, ultrafast demagnetisation \cite{Atxitia:2011ds,Sultan:2012bh} and switching\cite{Atxitia:2013ji}. However, true atomic scale variations of the magnetisation, as apparent in antiferromagnets, surfaces and interfaces, still require an atomistic approach. Additionally with the decreasing size of magnetic elements, finite size effects begin to play in increasing role in the physical properties of magnetic systems\cite{Hovorka:2012kx}.

In this article we present an overview of the common computational methods utilised in atomistic spin simulations and details of their implementation in the open-source \vampire software package\footnote{Details available from vampire.york.ac.uk}. Testing of the code is an essential part of ensuring the accuracy of the model and so we also detail important tests of the various parts of the model and compare them to analytic results while exploring some interesting physics of small magnetic systems.

\vampire is designed specifically with these problems in mind, and can easily simulate nanoparticles, multilayer films, interfacial mixing, surface anisotropy and roughness, core-shell systems, granular media and lithographically defined patterns, all with fully atomistic resolution. In addition truly realistic systems predicted by Molecular Dynamics simulations\cite{Dorfbauer:2006uq,Evans:2007tw,Evans:2006cf} can also be used giving unprecedented detail about the relationships between shape and structure and the magnetic properties of nanoparticles. In addition to these physical features \vampire also utilises the computing power of multiprocessor machines through parallelisation, allowing systems of practical interest to be simulated routinely, and large-scale problems on the 100+ nm length scale to be simulated on computing clusters.

\section{The atomistic spin model}
The physical basis of the atomistic spin model is the localisation of unpaired electrons to atomic sites, leading to an effective local atomistic magnetic moment. The degree of localisation of electrons has historically been a contentious issue in 3d metals\cite{Jiles:1991wd}, due to the magnetism originating in the outer electrons which are notionally loosely bound to the atoms. \textit{Ab initio} calculations of the electron density\cite{Schwarz:1984wn} show that in reality even in 'itinerant' ferromagnets the spin polarisation is well-localised to the atomic sites. Essentially this suggests that the bonding electrons are unpolarised, and after taking into account the bonding charge the remaining d-electrons form a well-defined effective localised moment on the atomic sites.

Magnetic systems are fundamentally quantum mechanical in nature since the electron energy levels are quantised, the exchange interaction is a purely quantum mechanical effect, and other important effects such as magnetocrystalline anisotropy arise from relativistic interactions of electronic orbitals with the lattice, which are the province of \textit{ab initio} models. In addition to these properties at the electronic level, the properties of magnetic materials are heavily influenced by thermal effects which are typically difficult to incorporate into standard density functional theory approaches. Therefore models of magnetic materials should combine the quantum mechanical properties with a robust thermodynamic formalism. The simplest model of magnetism using this approach is the Ising model\cite{Ising:1925em}, which allows the atomic moments one of two allowed states along a fixed quantisation axis. Although useful as a descriptive system, the forced quantisation is equivalent to infinite anisotropy, limiting the applicability of the Ising model in relation to real materials.
In the classical description the direction of the atomic moment is a continuous variable in 3D space allowing for finite anisotropies and dynamic calculations. In some sense the classical spin model is analogous to Molecular Dynamics, where the energetics of the system are determined primarily from quantum mechanics, but the time evolution and thermodynamic properties are treated classically.

\subsection{The classical spin Hamiltonian}
The Heisenberg spin model encapsulates the essential physics of a magnetic material at the atomic level, where the energetics of a system of interacting atomic moments is given by a \textit{spin} Hamiltonian (which neglects non-magnetic effects such the as the Coulomb term). The spin Hamiltonian $\mathcal{H}$ typically has the form:
\begin{equation}\label{eq:spinhamiltonian}
    \mathcal{H} = \mathcal{H}_{\mathrm{exc}} + \mathcal{H}_{\mathrm{ani}} + \mathcal{H}_{\mathrm{app}}
\end{equation}
denoting terms for the exchange interaction, magnetic anisotropy, and externally applied magnetic fields respectively.

The dominant term in the spin Hamiltonian is the Heisenberg exchange energy, which arises due to the symmetry of the electron wavefunction and the Pauli exclusion principle\cite{Jiles:1991wd} which governs the orientation of electronic spins in overlapping electron orbitals. Due to its electrostatic origin, the associated energies of the exchange interaction are around $1-2$ eV, which is typically up to 1000 times larger than the next largest contribution and give rise to magnetic ordering temperatures in the range 300-1300K. The exchange energy for a system of interacting atomic moments is given by the expression
\begin{equation}
    \mathcal{H}_{\mathrm{exc}} = -\sum_{i\ne j} \smJij \sms_i \cdot \sms_j
\label{eq:HJij}
\end{equation}
where $\smJij$ is the exchange interaction between atomic sites $i$ and
$j$, $\sms_i$ is a unit vector denoting the local spin moment direction and $\sms_j$ is the spin moment direction of neighbouring atoms. The unit vectors are taken from the actual atomic moment $\boldsymbol{\mu}_{\mathrm{s}}$ and given by $\sms_i = \boldsymbol{\mu}_{\mathrm{s}}/|\boldsymbol{\mu}_{\mathrm{s}}|$. It is important to note here the significance of the sign of $\smJij$. For ferromagnetic materials where neighbouring spins align in parallel, $\smJij > 0$, and for anti-ferromagnetic materials where the spins prefer to align anti-parallel $\smJij < 0$. Due to the strong distance dependence of the exchange interaction the sum in Eq.~\ref{eq:HJij} is often truncated to include nearest neighbours only. This significantly reduces the computational effort while being a good approximation for many materials of interest. In reality the exchange interaction can extend to several atomic spacings\cite{Mryasov:2005kj,Szunyogh:2011db}, representing hundreds of pairwise interactions.

In the simplest case the exchange interaction $\smJij$ is isotropic, meaning that the exchange energy of two spins depends only on their relative orientation. In more complex materials, the exchange interaction forms a tensor with components:
\begin{equation}
\smJijT = \begin{bmatrix} J_{xx}&J_{xy}&J_{xz}\\ J_{yx}&J_{yy}&J_{yz} \\ J_{zx}&J_{zy}&J_{zz} \end{bmatrix}\mathrm{,}
\label{eq:HJijM}
\end{equation}
which is capable of describing \textit{anisotropic} exchange interactions, such as two-ion anisotropy\cite{Mryasov:2005kj} and the Dzyaloshinskii-Moriya interaction (off-diagonal components of the exchange tensor). In the case of tensorial exchange $\mathcal{H}_{\mathrm{exc}}^{\mathrm{M}}$, the exchange energy is given by the product:
\begin{equation}
    \mathcal{H}_{\mathrm{exc}}^{\mathrm{M}} = -\sum_{i \ne j}\begin{bmatrix}S_x^i S_y^i S_z^i\end{bmatrix} \begin{bmatrix} J_{xx}&J_{xy}&J_{xz}\\ J_{yx}&J_{yy}&J_{yz} \\ J_{zx}&J_{zy}&J_{zz} \end{bmatrix} \begin{bmatrix}S_x^j \\S_y^j \\ S_z^j\end{bmatrix}\mathrm{.}
\label{eq:HJijSMS}
\end{equation}
Obtaining the components of the exchange tensor may be done phenomenologically, or via \textit{ab initio} methods such as the relativistic torque method\cite{rtm1,rtm2,rtm3,rtm4} or the spin-cluster expansion technique\cite{sce1,sce1b,Szunyogh:2011db,sce3}. The above expressions for the exchange energy also exclude higher-order exchange interactions such as three-spin and four-spin terms. In most materials the higher order exchange terms are significantly smaller than the leading term and can safely be neglected. 

While the exchange energy gives rise to magnetic ordering at the atomic level, the stability of a magnetic material is dominated by the magnetic anisotropy, or preference for the atomic moments to align along a preferred spatial direction. There are several physical effects which give rise to anisotropy, but the most important is the magnetocrystalline anisotropy (namely the preference for spin moments to align with particular crystallographic axes) arising from the interaction of atomic electron orbitals with the local crystal environment\cite{Skomski:2012uh,Bruno:1993wc}.

The simplest form of anisotropy is of the single-ion uniaxial type, where the magnetic moments prefer to align along a single axis, $\boldsymbol{e}$, often called the easy axis and is an interaction confined to the local moment. Uniaxial anisotropy is most commonly found in particles with elongated shape (shape anisotropy), or where the crystal lattice is distorted along a single axis as in materials such as hexagonal Cobalt and L1$_0$ ordered FePt. The uniaxial single ion anisotropy energy is given by the expression:
\begin{equation}
\mathcal{H}_{\mathrm{ani}}^{\mathrm{uni}} = -\smKu \sum_i
\left(\sms_i \cdot \boldsymbol{e} \right)^2
\end{equation}
where $\smKu$ is the anisotropy energy per atom. Materials with a cubic crystal structure, such as Iron and Nickel, have a different form of anisotropy known as cubic anisotropy. Cubic anisotropy is generally much weaker than uniaxial anisotropy, and has three principal directions which energetically are easy, hard and very hard magnetisation directions respectively. Cubic anisotropy is described by the expression:
\begin{equation}
\mathcal{H}_{\mathrm{ani}}^{\mathrm{cub}} =
\frac{k_{\mathrm{c}}}{2} \sum_i \left(S_x^4+S_y^4+S_z^4\right)
\end{equation}
where $\smKc$ is the cubic anisotropy energy per atom, and $S_x$,$S_y$, and $S_z$ are the $x$,$y$, and $z$ components of the spin moment $\sms$ respectively.

Most magnetic problems also involve interactions between the system and external applied fields, denoted as $\smH_{\mathrm{app}}$. External fields can arise in many ways, for example a nearby magnetic material, or as an effective field from an electric current. In all cases the applied field energy is simply given by:
\begin{equation}
\mathcal{H}_{\mathrm{app}} = - \sum_i \smmu \sms_i \cdot \smH_{\mathrm{app}}\mathrm{.}
\end{equation}

\subsection*{A note on magnetic units}
The subject of magnetic units is controversial due to the existence of multiple competing standards and historical origins.\cite{Jiles:1991wd} Starting from the atomic level however, the dimensionality of units is relatively transparent. Atomic moments are usually accounted for in multiples of the Bohr magneton ($\mu_{\mathrm{B}}$), the magnetic moment of an isolated electron, with units of Joules/Tesla. Given a number of atoms of moment $\smmu$ in a volume, the moment per unit volume is naturally in units of J/T/m$^3$, which is identical to the SI unit of A/m. However, the dimensionality (moment per unit volume) of the unit A/m is not as obvious as JT$^{-1}$m$^{-3}$, and so the latter form is used herein.

Applied magnetic fields are hence defined in Tesla, which comes naturally from the derivative of the spin Hamiltonian with respect to the local moment. The unit of Tesla for applied field is also beneficial for hysteresis loops, since the area enclosed a typical M-H loop is then given as an energy density (Joules/m$^3$). A list of key magnetic parameters and variables and their units are shown in Tab.~\ref{tab:parameters}.

\begin{table}[!htcdp]
\caption{Table of key variables and their units}
\begin{center}
\begin{tabular}{  l  c  l l }
\hline \hline
           &        &                         \\
  Variable & Symbol & \multicolumn{2}{c}{Unit} \\
\hline
  Atomic magnetic moment & $\smmu$  & Joules/Tesla & [JT$^{-1}$]\\
  Unit cell size         & $a$      & Angstroms    & [\AA] \\
  Exchange energy        & $\smJij$ & Joules/link  & [J] \\
  Anisotropy energy      & $\smKu$  & Joules/atom  & [J] \\
  Applied Field          & $\smH$   & Tesla & [T] \\
  Temperature            & $T$      & Kelvin       & [K] \\
  Time                   & $t$      & Seconds      & [s] \\
                         &          &              & \\
\end{tabular}

\begin{tabular}{  l  c  l }
  Parameter & Symbol & \multicolumn{1}{c}{Value} \\
\hline
  Bohr Magneton & $\muB$  & 9.2740 $\times 10^{-24}$ JT$^{-1}$ \\
  Gyromagnetic Ratio & $\gamma$  & 1.76 $\times 10^{11}$ T$^{-1}$s$^{-1}$ \\
  Permeability of Free Space & $\mu_{0}$ & $4\pi \times 10^{-7}$ T$^{2}$J$^{-1}$m$^3$ \\
  Boltzmann Constant & $\kB$    & $1.3807 \times 10^{-23}$ JK$^{-1}$ \\
                         &          &               \\
\hline \hline
\end{tabular}
\end{center}
\label{tab:parameters}
\end{table}

\section{System parameterisation and generation}
Unlike micromagnetic simulations where the magnetic system can be partitioned using either a finite difference or finite element discretisation, atomistic simulations generally require some \textit{a priori} knowledge of atomic positions. Most simple magnetic materials such as Fe, Co or Ni form regular crystals, while more complex systems such as oxides, antiferromagnets and Heusler alloys possess correspondingly complex atomic structures. For ferromagnetic metals, the details of atomic positions are generally less important due to the strong parallel orientation of moments, and so they can often (but not always) be represented using a simple cubic discretisation. In contrast, the properties of ferrimagnetic and antiferromagnetic materials are inherently tied to the atomic positions due to frustration and exchange interactions, and so simulation of these materials must incorporate details of the atomic structure.

In addition to the atomic structure of the material, it is also necessary to parameterise the terms of the spin Hamiltonian given by Eq.~\ref{eq:spinhamiltonian}, principally including exchange and anisotropy values but also with other terms. There are generally two ways in which this may be done: using experimentally determined properties or with a multiscale approach using \textit{ab initio} density functional theory calculations as input to the spin model.

\subsection*{Atomistic parameters from \textit{ab initio} calculations}
\textit{Ab initio} density functional theory (DFT) calculations utilise the Hohenberg-Kohn-Sham theory\cite{KS1,KS2} that the total energy $E$ of a system can be written solely in terms the electron density, $rho$. Thus, if the electron density is known then the physical properties of the system can be found. In practice, the both electron density and the spin density are used as fundamental quantities in the total energy expression for spin-polarised systems\cite{SP1}. In many implementations DFT-based methods only consider the outer electrons of a system, since the inner electrons play a minimal role in the bonding and also partially screen the effect of the nuclear core. These effects are approximated by a pseudopotential which determines the potential felt by the valence electrons. In all-electron methods, however, the core electron density is also relaxed.
By energy minimisation DFT enables the calculation of a wide range of properties, including lattice constants, and in the case of magnetic materials localised spin moments, magnetic ground state and the effective magnetocrystalline anisotropy. Standard software packages such as \textsc{vasp}\cite{vasp}, \textsc{castep}\cite{castep1,castep2} and \textsc{siesta}\cite{siesta}
 make such calculations readily accessible. At present determining site resolved properties such as anisotropy constants and pairwise exchange interactions is more involved and require \textit{ab initio} Green's functions techniques such as the fully relativistic Korringa-Kohn-Rostoker method\cite{skkr,skkr2} or the \textsc{lmto} method\cite{magneticforce2000,Pajda:2001ix} in conjunction with the magnetic force theorem\cite{rtm1}. An alternative approach for the calculation of exchange parameters is the utilisation of the generalised Bloch theorem for spin-spiral states in magnetic systems\cite{SS1} together with a Fourier transformation of $k$-dependent spin-spiral energies\cite{SS2,inducedmoments2013}.

A number of studies have determined atomic magnetic properties from first principles calculations by direct mapping onto a spin model, including the principal magnetic elements Co, Ni and Fe \cite{Pajda:2001ix}, metallic alloys including FePt\cite{Mryasov:2005kj}, IrMn\cite{Szunyogh:2009kf}, oxides\cite{Uhl:1999bz} and spin glasses\cite{Skubic:2009bv}, and also bilayer systems such as Fe/FePt.\cite{AasFePtFe2013} Such calculations give detailed insight into microscopic magnetic properties, including atomic moments, long-ranged exchange interactions, magnetocrystalline anisotropies (including surface and two-ion interactions) and other details not readily available from phenomenological theories. Combined with atomistic models it is possible to determine macroscopic properties such as the Curie temperature, temperature dependent anisotropies, and magnetic ground states, often in excellent agreement with experiment. However, the computational complexity of DFT calculations also means that the systems which can be simulated with this multi scale approach are often limited to small clusters, perfect bulk systems and 2D periodic systems, while real materials of course often contain a plethora of defects disrupting the long range order. Some studies have also investigated the effects of disorder in magnetic systems combined with a spin model mapping, such as dilute magnetic semiconductors\cite{disorder1} and metallic alloys\cite{disorder2}.

Magnetic properties calculated at the electronic level have a synergy with atomistic spin models, since the electronic properties can often be mapped onto a Heisenberg spin model with effective local moments. This multiscale electronic and atomistic approach avoids the continuum approximations of micromagnetics and treats magnetic materials at the natural atomic scale.

\subsection*{Atomistic parameters from macroscopic properties}
The alternative approach to multiscale atomistic/density-functional-theory simulations is to derive the parameters from experimentally determined values. This has the advantage of speed and lower complexity, whilst foregoing microscopic details of the exchange interactions or anisotropies. Another key advantage of generic parameters is the possibility of parametric studies, where parameters are varied explicitly to determine their importance for the macroscopic properties of the system, such as has been done for studies of surface anisotropy\cite{Yanes:2007gu} and exchange bias\cite{Evans:2011tb}.

Unlike micromagnetic simulations, the fundamental thermodynamic approach of the atomistic model means that all parameters must be determined for \emph{zero} temperature. The spin fluctuations then determine the intrinsic temperature dependence of the \textit{effective} parameters which are usually put into micromagnetic simulations as parameters. Fortunately determination of the atomic moments, exchange constants and anisotropies from experimental values is relatively straightforward for most systems.

\subsubsection*{Atomic spin moment}
The atomic spin moment $\smmu$ is related to the saturation magnetisation simply by:
\begin{equation}
\smmu = \frac{\MMs a^3}{n_{\mathrm{at}}}
\end{equation}
where $\MMs$ is the saturation magnetisation at 0K in JT$^{-1}$m$^{-3}$, $a$ is the unit cell size (m), and $n_{\mathrm{at}}$ is the number of atoms per unit cell. We also note the usual convention of expressing atomic moments in multiples or fractions of the Bohr magneton, $\muB$ owing to their electronic origin. Taking BCC Iron as an example, the zero temperature saturation magnetisation is 1.75 MA/m\cite{Danan:1968jw}, unit cell size of $a = 2.866$ \AA, this gives an atomic moment of 2.22 $\muB$/atom.
%

\subsubsection*{Exchange energy}
For a generic atomistic model with $z$ nearest neighbour interactions, the exchange constant is given by the mean-field expression:
\begin{equation}
\smJij = \frac{3 k_B \Tc}{\epsilon z}
\label{eq:JijTc}
\end{equation}
where $k_B$ is the Boltzmann constant and $\Tc$ is the Curie temperature $z$ is the number of nearest neighbours. $\epsilon$ is a correction factor from the usual mean-field expression which arises due to spin waves in the 3D Heisenberg model\cite{Garanin:351283} and is $\sim 0.86$. Because of this $\epsilon$ is also crystal structure and coordination number dependent, and so the calculated $\Tc$ will vary slightly according to the specifics of the system. For Cobalt with a $\Tc$ of 1388K and assuming a hexagonal crystal structure with $z = 12$, this gives a nearest neighbour exchange interaction $\smJij = 6.064 \times 10^{-21}$J/link.

\subsubsection*{Anisotropy energy}
The atomistic magnetocrystalline anisotropy $\smKu$ is derived from the macroscopic anisotropy constant $\MKu$ by the expression:
\begin{equation}
\smKu = \frac{\MKu a^3}{n_{\mathrm{at}}}
\end{equation}
where $\MKu$ in given in J/m$^3$. In addition to the atomistic parameters, it is also worth noting the analogous expressions for the anisotropy field $H_{\mathrm{a}}$ for a single domain particle:
\begin{equation}
H_{\mathrm{a}} = \frac{2\MKu}{\MMs} = \frac{2\smKu}{\smmu}
\end{equation}
where symbols have their usual meaning. At this point it is worth mentioning that the measured anisotropy is a free energy difference. While the intrinsic $\smKu$ remains (to a first approximation) temperature independent, at a non-zero temperature the free energy in the easy/hard directions is increased/decreased due to the magnetisation fluctuations. Thus the macroscopic anisotropy value decreases with increasing temperature, vanishing at $\Tc$. The thermodynamic basis of atomistic models makes them highly suitable for the investigation of such phenomena, as we show later.


Applying the preceding operations, parameters for the key ferromagnetic elements are given in Tab.~\ref{tab:constants}.

\subsubsection*{Ferrimagnets and antiferromagnets}
In the case of ferrimagnets and anti-ferromagnets the above methods for anisotropy and moment determination do not work due to the lack of macroscopic measurements, although the estimated exchange energies apply equally well to the N\'eel temperature provided no magnetic frustration (due to lattice symmetry) is present. In general, other theoretical calculations or formalisms are required to determine parameters, such as mean-field approaches\cite{Ostler:2011jf} or density functional theory calculations\cite{Szunyogh:2009kf}.


\begin{table*}[!htcdp]
\caption{Table of derived constants for the ferromagnetic elements Fe, Co, Ni and Gd.}
\begin{center}
\begin{tabular}{ l c c c c c c}
\hline \hline
   & Fe & Co & Ni & Gd & Unit \\
\hline
  Crystal structure          & bcc & hcp & fcc & hcp & -\\
  Unit cell size $a$         & 2.866 & 2.507 & 3.524 & 3.636 &\AA \\
  Interatomic spacing $r_{ij}$& 2.480 & 2.507 & 2.492 & 3.636 &\AA \\
  Coordination number $z$    & 8   & 12  & 12  & 12 & -\\
  Curie Temperature $\Tc$    & 1043 & 1388 & 631 & 293 & K \\
  Spin-wave MF correction\cite{Garanin:351283,Wysin:2000uf} $\epsilon$ & 0.766 & 0.790 & 0.790 & 0.790 & -\\
  Atomic spin moment $\smmu$ & 2.22 & 1.72 & 0.606 & 7.63 &$\mu_{\mathrm{B}}$\\
  Exchange energy $\smJij$   & 7.050 $\times 10^{-21}$ & 6.064 $\times 10^{-21}$ & 2.757 $\times 10^{-21}$ & 1.280 $\times 10^{-21}$ & J/link \\
  Anisotropy energy\cite{ColarietiTosti:2003iw} $k$  & 5.65 $\times 10^{-25}$ & 6.69 $\times 10^{-24}$ & 5.47 $\times 10^{-26}$ & 5.93 $\times 10^{-24}$ & J/atom\\
\hline  \hline
\end{tabular}
\end{center}
\label{tab:constants}
\end{table*}

\subsection*{Atomistic system generation}
In addition to determining the parameters of the spin Hamiltonian, an essential part of the atomistic model is the determination of the nuclear, or atomic, positions in the system. In the multiscale approach utilising \textit{ab initio} parameterisation of the system, the spin Hamiltonian is intrinsically tied to the atomic positions. The additional detail offered by first principles calculations is highly desirable even for perfect crystals and atomically sharp interfaces, however the computational complexity of the calculations limits the ability to parameterise a spin Hamiltonian for systems of extended defects over 10 nm+ length scales, including physical effects such as vacancies, impurities, dislocations and even amorphous materials.

For systems modelled using the nearest neighbour approximation, the atomic structures are much less restricted, allowing for simulations of material defects such as interface roughness\cite{Evans:2011tb}and intermixing\cite{Ho:eb}, magnetic multilayers\cite{raduPRX2013,Fan:2011es}, disordered magnetic alloys\cite{Ostler:2011jf}, surface\cite{Yanes:2007gu} and finite size effects\cite{Hovorka:2012kx}. \vampire includes extensive functionality to generate such systems, the details of which are included in Appendix~\ref{app:system-generation}. In addition to crystallographic and molecular systems\cite{SchroderPRL2005,SchroderPRL2005-2} it is also possible to investigate magnetic effects in disordered materials and nanoparticles by incorporating the results of Molecular Dynamics simulations\cite{Evans:2006cf,Evans:2007tw,Dorfbauer:2007vp}.

\section{Integration methods}
Although the spin Hamiltonian describes the energetics of the magnetic system, it provides no information regarding its time evolution, thermal fluctuations, or the ability to determine the ground state for the system. In the following the commonly utilised integration methods for atomistic spin models are introduced.

\subsection*{Spin Dynamics}
The first understanding of spin dynamics came from ferromagnetic resonance experiments, where the time dependent behaviour of a magnetic materials is described by the equation derived by Landau and Lifshitz
\cite{LandauLifshitz1935}, and in the modern form given by:
\begin{equation}\label{eqn:LL}
\frac{\partial \mathbf{m}}{\partial t} =
-\gamma[\mathbf{m} \times
\mathbf{H} + \alpha \mathbf{m} \times
(\mathbf{m} \times \mathbf{H})]
\end{equation}
where $\mathbf{m}$ is a unit vector describing the direction of the sample magnetisation, $\mathbf{H}$ is the effective applied acting on the sample, $\gamma$ is the gyromagnetic ratio and $\alpha$ is a phenomenological damping constant which is a property of the material. The physical origin of the Landau-Lifshitz (LL) equation arises due to two distinct physical effects. The precession of the magnetisation (first term in Eq.~\ref{eqn:LL}) arises due to the quantum mechanical interaction of an atomic spin with an applied field. The relaxation of the magnetisation (second term in Eq.~\ref{eqn:LL}) is an elementary formulation of energy transfer representing the coupling of the magnetisation to a heat bath which aligns the magnetisation along the field direction with a characteristic coupling strength determined by $\alpha$. In the LL equation the relaxation rate of the magnetisation to the field direction is a linear function of the damping parameter, which was shown by Gilbert to yield incorrect dynamics for materials with high damping\cite{Gilbert1955}. Subsequently Gilbert introduced critical damping, with a maximum effective damping for $\alpha = 1$, to arrive at the Landau-Lifshitz-Gilbert (LLG) equation. Although initially derived to describe the dynamics of the macroscopic magnetisation of a sample, the LLG is the standard equation of motion used in numerical micromagnetics, describing the dynamics of small magnetic elements.

The same equation of motion can also be applied at the atomistic level. The precession term arises quantum mechanically for atomic spins and the relaxation term now describes direct angular momentum transfer between the spins and the heat bath, which includes contributions from  the lattice\cite{Karakurt:2007cz} and the electrons\cite{Fahnle:2010ev}. A distinction between the macroscopic LLG and the atomistic LLG now appears in terms of the effects included within the damping parameter. In the macroscopic LLG, $\alpha$ includes all contributions, both intrinsic (such as spin-lattice and spin-electron interactions) and extrinsic (spin-spin interactions arising from demagnetisation fields, surface defects\cite{Dobin:2004vq}, doping\cite{Ellis:2012bb} and temperature\cite{Garanin:1997uf}), while the atomistic LLG only includes the local intrinsic contributions. To distinguish the different definitions of damping we therefore introduce a microscopic damping parameter $\lambda$. Although the form of the LLG is identical for atomistic and macroscopic length scales, the microstructural detail in the atomistic model allows for calculations of the effective damping including extrinsic effects, such as rare-earth doping\cite{Ellis:2012bb}. Including a microscopic damping $\lambda$ the atomistic Landau-Lifshitz-Gilbert equation is given by
\begin{equation}\label{eqn:LLG}
\frac{\partial \sms_i }{\partial t} =
-\frac{\gamma}{(1+\lambda^{2})}[\sms_i \times
\smH^i_{\mathrm{eff}} + \lambda \sms_i \times
(\sms_i \times \smH^i_{\mathrm{eff}})]
\end{equation}
where $\sms_i$ is a unit vector representing the direction of the magnetic spin moment of site $i$, $\gamma$ is the gyromagnetic ratio and $\smH^i_{\mathrm{eff}}$ is the net magnetic field on each spin. The atomistic LLG equation describes the interaction of an atomic spin moment $i$ with an effective magnetic field, which is obtained from the negative first derivative of the complete spin Hamiltonian, such that:
\begin{equation}\label{eqn:Heff}
  \smH^i_{\mathrm{eff}} = -\frac{1}{\smmu}\frac{\partial \mathcal{H}}{\partial \sms_i}
\end{equation}
where $\smmu$ is the local spin moment. The inclusion of the spin moment within the effective field is significant, in that the field is then expressed in units of Tesla, given a Hamiltonian in Joules. Given typical energies in the Hamiltonian of 10 $\mu$eV - 100 meV range. This gives fields typically in the range 0.1 - 1000 Tesla, given a spin moment of the same order as the Bohr magneton ($\muB$).

\subsubsection*{Langevin Dynamics}
In its standard form the LLG equation is strictly only applicable to simulations at zero temperature. Thermal effects cause thermodynamic fluctuations of the spin moments which at sufficiently high temperatures are stronger than the exchange interaction, giving rise to the ferromagnetic-paramagnetic transition. The effects of temperature can be taken into account by using Langevin Dynamics, an approach developed by Brown \cite{WFBrown1979}. The basic idea behind Langevin Dynamics is to assume that the thermal fluctuations on each atomic site can be represented by a Gaussian white noise term. As the temperature is increased, the width of the Gaussian distribution increases, thus representing stronger thermal fluctuations. The established Langevin Dynamics method is widely used
for spin dynamics simulations and incorporates an effective thermal field
into the LLG equation to simulate thermal effects \cite{GarciaPalacios:1998wz,Lyberatos:1999da,Nowak:2005eg}. The thermal fluctuations are represented by a gaussian distribution $\boldsymbol{\Gamma}(t)$ in three dimensions with a mean of zero. At each time step the instantaneous thermal field on each spin $i$ is given by:
\begin{equation}
\smH^{i}_{\mathrm{th}} = \boldsymbol{\Gamma}(t) \sqrt{\frac{2 \lambda \kB T}{\gamma \smmu \Delta t}}
\end{equation}
where  $\kB$ is the Boltzmann constant, $T$ is the system
temperature, $\lambda$ is the Gilbert damping parameter, $\gamma$
is the absolute value of the gyromagnetic ratio, $\smmu$ is the
magnitude of the atomic magnetic moment, and $\Delta t$ is the integration time step. The effective field for application in the LLG equation with Langevin Dynamics then reads:
\begin{equation}\label{eqn:Heffth}
  \smH^i_{\mathrm{eff}} = -\frac{1}{\smmu}\frac{\partial \mathcal{H}}{\partial \sms_i} + \smH^{i}_{\mathrm{th}}\mathrm{.}
\end{equation}

Given that for each time step three Gaussian distributed random numbers are required for every spin, efficient generation of such numbers is essential. \vampire makes use the Mersenne Twister\cite{Matsumoto:1998jt} uniform random number generator and the Ziggurat method\cite{Marsaglia:2000wj} for generating the Gaussian distribution.

It is useful at the this point to address the applicability of the atomistic LLG to slow and fast problems respectively. In reality the thermal and magnetic fluctuations are correlated at the atomic level, arising from the dynamic interactions between the atoms and lattice/electron system. These correlations may be important in terms of the thermal fluctuations experienced by the atomistic spins. In the conventional Langevin dynamics approach described above, the noise term is completely uncorrelated in time and space. For short timescales however, the thermal fluctuations are correlated in time, and so the noise is \textit{Coloured}\cite{Atxitia:2009io}. The effect of the Coloured noise is to lessen the effect of sudden temperature changes on the magnetisation dynamics. However, the existence of ultrafast magnetisation dynamics\cite{Beaurepaire:1996es,Radu:2011krb}, and that it is driven by a thermal rather than quantum mechanical process\cite{Schellekens:2013ci}, requires that the effective correlation time is short, with an upper bound of around 1 fs. Given that the correlation time is close to the integration timestep, the applicability of the LLG to problems with timescales $\geq 1$ fs is sound. There will be a point however where the LLG is no longer valid, where direct simulation of the microscopic damping mechanisms will be necessary. Progress has been made in linking molecular dynamics and spin models \cite{Grossmann:1996vi,Karakurt:2007cz,Ma:2012hp} which enables the simulation of spin-lattice interactions, which is particularly relevant for slow problems, such as conventional magnetic recording where switching occurs over ns timescales. However, it is also essential to consider spin-electron effects\cite{Fahnle:2010ev,Battiato:2010br} necessary for ultrafast demagnetisation processes, although the physical origins are still currently debated\cite{Schellekens:2013em}.

\begin{figure*}[!htb]
\center
\includegraphics[width=8cm, trim=10 0 10 0]{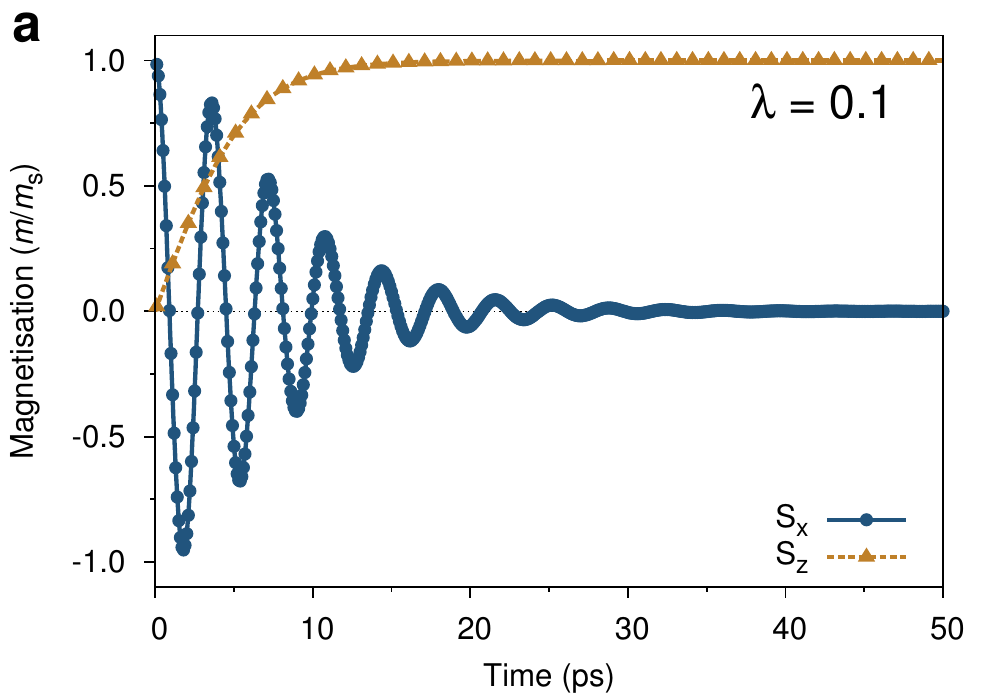}
\hspace{0.5cm}
\includegraphics[width=8cm, trim=10 0 10 0]{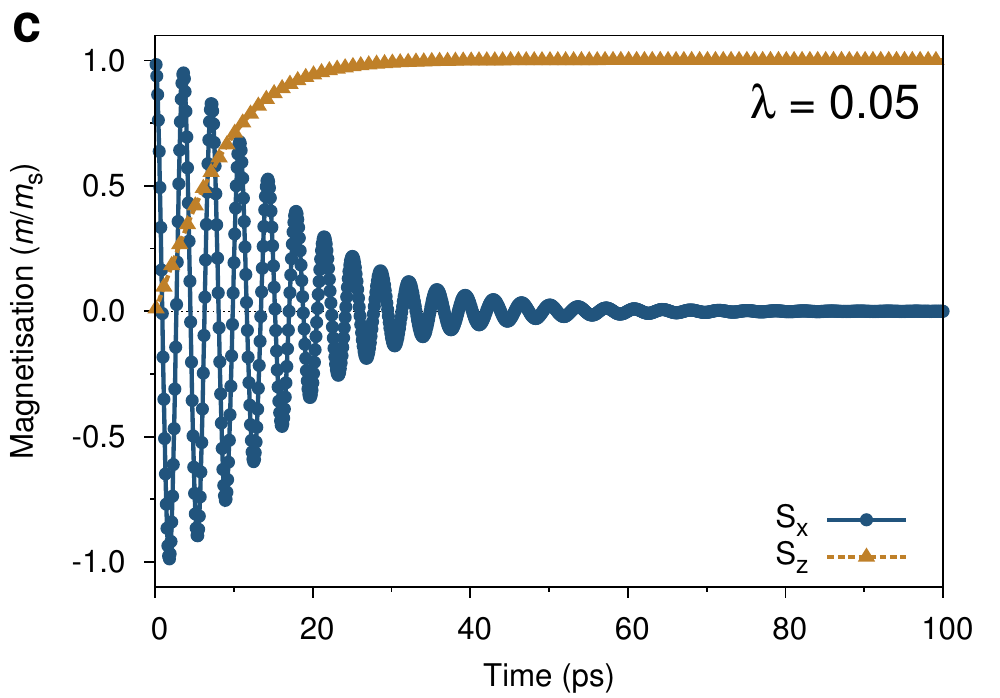}
\includegraphics[width=8cm, trim=10 0 10 0]{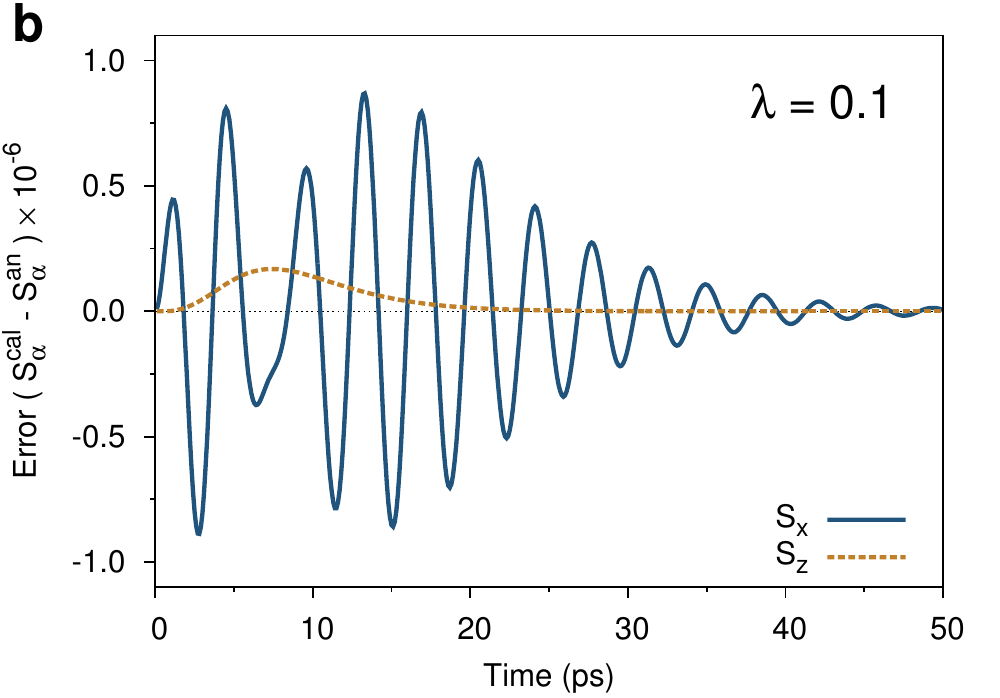}
\hspace{0.5cm}
\includegraphics[width=8cm, trim=10 0 10 0]{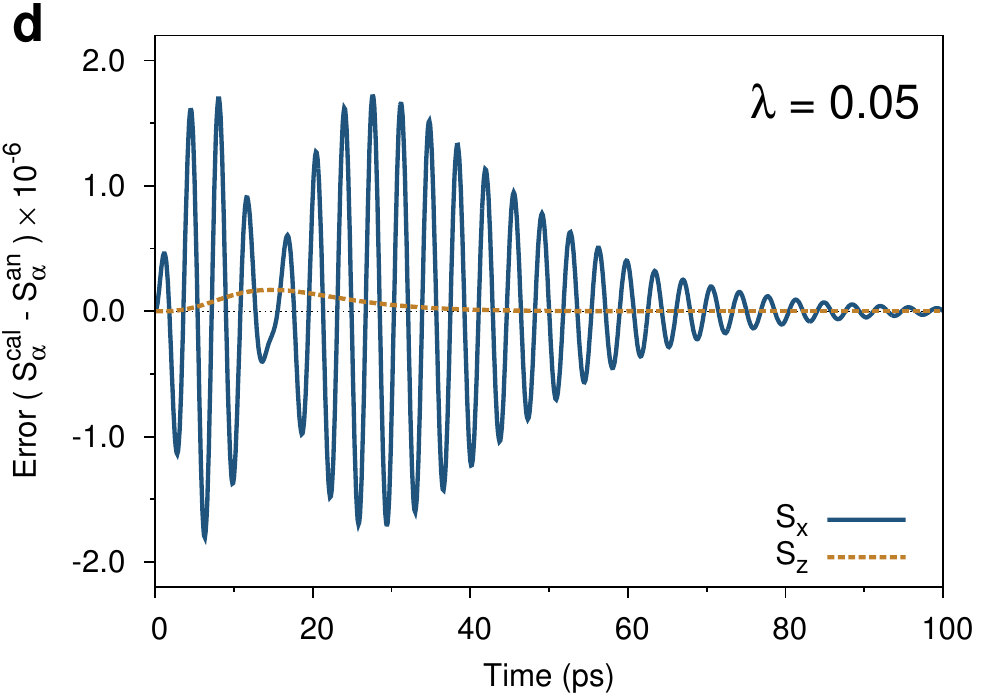}
\caption{Time evolution of a single isolated spin in an applied field of 10T and time step of 1 fs. magnetisation traces \textbf{a} and \textbf{c} show relaxation of the magnetisation to the $z$-direction and precession of the $x$ component (the $y$-component is omitted for clarity) for damping constants $\lambda = 0.1$ and $\lambda = 0.05$ respectively. The points are the result of direction integration of the LLG and the lines are the analytical solution plotted according to Eq.~\ref{eq:analytic_LLG}. Panels \textbf{b} and \textbf{d} show the corresponding error traces (difference between the expected and calculated spin components) for the two damping constants for \textbf{a} and \textbf{c} respectively. For $\lambda=0.1$ the error is below $10^{-6}$, while for lower damping the numerical error increases significantly due to the increased number of precessions, highlighting the damping dependence of the integration time step. (Colour online).}
\label{fig:heuntest}
\end{figure*}

\subsubsection*{Time Integration of the LLG Equation}
In order to determine the time evolution of a system of spins, it is necessary to solve the stochastic LLG equation, as given by Eqs.~\ref{eqn:LLG} and \ref{eqn:Heffth}, numerically. The choice of solver is limited due to the stochastic nature of the equations. Specifically, it is necessary to ensure convergence to the Stratonovich solution. This has been considered in detail by Garcia-Palacios\cite{GarciaPalacios:1998wz}, but the essential requirement \cite{Berkov2002} is that the solver enforces the conservation of the magnitude of the spin, either implicitly or by renormalisation. The most primitive integration scheme uses Euler's method, which assumes a linear change in the spin direction in a single discretised time step, $\Delta t$. An improved integration scheme, known as the Heun method \cite{GarciaPalacios:1998wz} is commonly used, which allows the use of larger time steps because of its use of a predictor-corrector algorithm. Other more advanced integration schemes have been suggested, such as the midpoint method rule\cite{dAquino:2005fx} and modified predictor-corrector midpoint schemes\cite{Ellis:2012bb,Mentink:2010vk}. The principal advantage of the midpoint scheme is that the length of the spin vector is preserved during the integration which allows larger time steps to be used. However for the midpoint scheme the significant increase in computational complexity outweighs the benefits of larger time steps\cite{Mentink:2010vk}. Modified predictor-corrector schemes\cite{Mentink:2010vk,Ellis:2012bb} reduce the computational complexity of the midpoint scheme, but with a loss of accuracy, particularly in the time-dependent dynamics\cite{Ellis:2012bb}. For valid integration of the stochastic LLG equation it is also necessary for the numerical scheme to converge to the Stratonovich solution\cite{GarciaPalacios:1998wz,dAquino:2006kj}. Although proven for the midpoint and Heun numerical schemes, the validity of the predictor-corrector midpoint schemes for the stochastic LLG have yet to be confirmed. On balance the Heun scheme, despite its relative simplicity, is sufficiently computationally efficient that it is still the most widely used integration scheme for stochastic magnetisation dynamics, and so we proceed to describe its implementation in detail.

In the Heun method, the first (predictor) step calculates the new spin direction, $\sms_i'$, for a given effective field $\smH_{\mathrm{eff}}^i$ by performing a standard Euler integration step, given by:
\begin{equation}
\sms_i'= \sms_i + \Delta \sms \Delta t
\end{equation}
where
\begin{equation}
\Delta \sms = - \frac{\gamma}{(1+\lambda^{2})}[\sms_i
\times \smH_{\mathrm{eff}}^i + \lambda \sms_i \times
(\sms_i \times \smH_{\mathrm{eff}}^i)]\mathrm{.}
\end{equation}
The Heun scheme does not preserve the spin length and so it is essential to renormalise the spin unit vector length $\sms_i$ after both the predictor and corrector steps to ensure numerical stability and convergence to the Stratanovich solution. After the first step the effective field must be re-evaluated as the intermediate spin positions have now changed. It should be noted that the random thermal field does not change between steps. The second (corrector) step then uses the predicted spin position and revised effective field $\smH_{\mathrm{eff}}^{i'}$ to calculate the final spin position, resulting in a complete integration step given by:

\begin{equation}
\sms_i^{t + \Delta t} = \sms_i +
\frac{1}{2}\left[\Delta \sms + \Delta
\sms' \right]\Delta t
\end{equation}

where

\begin{equation}
\Delta \sms' =
-\frac{\gamma}{(1+\lambda^{2})}[\sms_i' \times
\smH^{i'}_{\mathrm{eff}} + \lambda \sms_i' \times
(\sms_i' \times \smH^{i'}_{\mathrm{eff}})]\mathrm{.}
\end{equation}

The predictor step of the integration is performed on every spin in the system before proceeding to evaluate the corrector step for every spin. This is then repeated many times so that the time evolution of the system can be simulated. Although the Heun scheme was derived specifically for a stochastic equation with multiplicative noise, in the absence of the noise term the Heun method reduces to a standard second order Runge-Kutta method\cite{HeunzeroK}. In order to test the implementation of the Heun integration scheme, it is possible to compare the calculated result with the analytical solution for the LLG. For the simple case of a single spin initially along the $x$-axis in an applied field along the $z$-axis, the time evolution\cite{Hannay} is given by:
\begin{eqnarray}
S_{x} (t) &=& \sech \left(\frac{\lambda \gamma H}{1+\lambda^2}t\right)\cos\left(\frac{\gamma H}{1+\lambda^2}t\right) \nonumber \\
S_{y} (t) &=& \sech \left(\frac{\lambda \gamma H}{1+\lambda^2}t\right)\sin\left(\frac{\gamma H}{1+\lambda^2}t\right) \nonumber \\
S_{z} (t) &=& \tanh \left(\frac{\lambda \gamma H}{1+\lambda^2}t\right)\mathrm{.}
\label{eq:analytic_LLG}
\end{eqnarray}
The expected and simulated time evolution for a single spin with $H$ = 10T, $\Delta t = 1 \times 10^{-15}$ s and $\lambda = 0.1, 0.05$ is plotted in Fig~\ref{fig:heuntest}. Superficially the simulated and expected time evolution agree very well, with errors around $10^{-6}$. The error gives a characteristic trace the size and shape of which is indicative of a correct implementation of the Heun integration scheme.

\begin{figure}[!t]
\center
\includegraphics[width=8cm, trim=10 00 10 0]{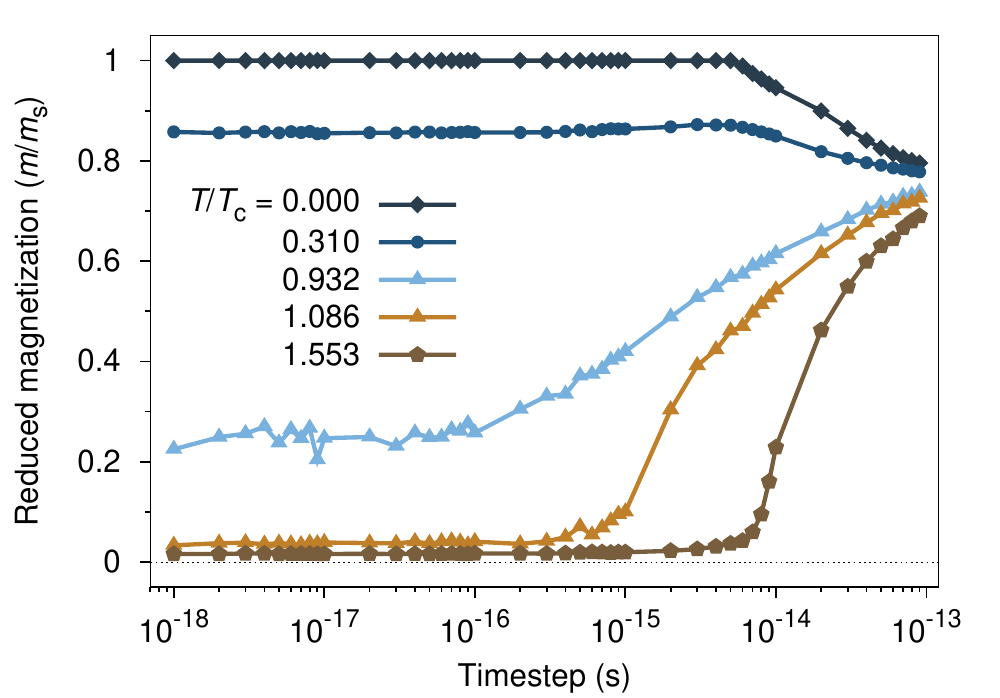}
\caption{Time step dependence of the mean magnetisation for different reduced temperatures for the heun integration scheme. Low $(T << \Tc)$ and high $(T >> \Tc)$ temperatures integrate accurately with a 1fs timestep, but in the vicinity of $\Tc$ a timestep of around $10^{-16}$ is required for this system. (Colour online.)}
\label{fig:dttest}
\end{figure}

Ideally one would like to use the largest time step possible so as to simulate systems for the longest time. For micromagnetic simulations at zero temperature, the time step is a well defined quantity since the largest field (usually the exchange term) essentially defines the precession frequency. However, for atomistic simulations using the stochastic LLG equation with Langevin dynamics, the effective field becomes temperature dependent. The consequence of this is that for atomistic models the most difficult region to integrate is in the immediate vicinity of the Curie point. Errors in the integration of the system will be apparent from a non-converged value for the average magnetisation. This gives a relatively simple case which can then be used to test the stability of integration schemes for the stochastic LLG model. A plot of the mean magnetisation as a function of temperature is shown in Fig.~\ref{fig:dttest} for a representative system consisting of $22 \times 22 \times 22$ unit cells with generic material parameters of FePt with an fcc crystal structure, nearest neighbour exchange interaction of $J_{ij} = 3.0 \times 10^{-21}$ J/link and uniaxial anisotropy of $1.0 \times 10^{-23}$ J/atom. The system is first equilibrated for 10 ps at each temperature and then the mean magnetisation is calculated over a further 10 ps.

First, comparing the effect of temperature on the minimum allowable time step, the data show that for low temperatures reasonably large time steps of $1 \times 10^{-15}$ give the correct solution of the LLG equations, while near the Curie point (690 K) the deviations from the correct equilibrium value are significant. Consequently for simulations studying high temperature reversal processes time steps of $1 \times 10^{-16}$ s are necessary. It should be noted that the time steps which can be used are material dependent - specifically if a material with higher Curie temperature is used then the usable time steps will be correspondingly lower due to the increased exchange field.

From a practical perspective a significant advantage of the spin dynamics method is the ability to parallelise the integration system by domain decomposition, details of which are given in Appendix~\ref{parallelisation}. This allows the efficient simulation of relatively large systems consisting of tens or hundreds of grains or nano structures with dimensions greater than 100 nm for nanosecond timescales, with typical numbers of spins in the range $10^6-10^8$.

\subsection*{Monte Carlo Methods}
While spin dynamics are particularly useful for obtaining dynamic information about the magnetic properties or reversal processes for a system, they are often not the optimal method for determining the equilibrium properties, for example the temperature dependent magnetisation. The Monte Carlo Metropolis algorithm\cite{Metropolis:1953in} provides a natural way to simulate temperature effects where dynamics are not required due to the rapid convergence to equilibrium and relative ease of implementation.

The Monte Carlo metropolis algorithm for a classical spin system proceeds as follows. First a random spin $i$ is picked and its initial spin direction $\sms_i$ is changed randomly to a new trial position $\sms_i'$, a so-called trial move. The change in energy $\Delta E = E(\sms_i') - E(\sms_i)$ between the old and new positions is then evaluated, and the trial move is then accepted with probability
\begin{equation}
    P = \mathrm{exp} \left({-\frac{\Delta E}{\kB T}}\right)\mathrm{.}
\label{eq:AMCP}
\end{equation}
by comparison with a uniform random number in the range 0-1. Probabilities greater than 1, corresponding with a reduction in energy, are accepted unconditionally. This procedure is then repeated until $N$ trial moves have been attempted, where $N$ is the number of spins in the complete system. Each set of $N$ trial moves comprises a single Monte Carlo step.

The nature of the trial move is important due to two requirements of any Monte Carlo algorithm: ergodicity and reversibility. Ergodicity expresses the requirement that all possible states of the system are accessible, while reversibility requires that the transition probability between two states is invariant, explicitly $P(\sms_i \rightarrow \sms_i') = P(\sms_i' \rightarrow \sms_i)$. From Eq.~\ref{eq:AMCP} reversibility is obvious since the probability of a spin change depends only on the initial and final energy. Ergodicity is easy to satisfy by moving the selected spin to a random position on the unit sphere, however this has an undesirable consequence at low temperatures since large deviations of spins from the collinear direction are highly improbable due to the strength of the exchange interaction. Thus at low temperatures a series of trial moves on the unit sphere will lead to most moves being rejected. Ideally a move acceptance rate of around 50\% is desired, since very high and very low rates require significantly more Monte Carlo steps to reach a state representative of true thermal equilibrium.

\begin{figure}[!tb]
\begin{center}
\includegraphics[width=8cm]{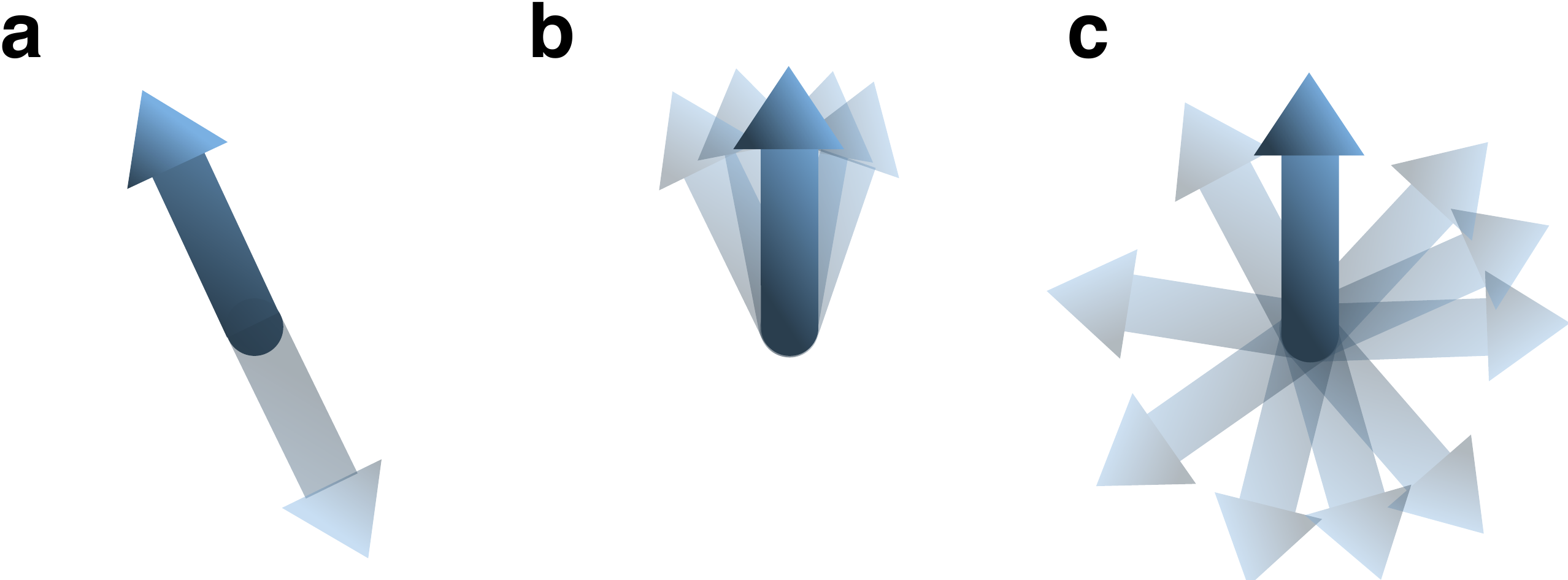}
\caption[]{Schematic showing the three principal Monte Carlo moves: (\textbf{a}) spin flip; (\textbf{b}) gaussian; and (\textbf{a}) random. (Colour online.)}      \label{fig:MCschematic}
\end{center}
\end{figure}

One of the most efficient Monte Carlo algorithms for classical spin models was developed by Hinzke and Nowak\cite{Hinzke:1999ud}, involving a combinational approach using a mixture of different trial moves. The principal advantage of this method is the efficient sampling of all available phase space while maintaining a reasonable trial move acceptance rate. The Hinzke-Nowak method
utilises three distinct types of move: spin-flip, Gaussian and random, as illustrated schematically in Fig.~\ref{fig:MCschematic}.

The spin-flip move simply reverses the direction of the spin such that $\sms_i' = - \sms_i$ to explicitly allow the nucleation of a switching event. The spin flip move is identical to a move in Ising spin models. It should be noted that spin flip moves do not by themselves satisfy ergodicity in the classical spin model, since states perpendicular to the initial spin direction are inaccessible. However, when used in combination with other ergodic trial moves this is quite permissible. The Gaussian trial move takes the initial spin direction and moves the spin to a point on the unit sphere in the vicinity of the initial position according to the expression
\begin{equation}
\sms'_i = \frac{\sms_i + \sigma_{\mathrm{g}}\boldsymbol{\Gamma}}{|\sms_i + \sigma_{\mathrm{g}}\boldsymbol{\Gamma}|}
\label{eq:MCgaussian}
\end{equation}
where $\boldsymbol{\Gamma}$ is a Gaussian distributed random number and $\sigma_{\mathrm{g}}$ is the width of a cone around the initial spin $\sms_i$. After generating the trial position $\sms_i'$ the position is normalised to yield a spin of unit length. The choice of a Gaussian distribution is deliberate since after normalisation the trial moves have a uniform sampling over the cone. The width of the cone is generally chosen to be temperature dependent and of the form
\begin{equation}
\sigma_{\mathrm{g}} = \frac{2}{25}\left(\frac{\kB T}{\muB}\right)^{1/5}\mathrm{.}
\label{eq:MCgaussianSigma}
\end{equation}
The Gaussian trial move thus favors small angular changes in the spin direction at low temperatures, giving a good acceptance probability for most temperatures.

The final random trial move picks a random point on the unit sphere according to
\begin{equation}
\sms'_i = \frac{\boldsymbol{\Gamma}}{|\boldsymbol{\Gamma}|}
\label{eq:MCrandom}
\end{equation}
which ensures ergodicity for the complete algorithm and ensures efficient sampling of the phase space at high temperatures. For each trial step one of these three trial moves is picked randomly, which in general leads to good algorithmic properties.

\begin{figure}[!t]
\center
\includegraphics[width=8cm]{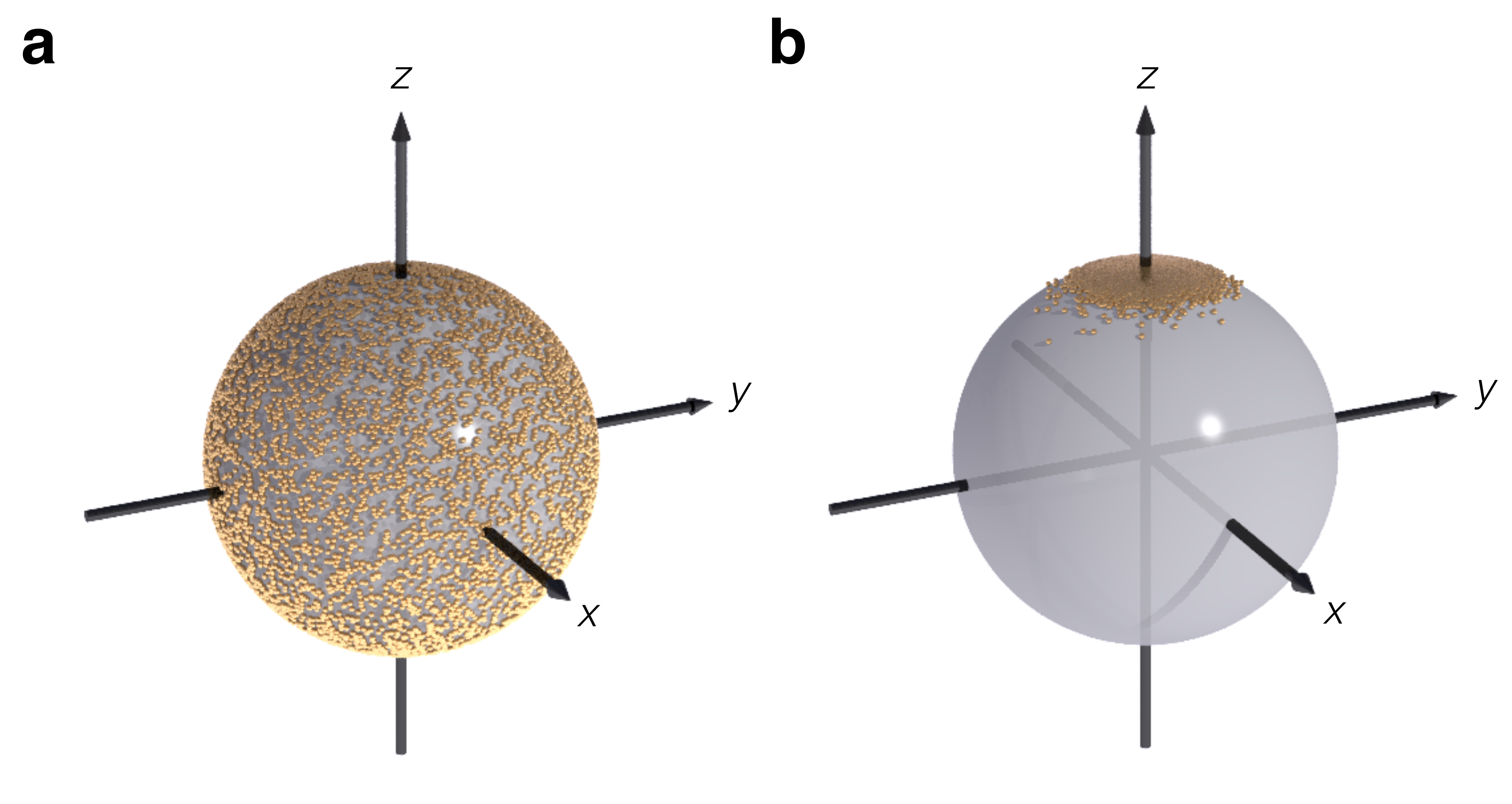}
\caption{Visualisation of Montecarlo sampling on the unit sphere for (\textbf{a}) random and (\textbf{b}) Gaussian sampling algorithms at $T = 10$ K. The dots indicate the trial moves. The random algorithm shows a uniform distribution on the unit sphere, and no preferential biasing along the axes. The Gaussian trial moves are clustered around the initial spin position, along the $z$-axis. (Colour online.)}
\label{fig:mc-sampling}
\end{figure}

To verify that the random sampling and Gaussian trial moves give the expected behaviour, a plot of the calculated trial moves on the unit sphere for the different algorithms is shown in Fig.~\ref{fig:mc-sampling}. The important points are that the random trial move is uniform on the unit sphere, and that the Gaussian trial move is close to the initial spin direction, along the $z$-axis in this case.

\begin{figure}[!t]
\begin{center}
\includegraphics[width=8cm,trim= 15 10 15 0]{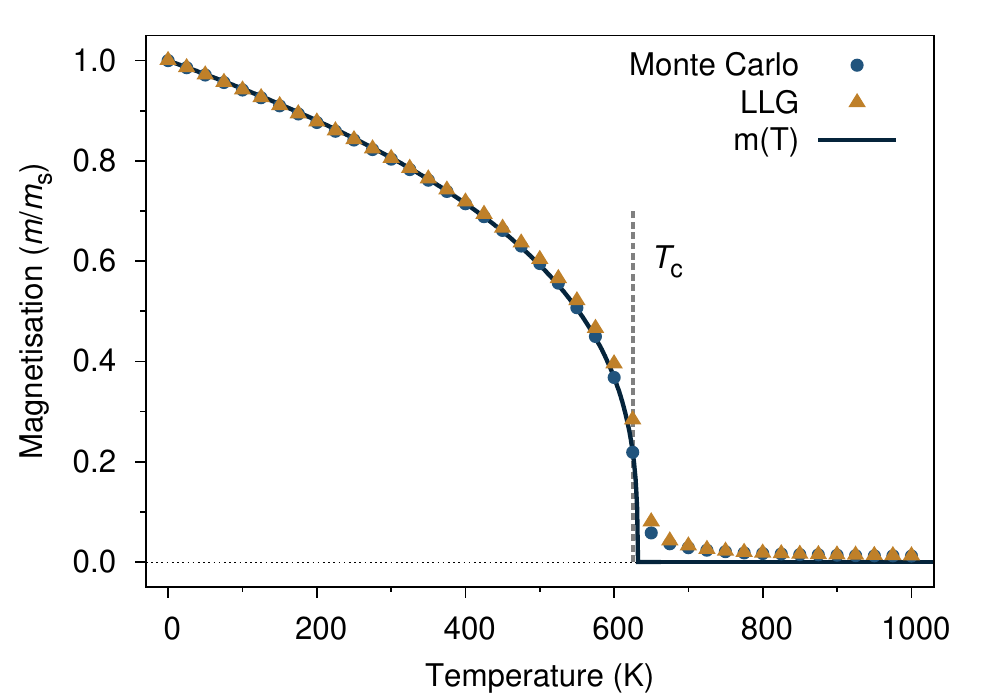}
\caption[]{Comparative simulation of temperature dependent magnetisation for Monte Carlo and LLG simulations. Simulation parameters assume a nearest neighbour exchange of $6.0 \times 10^{-21}$ J/link with a simple cubic crystal structure, periodic boundary conditions and 21952 atoms. The Monte Carlo simulations use 50,000 equilibration and averaging steps, while the LLG simulations use 5,000,000 equilibration and averaging steps with critical damping $(\lambda=1)$ and a time step of 0.01 fs. The value of $\Tc \sim 625K$ calculated from Eq.~\ref{eq:JijTc} is shown by the dashed vertical line. The temperature-dependent magnetisation is fitted to the expression $m(T) = (1-T/\Tc)^{\beta}$ (shown by the solid line) which yields a fitted $\Tc=631.82$ K and and exponent $\beta=0.334297$. (Colour online.)}
\label{fig:MCLLG}
\end{center}
\end{figure}

At this point it is worthwhile considering the relative efficiencies of Monte Carlo and spin dynamics for calculating equilibrium properties. Fig.~\ref{fig:MCLLG} shows the simulated temperature dependent magnetisation for a test system using both LLG spin dynamics and Monte Carlo methods. Agreement between the two methods is good, but the spin dynamics simulation takes around twenty times longer to compute due to the requirements of a low time step and slower convergence to equilibrium. However, Monte Carlo algorithms are notoriously difficult to parallelise, and so for larger systems LLG spin dynamic simulations are generally more efficient than Monte Carlo methods.

\section{Test Simulations}
Having outlined the important theoretical and computational methods for the atomistic simulation of magnetic materials, we now proceed to detail the tests we have refined to ensure the correct implementation of the main components of the model. Such tests are particularly helpful to those wishing to implement these methods. Similar tests developed for micromagnetic packages\cite{mmagstandardproblems} have proven an essential benchmark for the implementation of improved algorithms and codes with different capabilities but the same core functionality.

\subsection*{Angular variation of the Coercivity}
Assuming a correct implementation of an integration scheme as described in the previous section, the first test case of interest is the correct implementation of uniaxial magnetic anisotropy. For a single spin in an applied field and at zero temperature, the behaviour of the magnetisation is essentially that of a Stoner-Wohlfarth particle, where the angular variation of the coercivity, or reversing field, is well known\cite{Stoner:1948ur}. This test serves to verify the static solution for the LLG equation by ensuring an easy axis loop gives a coercivity of $H_k = 2 \smKu / \smmu$ as expected analytically. For this problem the Hamiltonian reads
\begin{equation}
    \mathcal{H} = -\smKu S_z^2 -\smmu \sms \cdot \smH_{\mathrm{app}}
\end{equation}
where $\smKu$ is the on-site uniaxial anisotropy constant and $\smH_{\mathrm{app}}$ is the external applied field. The spin is initialised pointing along the applied field direction, and then the LLG equation is solved for the system, until the net torque on the system $\sms \times \smH_{\mathrm{eff}} \leq 10^{-6}$ T, essentially a condition of \textit{local} minimum energy.

\begin{figure}[!tb]
   \begin{center}
      \includegraphics[width=8cm, trim= 15 10 15 0]{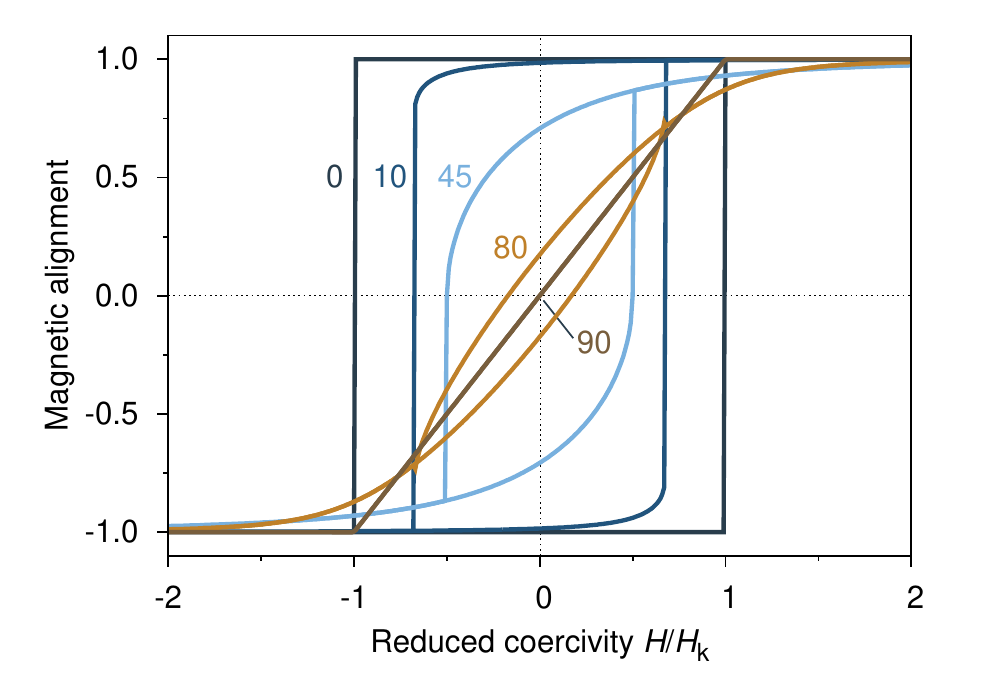}
      \caption[]{Plot of alignment of magnetisation with the applied field for different angles of from the easy axis. The $0^{\circ}$ and $90^{\circ}$ loops were calculated for very small angles from the easy and hard axes respectively, since in the perfectly aligned case the net torque is zero and no change of the spin direction occurs. (Colour online.)}
      \label{fig:mdoth}
   \end{center}
\end{figure}

The field strength is decreased from saturation in steps of 0.01 $H/H_k$ and solved again until the same condition is reached. A plot of the calculated alignment of the magnetisation to the applied field ($\sms \cdot \smH_{\mathrm{app}}$) for different angles from the easy axis is shown in Fig.~\ref{fig:mdoth}. The calculated hysteresis curve conforms exactly to the Stoner-Wohlfarth solution.

\subsection*{Boltzmann distribution for a single spin}
\begin{figure}[!tb]
   \begin{center}
      \includegraphics[width=8cm, trim= 15 10 15 0]{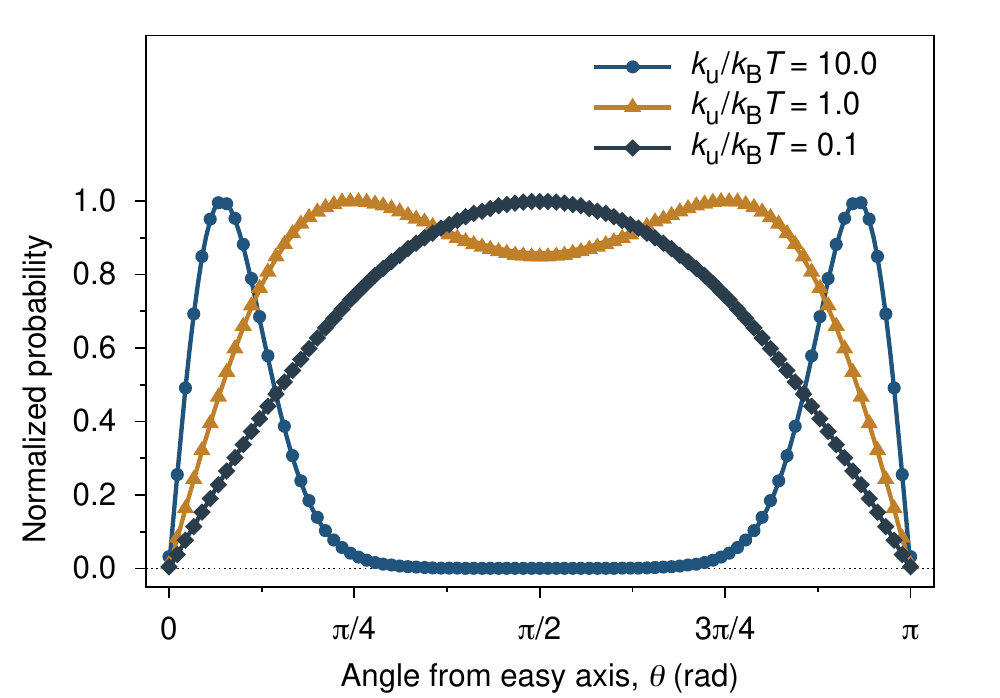}
      \caption[]{Calculated angular probability distribution for a single spin with anisotropy for different effective temperatures $\smKu/\kB T$. The lines show the analytic solution given by Eq.~\ref{eqn:boltz}. (Colour online.)}
      \label{fig:boltz}
   \end{center}
\end{figure}

To quantitatively test the thermal effects in the model and the correct implementation of the stochastic LLG or Monte Carlo integrators, the simplest case is that of the Boltzmann distribution for a single spin with anisotropy (or applied field), where the probability distribution is characteristic of the temperature and the anisotropy energy. The Boltzmann distribution is given by:
\begin{equation}\label{eqn:boltz}
    P(\theta) \propto \sin \theta \exp\left(-\frac{\smKu \sin^2\theta}{k_{\mathrm{B}} T}\right)
\end{equation}
where $\theta$ is the angle from the easy axis. The spin is initialised along the easy axis direction and the system is allowed to evolve for $10^8$ time steps after equilibration, recording the angle of the spin to the easy axis at each time. Since the anisotropy energy is symmetric along the easy axis, the probability distribution is reflected and summed about $\pi/2$, since at low temperatures the spin is confined to the upper well ($\theta < \pi/2$). Fig.~\ref{fig:boltz} shows the normalised probability distribution for three reduced temperatures.

The agreement between the calculated distributions is excellent, indicating a correct implementation of the stochastic LLG equation.

\subsection*{Curie temperature}
Tests such as the Stoner-Wohlfarth hysteresis or Boltzmann distribution are helpful in verifying the mechanical implementation of an algorithm for a single spin, but interacting systems of spins present a significant challenge in that no analytical solutions exist. Hence it is necessary to calculate some well defined macroscopic property which ensures the correct implementation of interactions in a system. The Curie temperature $T_\mathrm{c}$ of a nanoparticle is primarily determined by the strength of the exchange interaction between spins and so makes an ideal test of the exchange interaction. As discussed previously the bulk Curie temperature is related to the exchange coupling by the mean field expression given in Eq.~\ref{eq:JijTc}. However, for nanoparticles with a reduction in coordination number at the surface and a finite number of spins, the Curie temperature and \textit{criticality} of the temperature dependent magnetisation will vary significantly with varying size\cite{Hovorka:2012kx}.

\begin{figure}[!tb]
\includegraphics[width=8cm, trim= 15 10 15 0]{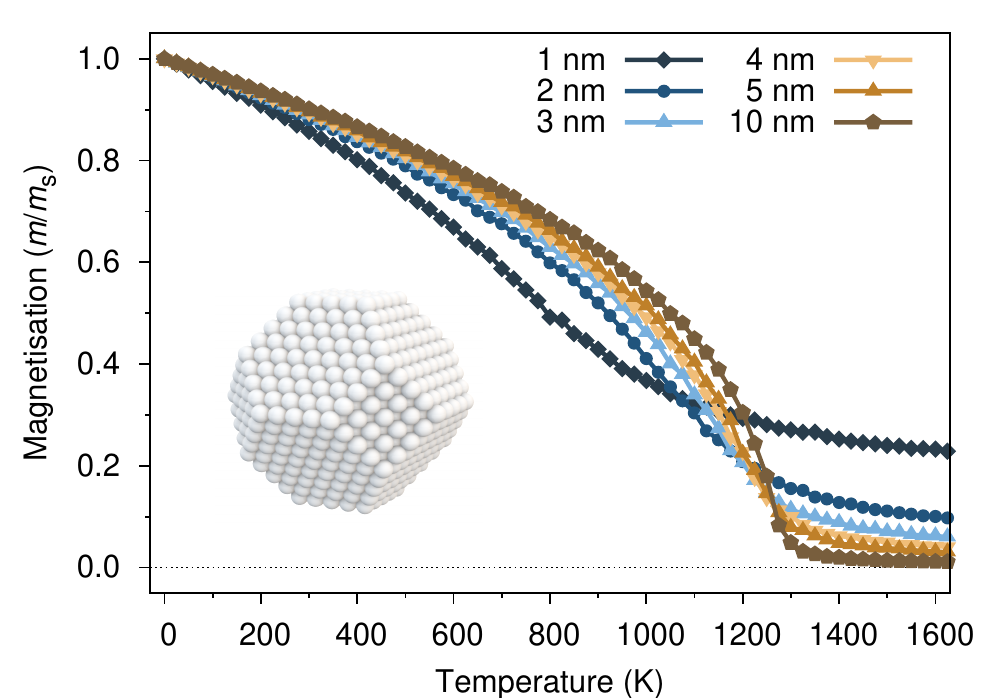}
\caption{Calculated temperature dependent magnetisation and Curie temperature for truncated octahedron nanoparticles with different size. A visualisation of a 3nm diameter particle is inset. (Colour online.)}
\label{fig:Tc-size}
\end{figure}

To investigate the effects of finite size and reduction in surface coordination on the Curie temperature, the equilibrium magnetisation for different sizes of truncated octahedron nanoparticles was calculated as a function of temperature. The Hamiltonian for the simulated system is
\begin{equation}
    \mathcal{H} = -\sum_{i\ne j} \smJij \sms_i \cdot \sms_j
\end{equation}
where $\smJij = 5.6 \times 10^{-21}$ J/link, and the crystal structure is face-centred-cubic, which is believed to be representative of Cobalt nanoparticles. Given the relative strength of the exchange interaction, anisotropy generally has a negligible impact on the Curie temperature of a material, and so the omission of anisotropy from the Hamiltonian is purely for simplicity. The system is simulated using the Monte Carlo method with 10,000 equilibration and 20,000 averaging steps. The system is heated sequentially in 10K steps, with the final state of the previous temperature taken as the starting point of the next temperature to minimise the number of steps required to reach thermal equilibrium. The mean temperature dependent magnetisation for different particle sizes is plotted in Fig.~\ref{fig:Tc-size}.

From Eq.~\ref{eq:JijTc} the expected Curie temperature is 1282K, which is in agreement with the results for the 10 nm diameter nanoparticle. For smaller particle sizes the magnetic behaviour close to the Curie temperature loses its criticality, making $T_C$  difficult to determine. Traditionally the Curie point is taken as the maximum of the gradient $dm/dT$\cite{Hovorka:2012kx}, however this significantly underestimates the actual temperature at which magnetic order is lost (which is, by definition, the Curie temperature). Other estimates of the Curie point such as the divergence in the susceptibility are probably a better estimate for finite systems, but this is beyond the scope of the present article. Another effect visible for very small particle sizes is the appearance of a magnetisation above the Curie point, an effect first reported by Binder\cite{BinderPLA1969}. This arises from local moment correlations which exist above $T_C$. It is an effect only observable in nanoparticles where the system size is close to the magnetic correlation length.  

\subsection*{Demagnetising fields}
For systems larger than the single domain limit\cite{Kazantseva:2008ja} and systems which have one dimension significantly different from another, the demagnetisation field can have a dominant effect on the macroscopic magnetic properties. In micromagnetic formalisms implemented in software packages such as \textsc{oommf}\cite{oommf}, \textsc{magpar}\cite{Scholz:2003ty} and \textsc{nmag}\cite{Fischbacher:ii}, the calculation of the demagnetisation fields is calculated accurately due to the routine simulation of large systems where such fields dominate. Due to the long-ranged interaction the calculation of the demagnetisation field generally dominates the compute time and so computational methods such as the fast-fourier-transform\cite{Yuan:1992ue,Berkov:1993th} and multipole expansion\cite{Kurzak:2005ur} have been developed to accelerate their calculation.

In large-scale atomistic calculations, it is generally sufficient to adopt a micromagnetic discretisation for the demagnetisation fields, since they only have a significant effect on nanometer length scales\cite{Boerner2005}. Additionally due to the generally slow variation of magnetisation, the timescales associated with the changes in the demagnetisation field are typically much longer than the time step for atomistic spins. Here we present a modified finite difference scheme for calculating the demagnetisation fields, described as follows.

\begin{figure}[!tb]
\center
\includegraphics[width=8cm,trim=0 0 0 0]{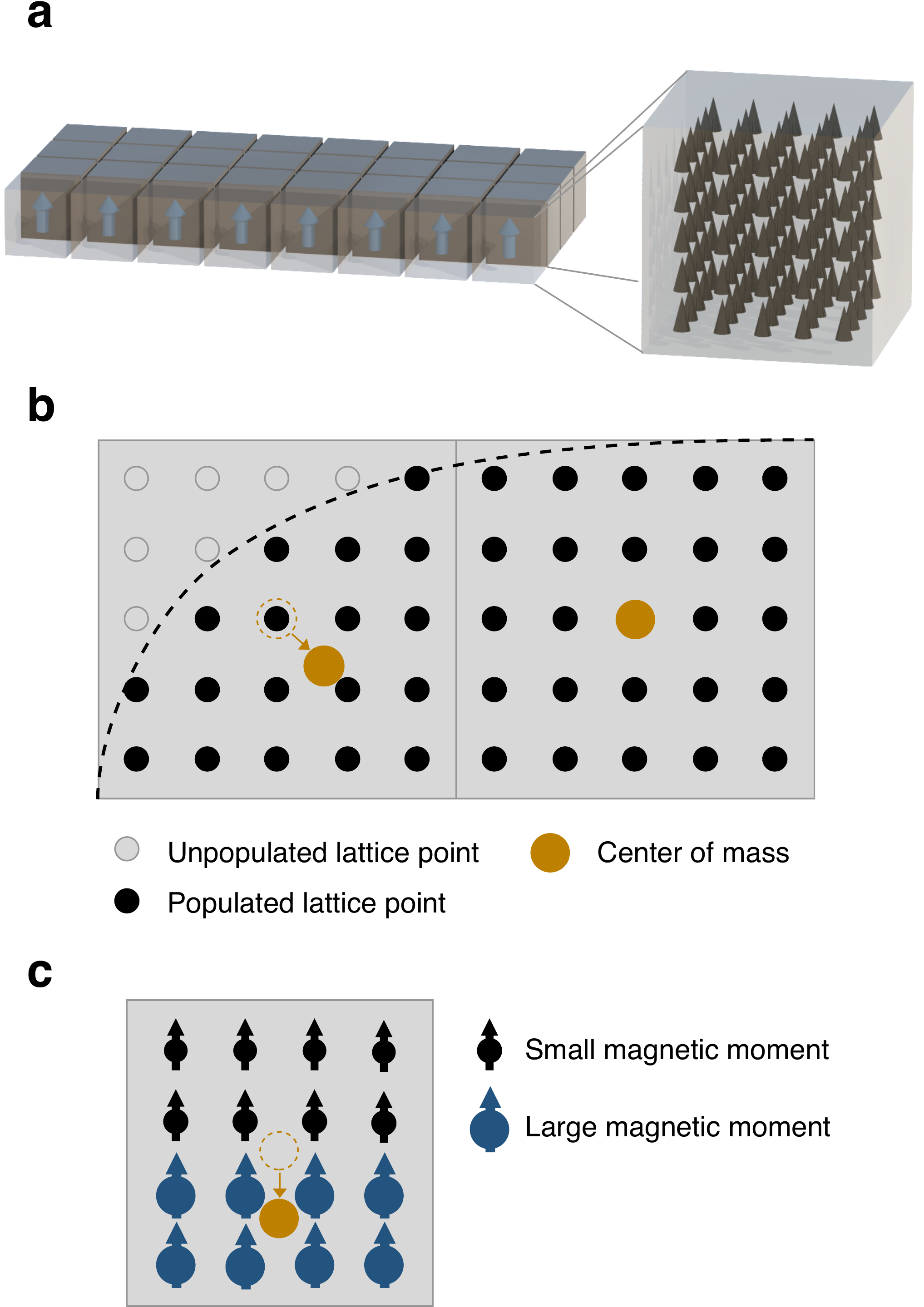}
\caption{\textbf{a}). Visualisation of the macrocell approach used to calculate the demagnetisation field, with the system discretised into cubic macrocells. Each macrocell consists of several atoms, shown schematically as cones. \textbf{b}). Schematic of the macrocell discretisation at the curved surface of a material, indicated by the dashed line. The mean position of the atoms within the macrocell defines the centre of mass where the effective macrocell dipole is located. \textbf{c}). Schematic of a macrocell consisting of two materials with different atomic moments. Since the magnetisation is dominated by one material, the magnetic centre of mass moves closer to the material with the higher atomic moments. (Colour online.)}
\label{fig:demag}
\end{figure}

The complete system is first discretised into macrocells with a fixed cell size, each consisting of a number of atoms, as shown in Fig.~\ref{fig:demag}(a). The cell size is freely adjustable from atomistic resolution to multiple unit cells depending on the accuracy required. The position of each macrocell $p_\mathrm{mc}$ is determined from the magnetic `centre of mass' given by the expression
\begin{equation}
p^{\alpha}_{\mathrm{mc}} = \frac{\sum_i^n \smmu p_i^{\alpha}}{\sum_i^n \smmu}
\label{eq:macrocellpos}
\end{equation}
where $n$ is the number of atoms in the macrocell and $\alpha$ represents the spatial dimension $x,y,z$. For a magnetic material with the same magnetic moment
at each site, Eq.~\ref{eq:macrocellpos} corrects for partial occupation of a macrocell by using the mean atomic position as the origin of the macrocell dipole, as shown in Fig.~\ref{fig:demag}(b). For a sample consisting of two materials with different atomic moments, the `magnetic centre of mass' is closer to the atoms with the higher atomic moments, as shown in Fig.~\ref{fig:demag}(c). This modified micromagnetic scheme gives a good approximation of the demagnetisation field without having to use computationally costly atomistic resolution calculation of the demagnetisation field.

The total \emph{moment} in each macrocell $m_{\mathrm{mc}}$ is calculated from the vector sum of the atomic moments within each cell, given by
\begin{equation}
m_{\mathrm{mc}}^{\alpha} = \sum_i^n \smmu S_i^{\alpha}\mathrm{.}
\end{equation}
Depending on the particulars of the system, the macrocell moments can vary significantly depending on position, composition and temperature. At elevated temperatures close to the Curie point, the macrocell magnetisation becomes small, and so the effects of the demagnetising field are much less important. Similarly in compensated ferrimagnets consisting of two competing sublattices the overall macrocell demagnetisation can also be small again leading to a reduced influence of the demagnetising field.

The demagnetisation field within each macrocell $p$ is given by
\begin{equation}
\smH_{\mathrm{demag}}^{\mathrm{mc},p} = \frac{\mu_0}{4\pi}\left(\sum_{p \ne q}\frac{3(\mmc^q\cdot \hat{\mathbf{r}})\hat{\mathbf{r}}-\mmc^q}{r^3}\right) - \frac{\mu_0}{3}\frac{\mmc^p}{V_{\mathrm{mc}}^p}
\label{eq:demag}
\end{equation}
where $r$ is the separation between dipoles $p$ and $q$, $\hat{\mathbf{r}}$ is a unit vector in the direction $p \rightarrow q$, and $V_{\mathrm{mc}}^p$ is the volume of the macrocell $p$. The first term in Eq.~\ref{eq:demag} is the usual dipole term arising from all \emph{other} macrocells in the system, while the second term is the self-demagnetisation field of the macrocell, taken here as having a demagnetisation factor $^1/_3$. Strictly this is applicable only for the field at the centre of a cube. However, the non-uniformity of the field inside a uniformly magnetised cube is not large and the assumption of a uniform demagnetisation field is a reasonable approximation. The self-demagnetisation term is often neglected in the literature, but in fact is essential when calculating the field inside a magnetic material. Once the demagnetisation field is calculated for each macrocell, this is applied uniformly to all atoms as an effective field within the macrocell.  It should be noted however that the macrocell size cannot be larger than the smallest sample dimension, otherwise significant errors in the calculation of the demagnetising field will be incurred.

The volume of the macrocell $V_{\mathrm{mc}}$ is an effective volume determined from the number of atoms in each cell and given by
\begin{equation}
V_{\mathrm{mc}} = n^{\mathrm{a}}_{\mathrm{mc}} V_{\mathrm{atom}} = n^{\mathrm{a}}_{\mathrm{mc}} \frac{V_{\mathrm{uc}}}{n^{\mathrm{a}}_{\mathrm{uc}}}
\label{eq:mc_volume}
\end{equation}
where $n^{\mathrm{a}}_{\mathrm{mc}}$ is the number of atoms in the macrocell, $n^{\mathrm{a}}_{\mathrm{uc}}$ is the number of atoms in the unit cell and $V_{\mathrm{uc}}$ is the volume of the unit cell. The macrocell volume is necessary to determine the magnetisation (moment per volume) in the macrocell. For unit cells much smaller than the system size, Eq.~\ref{eq:mc_volume} is a good approximation, however for a large unit cell with significant free space, for example a nanoparticle in vacuum, the free space contributes to the effective volume which reduces the effective macrocell volume.

\subsubsection*{Demagnetising field of a platelet}
\begin{figure}[!tb]
\center
\includegraphics[width=8cm,trim=0 0 0 0]{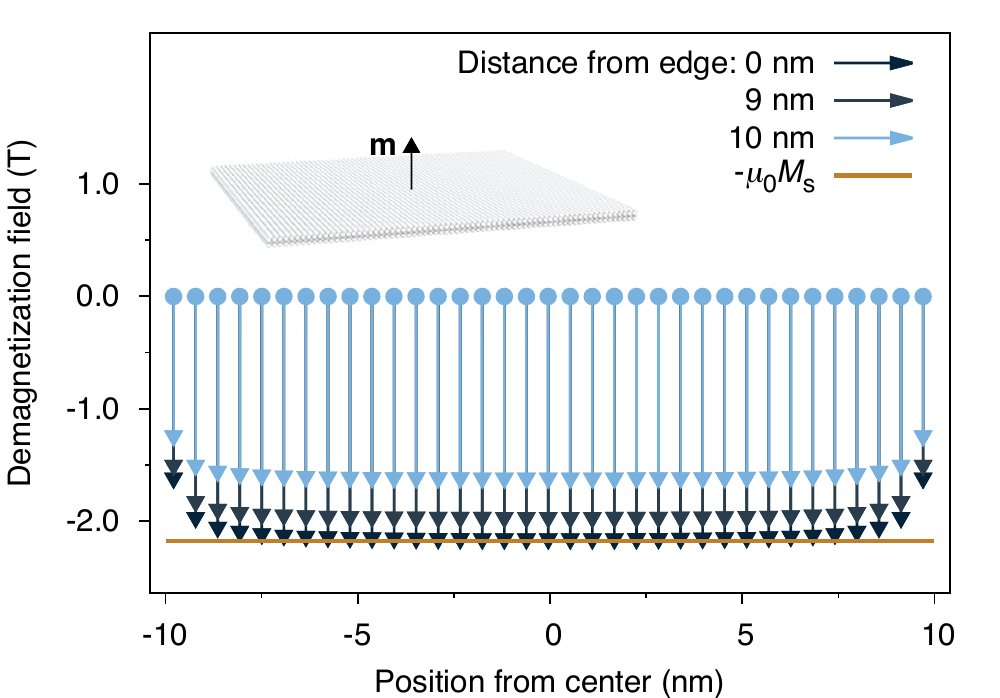}
\caption{Calculated cross-section of the demagnetisation fields in a 20 nm $\times$ 20 nm $\times$ 1 nm platelet (visualisation inset) with magnetisation perpendicular to the film plane. A macrocell size of 2 unit cells is used. In the centre of the film the calculated demagnetisation field is $-2.236$ T which compares well to the analytic solution of $\smH_{\mathrm{demag}} = -\mu_0 M = -2.18$ T. Note that the 2D grid used slightly overestimates the demagnetisation field. (Colour online.)}
\label{fig:demag-platelet}
\end{figure}

To verify the implementation of the demagnetisation field calculation it is necessary to compare the calculated fields with some analytic solution. Due to the complexity of demagnetisation fields analytical solutions are only available for simple geometric shapes such cubes and cylinders\cite{Aharoni:1998ku}, however for an infinite perpendicularly magnetised platelet the demagnetisation field approaches the magnetic saturation $-\mu_0 M$. To test this limit we have calculated the demagnetising field of a 20 nm $\times$ 20 nm $\times$ 1 nm platelet as shown in Fig.~\ref{fig:demag-platelet}. In the centre of the film agreement with the analytic value is good, while at the edges the demagnetisation field is reduced as expected.

\begin{figure*}[!tb]
\center
\includegraphics[width=8cm,trim=0 0 0 0]{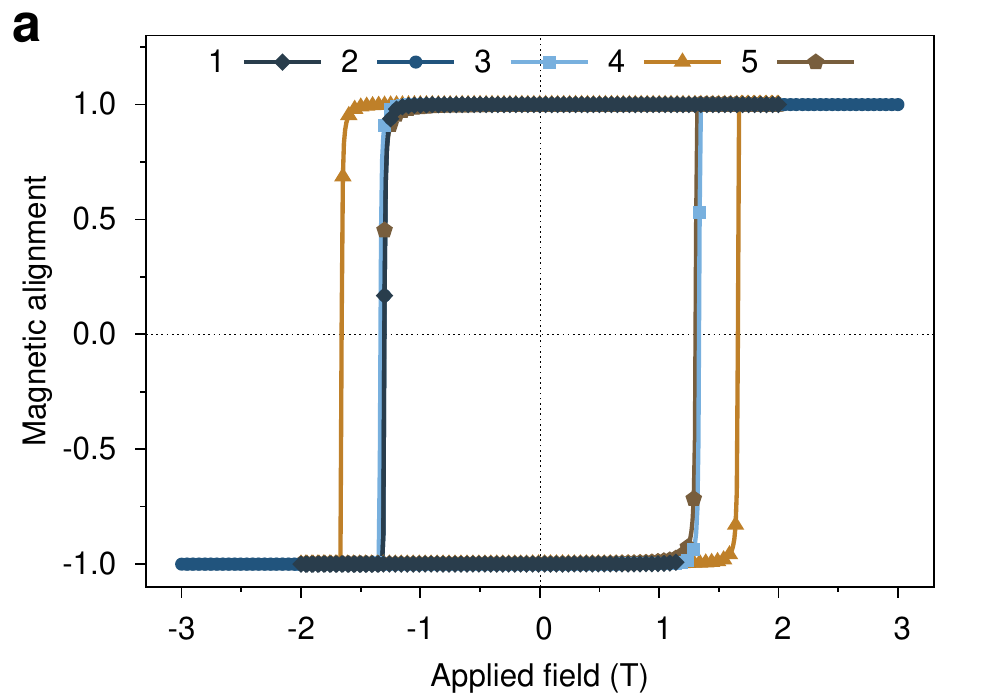}
\hspace{0.5cm}
\includegraphics[width=8cm,trim=0 0 0 0]{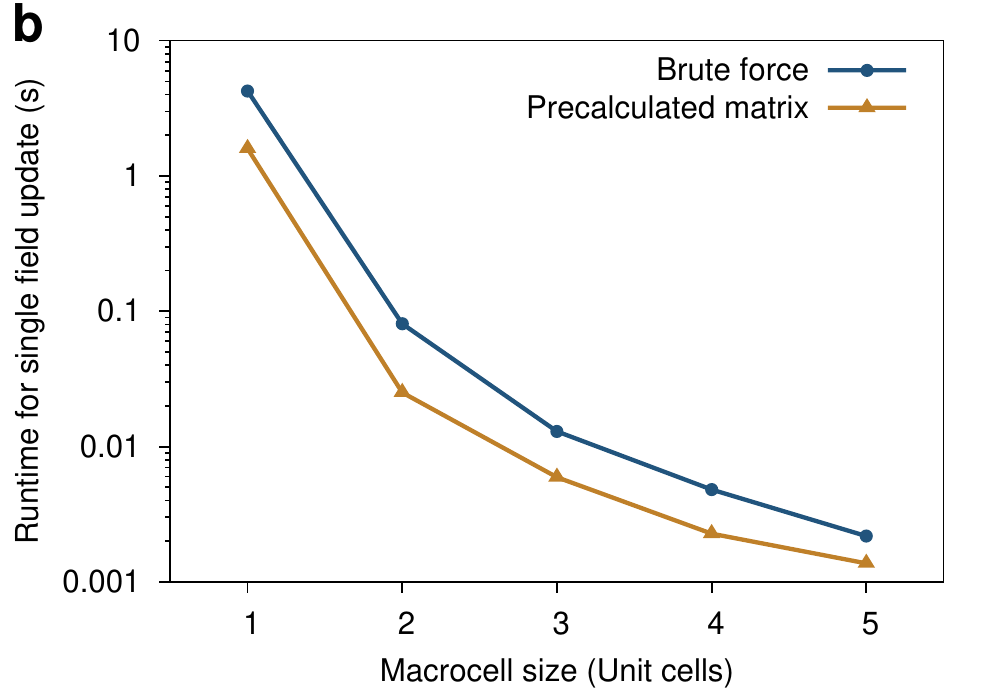}
\includegraphics[width=8cm,trim=0 0 0 0]{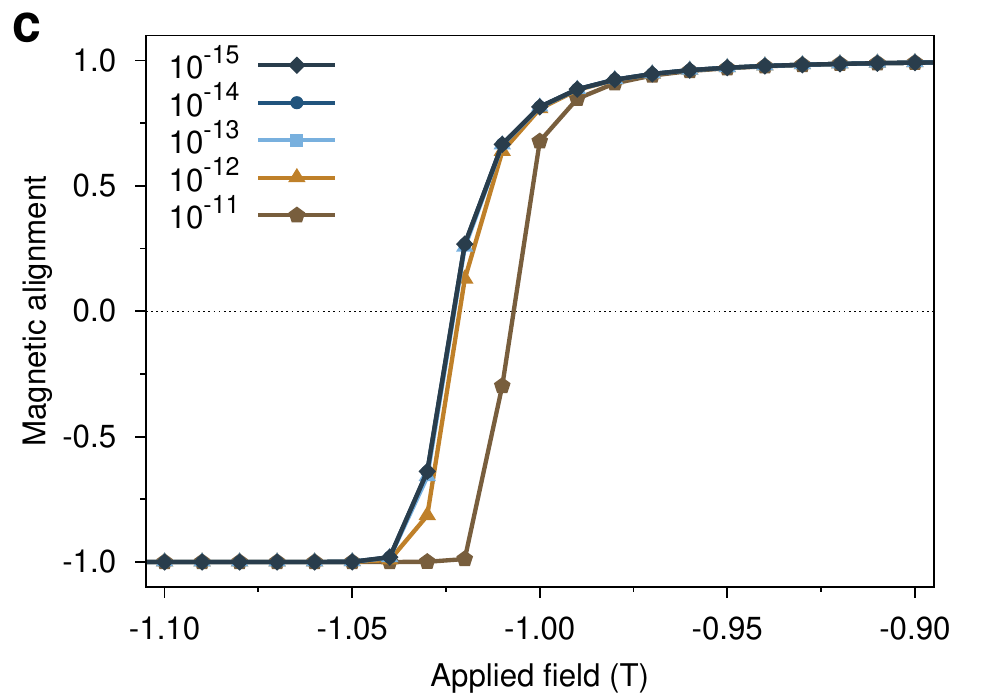}
\hspace{0.5cm}
\includegraphics[width=8cm,trim=0 0 0 0]{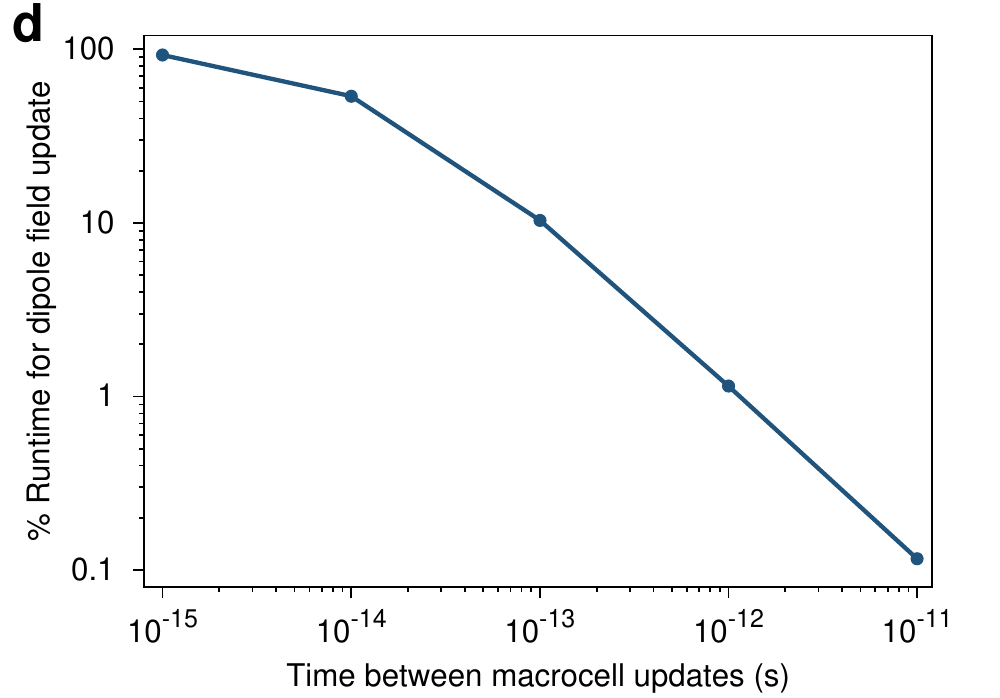}
\caption{Simulated hysteresis loops and computational efficiency for a 20 nm $\times$ 1nm nanodot for different cell sizes (multiples of unit cell size) (a,b) and update rates (seconds between update calculations) (c,d). (Colour online.)}
\label{fig:demag-performance}
\end{figure*}

\subsubsection*{Performance characteristics}
In micromagnetic simulations, calculation of the demagnetisation field usually dominates the runtime of the code and generally it is preferable to have as large a cell size as possible. For atomistic calculations however, additional flexibility in the frequency of updates of the demagnetisation field is permitted due to the short time steps used and the fact that the magnetisation is generally a slowly varying property.

To investigate the effects of different macrocell sizes and the time taken between updates of the demagnetisation field we have simulated hysteresis loops of a nanodot of diameter 40 nm and height of 1.4 nm. Fig~\ref{fig:demag-performance}(a) shows hysteresis loops calculated for different macrocell sizes for the nanodot. For most cell sizes the results of the calculation agree quite well, however, for a cell size of 4 unit cells, the calculated coercivity is significantly larger, owing to the creation of a flat macrocell (with dimensions $4 \times 4 \times 1$ unit cells). This illustrates that for systems with small dimensions, extra care must be taken when specifying the macro cell size in order to avoid non-cubic cells. In general, the problem with asymmetric macrocells is not trivial to solve within the finite difference formalism, since the problem arises due to a modification of both the intracell and intercell contributions to the demagnetising field.

Fig~\ref{fig:demag-performance}(b) shows the runtime for a single update of the demagnetising field on a single CPU for different macrocell size discretisations. Noting the logarithmic scale for the simulation time, single unit cell discretisations are computationally costly while not giving significantly better results than larger macrocell discretisations. Although the demagnetisation field calculation is an $n_{\mathrm{mc}}^2$ problem, it is possible to pre-calculate the distances between the macrocells at the cost of increased memory usage. Due to the computational cost of calculating the position vectors, this method is often much faster than the brute force calculation. However, due to the fact that memory usage increases proportionally to $n_{\mathrm{mc}}^2$, fine discretisations for large systems can require many GBs of memory.

By collating terms in Eq.~\ref{eq:demag} it is possible to construct the following matrix $\textbf{M}_{pq}$ for each pairwise interaction:
\begin{widetext}
\begin{equation}
\textbf{M}_{pq} = \begin{bmatrix}
(3r_x r_x-1)/r_{pq}^3 -1/3  &  3r_x r_y               &      3r_xr_z \\
 3r_xr_y             &  (3r_yr_y-1)/r_{pq}^3 -1/3 &      3r_yr_z   \\
 3r_xr_z             &  3r_yr_z               &  (3r_zr_z-1)/r_{pq}^3 -1/3)
\end{bmatrix}
\end{equation}
\end{widetext}
where $r_x$, $r_y$, $r_z$ are the components of the unit vector in the direction $p \rightarrow q$, and $r_{pq}$ is the separation of macrocells. Since the matrix is symmetric along the diagonal only 6 numbers need to be stored in memory. The total demagnetisation field for each macrocell $p$ is then given by:
\begin{equation}
\smH_{\mathrm{demag}}^{\mathrm{mc},p} = \frac{\mu_0}{4\pi}\left(\sum_{p \ne q}\textbf{M}_{pq} \cdot \mmc^q\right) - \frac{\mu_0}{3}\frac{\mmc^p}{V_{\mathrm{mc}}^p}\mathrm{.}
\label{eq:demag-fast}
\end{equation}
The relative performance of the matrix optimisation is plotted for comparison in Fig~\ref{fig:demag-performance}(b), showing a significant reduction in runtime. Where the computer memory is sufficiently large, the recalculated matrix should always be employed for optimal performance.

In addition to variable macrocell sizes, due to the small time steps employed in atomistic models and that the magnetisation is generally a slowly varying property, it is not always necessary to update the demagnetisation fields every single time step. Hysteresis loops for different times between updates of the demagnetisation field are plotted in Fig~\ref{fig:demag-performance}(c). In general hysteresis calculations are sufficiently accurate with a picosecond update of the demagnetising field, which significantly reduces the computational cost.

In general good accuracy for the demagnetising field calculation can be achieved with coarse discretisation and infrequent updates, but fast dynamics such as those induced by laser excitation require much faster updates, or simulation of domain wall processes in high anisotropy materials requires finer discretisations to achieve correct results.

\subsubsection*{Demagnetising field in a prolate ellipsoid}
Since the macrocell approach works well in platelets and nanodots, it is also interesting to apply the same method to a slightly more complex system: a prolate ellipsoid. An ellipsoid adds an effective shape anisotropy due to the demagnetisation field, and so for a particle with uniaxial magnetocrystalline anisotropy along the elongated direction ($z$), the calculated coercivity should increase according to the difference in the demagnetisation field along $x$ and $z$, given by:
\begin{equation}
H_{\mathrm{dm}}^{\mathrm{shape}} = +\Delta N \mu_{0} \MMs
\end{equation}
where $\Delta N = N_z - N_x$. The demagnetising factors $N_x$, $N_y$, and $N_z$ are known analytically for various ellipticities\cite{Osborn:1945to}, and here we assume $c/a = b/a = 0.5$, where $a$, $b$, and $c$ are the extent of the ellipsoid along $z$, $x$ and $y$ respectively.

\begin{figure}[!tb]
\center
\includegraphics[width=8cm,trim=0 0 0 0]{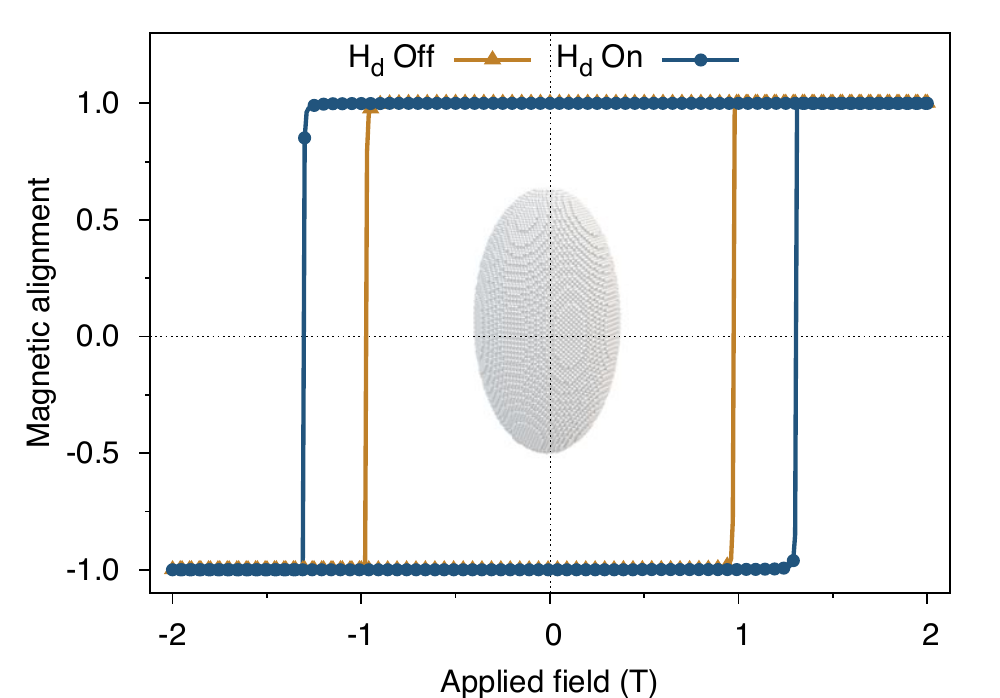}
\caption{Simulated hysteresis loops for an ellipsoidal nanoparticle with an axial ratio of 2 showing the effect of the demagnetising field calculated with the macrocell approach. A visualisation of the simulated particle is inset. (Colour online.)}
\label{fig:demag-ellipsoid}
\end{figure}

To verify the macrocell approach gives the same expected increase of the coercivity we have simulated a generic ferromagnet with atomic moment 1.5 $\muB$, an FCC crystal structure with lattice spacing 3.54 \AA and anisotropy field of $H_{\mathrm{a}}= 1$T. The particle is cut from the lattice in the shape of an ellipsoid, of diameter 10 nm and height of 20 nm, as shown inset in Fig~\ref{fig:demag-ellipsoid}. A macrocell size of 2 unit cells is used, which is updated every 100 time steps (0.1 ps).

As expected the coercivity increases due to the shape anisotropy. From Ref.~\onlinecite{Osborn:1945to} the expected increase in the coercivity is $H_{\mathrm{dm}}^{\mathrm{shape}}=0.37$ T which compares well to the simulated increase of $0.33$ T.

\section{Parallel implementation and scaling}
Although the algorithms and methods discussed in the preceding sections describe the mechanics of atomistic spin models, it is important to note finally the importance of parallel processing in simulating realistic systems which include many-particle interactions, or nano patterned elements with large lateral sizes. Details of the parallelisation strategies which have been adopted to enable the optimum performance of \vampire for different problems are presented in Appendix~\ref{parallelisation}. In general terms the parallelisation works by subdividing the simulated system into sections, with each processor simulating part of the complete system. Spin orientations at the processor boundaries have to be exchanged with neighbouring processors to calculate the exchange interactions, which for small problems and large numbers of processors can significantly reduce the parallel efficiency. The use of latency hiding, where the local spins are calculated in parallel with the inter-processor communications, is essential to ensure good scaling for these problems.

\begin{figure}[!t]
\center
\includegraphics[width=8cm, trim=10 0 10 0]{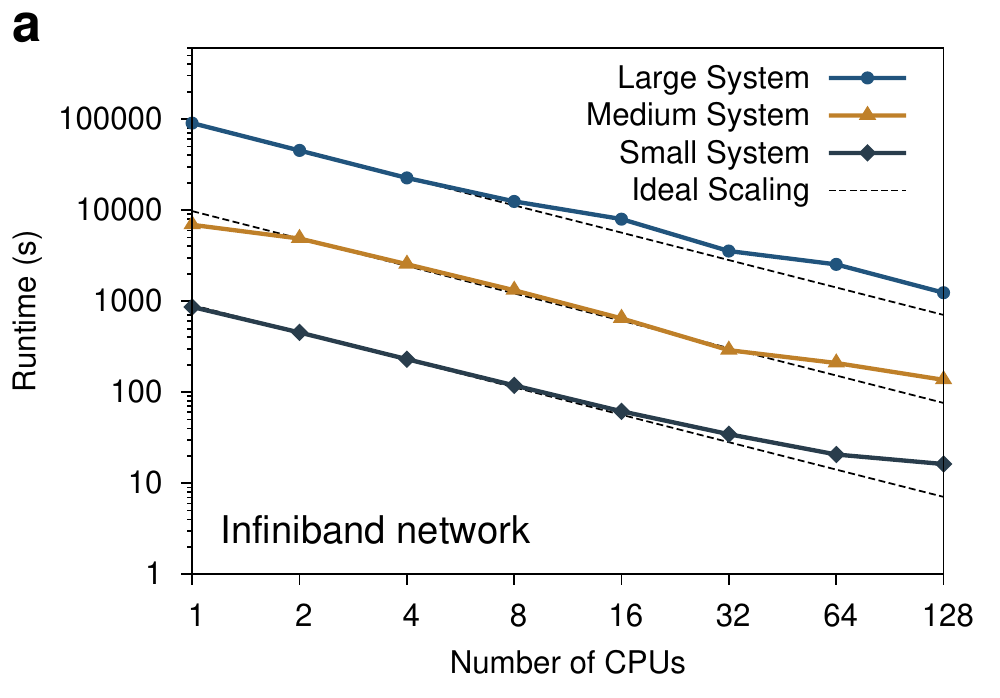}
\includegraphics[width=8cm, trim=10 0 10 0]{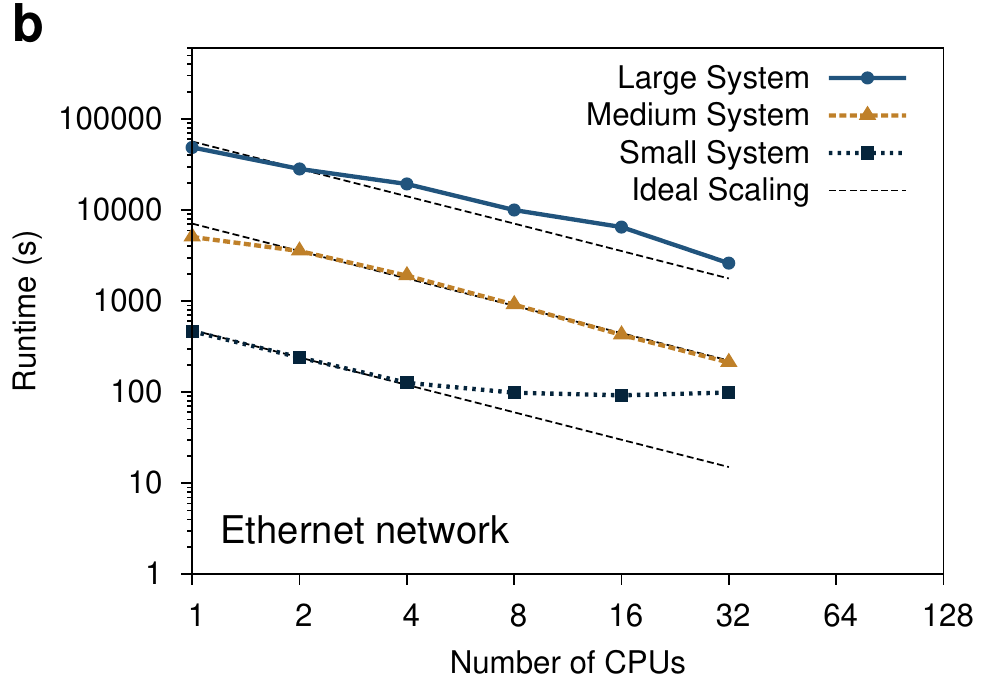}
\caption{Runtime scaling of \vampire for three different problem sizes on the infiniband network (a) and Ethernet network (b), normalised to the runtime for 2 cores for each problem size. (Colour online.)}
\label{fig:scaling}
\end{figure}

To demonstrate the performance and scalability of \vampire, we have performed tests for three different system sizes: small (10628 spins), medium ($8 \times 10^5$ spins), and large ($8 \times 10^6$ spins). We have access to two beowulf-class clusters; one with 8 cores/node with an Infiniband 10 Gbps low-latency interconnect, and another with 4 cores/node with a Gigabit Ethernet interconnect. For parallel simulations the interconnect between the nodes can be a limiting factor for increasing performance with increasing numbers of processors, since as more processors are added, each has to do less work per time step. Eventually network communication will dominate the calculation since processors with small amounts of work require the data from other processors in shorter times, leading to a drop in performance. The scaling performance of the code for 100,000 time steps on both machines is presented in Fig.~\ref{fig:scaling}.

The most challenging case for parallelisation is the small system size, since a significant fraction of the system must be communicated to other processors during each timestep. On the Ethernet network  system for the smallest system size reasonable scaling is seen only for 4 CPUs due to the high latency of the network. However larger problems are much less sensitive to network latency do to latency hiding, and show excellent scalability up to 32 CPUs. Essentially this means that larger problems scale much better than small ones, allowing more processors to be utilised. This is of course well known for parallel scaling problems, but even relatively modest systems consisting of $10^5$ spins show significant improvements with more processors.

For the system with the low-latency Infiniband network, excellent scalability is seen for all problems up to 64 CPUs. Beyond 64 CPUs the reduced scalability for all problem sizes is likely due to a lack of network bandwidth. The bandwidth requirements are similar for all problem sizes, since smaller problems complete more time steps in a given period of time and so have to send several sets of data to other processors. Nevertheless improved performance is seen with increasing numbers of CPUs allowing for continued reductions in compute time. Although not shown, initial tests on an IBM Blue Gene class system have demonstrated excellent scaling of \vampire up to 16,000 CPUs, allowing the real possibility for atomistic simulations with lateral dimensions of micrometers. Additional scaling tests for systems including calculation of the demagnetising field and a long-ranged exchange interaction are presented in Appendix~\ref{parallelisation}.

\section{Conclusions and perspectives}
We have described the physical basis of the rapidly developing field of atomistic spin models, and given examples via its implementation in the form of the \vampire code. Although the basic formalism underpinning atomistic spin models is well established, ongoing developments in magnetic materials and devices means that new approaches will need to be developed to simulate a wider range of physical effects at the atomistic scale. One of the most important phenomena is spin transport and magnetoresistance which is behind an emergent field of spin-electronics, or spintronics. Simulation of spin transport and spin torque switching is already in development, and must be included in atomistic level models in order to simulation of a wide range of spintronic materials and devices, including read sensors and MRAM (magnetic random access memory). Other areas of interest include ferroelectrics, the spin Seebeck effect\cite{Hinzke:2011gp}, and Coloured noise\cite{Atxitia:2009io} where simulation capabilities are desirable, and incorporation of these effects are planned in future. In addition to modelling known physical effects, it is hoped that improved models of damping incorporating phononic and electronic mechanisms will be developed which enable the study of magnetic properties of materials at sub-femtosecond timescales.

The ability of atomistic models to incorporate magnetic parameters from density functional theory calculations is a powerful combination which allows complex systems such as alloys, surfaces and defects to be accurately modelled. This multiscale approach is essential to relate microscopic quantum mechanical effects to a macroscopic length scale accessible to experiment. Such a multiscale approach leads to the possibility of simulation driven technological development, where the magnetic properties of a complete device can be predicted and optimised through a detailed understanding of the underlying physics. Due to the potential of multiscale simulations, it is planned in future to develop links to existing DFT codes such as \textsc{castep}\cite{castep1,castep2} to allow easier integration of DFT parameters and atomistic spin models.

The computational methods presented here provide a sound basis for atomistic simulation of magnetic materials, but further improvements in both algorithmic and computational efficiency are of course likely. One area of potential computational improvement is GPGPU (general purpose graphics processing unit) computation, which utilises the massively parallel nature of GPUs to accelerate simulations, with speed ups over a single CPU of 75 times routinely reported. With several supercomputers moving to heterogenous computing architectures utilising both CPUs and GPUs, supporting GPGPU computation is likely to be important in future, and an implementation in our \vampire code is currently planned. In terms of algorithmic improvements it should be noted that the Heun numerical scheme although simple is relatively primitive by modern standards, and moving to a midpoint scheme may allow for larger time steps than currently to be used.

With the continuing improvements in computer power, atomistic simulations have become a viable option for the routine simulation of magnetic materials. With the increasing complexity of devices and material properties, atomistic models are a significant and important development. While micromagnetic models remain a powerful tool for the simulation and design of devices, the limitations of the (continuum) micromagnetic formalism are increasingly exposed by its failure to deal with the complex physics of elevated temperatures, ultrafast magnetisation processes and interfaces. While micromagnetics will remain the computational model of choice for large scale and coarse-grained applications, the ability to accurately model the effects of microscopic details, temperature effects and ultrafast dynamics make atomistic models an essential tool to understand the physics of magnetic materials at the forefront of of the field.

\begin{acknowledgments}
The authors wish to thank Jerome Jackson for helpful discussions particularly with regard to the parallel implementation of the code; Thomas Schrefl for insisting on the name of \vampire; Matt Probert, Phil Hasnip, Joe Barker, Uli Nowak and Ondrej Hovorka for helpful discussions; and Laszlo Szunyogh, Phivos Mavropoulos and Stefan Bl{\"u}gel for assistance with the section on the \textit{ab initio} parameterisation of the atomistic model. We also acknowledge the financial support from the EU Seventh Framework Programme grant agreement No. 281043 \textsc{femtospin} and Seagate Technology Inc. This work made use of the facilities of N8 \textsc{hpc} provided and funded by the N8 consortium and \textsc{epsrc} (Grant No. EP/K000225/1) co-ordinated by the Universities of Leeds and Manchester and the \textsc{epsrc} Small items of research equipment at University of York \textsc{energy} (Grant No. EP/K031589/1).
\end{acknowledgments}

\appendix

\section{Code structure and design philosophy}
In addition to implementing the necessary computational methods for magnetic atomistic calculations, it is also important to provide a framework structure for the code, where new additions in the form of features or improvements can be made with minimal disturbance to other sections of the code. Equally important for intensive computational problems is ensuring high performance of the code so that simulations proceed as rapidly as possible.

In \vampire this is achieved through hybrid coding using a mixture of object-oriented and functional programming styles. Object-oriented programming is widely used in modern software projects as a level of abstraction around the data, or objects, in the code. This abstraction makes it easy to store information about an element, for example an atom, as a single unified data type, known as a class. One significant caveat with object-oriented code is that it is generally hard to optimise for maximum performance. High performance codes generally utilise a different coding approach known as functional programming, where the focus is on functions which operate on numerous data sets. However the organisation of data into large blocks in functional programming generally makes it harder to organise the data. \vampire therefore makes use of both methodologies to gain the benefits of object-oriented design during the initialisation phase, while for the performance-critical parts of the code the data is re-arranged to use a functional style for maximum performance. Due to the requirements of high performance, object-oriented design and parallelisation, the C++ programming language was chosen for all parts of the code. The popularity of the C++ language also allows for easy future integration of other libraries, such as \textsc{nvidia}'s \textsc{cuda} framework for utilising graphics processing units. For portability the code also has a minimal dependence on external libraries and also conforms to the published standard allowing simple compilation on the widest possible variety of computer architectures.

In addition to the low-level structure described in terms of object-oriented and functional programming styles, the code is also designed in a modular fashion so that most of the mechanistic operations (such as the parallelisation and data analysis) are separated from high level functions which control the various simulation types. This enables users to easily add new simulation types or physical effects to the code, without having to be concerned with the inner workings.

\section{Atomistic system generation in VAMPIRE}\label{app:system-generation}
\vampire has a number of dedicated functions for generating atomic system within the nearest neighbour approximation. The principal advantage of the nearest neighbour approximation is its simplicity and ability to consider a wide range of physical effects such as finite size, surfaces, ferri and antiferromagnets, disordered systems etc. with relative ease. \vampire also includes in-built particle structures to enable generation of systems with simple geometric shapes such as spheres, cylinders, truncated octahedra and cubes.

\begin{figure*}[!t]
\center
\includegraphics[width=16cm]{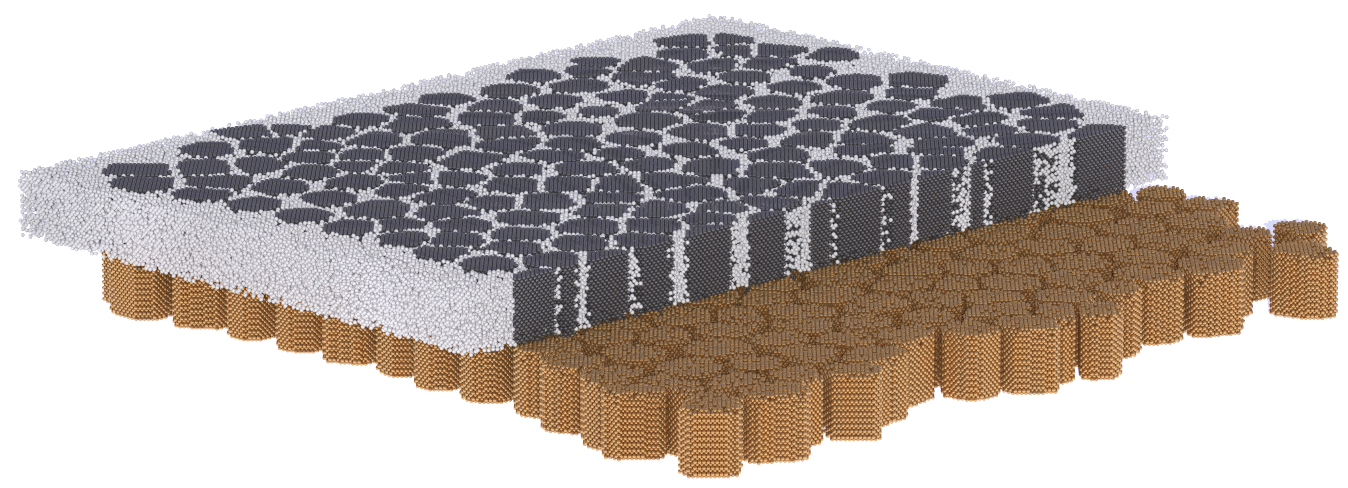}
\caption{Visualisation of a magnetic recording medium generated using \vampire. The medium consists of magnetically hard and soft layers with interfacial mixing of atoms between the layers. The material is granular in nature, and so a voronoi tessellation as overlaid on top the layers to form the isolated magnetic grains. Finally, a dilute intermixing layer is applied between the grains representing the diffusion of magnetic atoms into the SiO$_2$ between the grains, as seen in real media. (Colour online.)}
\label{fig:recording}
\end{figure*}

The first step is to generate a crystal lattice of the desired type and dimensions sufficiently large to incorporate the complete simulated system. For the nearest neighbour approximation the Hamiltonian is generally only well defined for a single unified crystal structure, and therefore such generic simulations require a single crystal from which the system is cut. More complex structures are readily simulated, however the user must define the complete Hamiltonian for the system, taking into account the realistic interfaces between different crystals.  \vampire uses the unit cell as the essential building block of the atomic structure, since the exchange interactions of atoms between neighbouring unit cells are known before the structure is generated. The global crystal is generated by replicating the basic unit cell on a grid in $x$,$y$ and $z$. This bare crystal structure is then cut into the desired geometry, for example a single nanoparticle, voronoi granular structure, or a user defined 2D geometry by removing atoms from the complete generated crystal. Atoms within this geometry are then assigned to one or more materials as desired (each material having different magnetic properties such as atomic spin moments, anisotropy or exchange interactions), generating the complete atomic system. The assignment of different parts of the system to different materials enables the easy simulation of multilayers and core-shell nanoparticles, as well as combinations of these for systems such as multilayer magnetic recording media. As an example, Fig.~\ref{fig:recording} shows a visualisation of a multilayer magnetic recording media generated using \vampire.

Once the structure is defined the exchange interactions for all atoms in the are calculated from a list of nearest neighbour interactions for the defined unit cell. Since each cell on the grid contains a fixed number of atoms, and the exchange interactions of those atoms with other neighbouring cells is known relative to the local cell, the interaction list is trivial to generate. For computational efficiency the final interaction list is then stored as a 1D linked list.

\section{Parallelisation strategies}\label{parallelisation}
A consistent trend among computers today is the drive towards parallel architectures, designed to improve overall performance in a consistent and scalable fashion. The downside of this approach is that the software must be specially modified to take advantage of the hardware, which still presents a significant challenge. In order to make the best use of parallel computers, we have adopted a number of distinct parallelisation strategies. This approach means that for any given problem, an optimal strategy can be utilised to achieve maximum performance.

\subsubsection*{Statistical Parallelism}
The most trivial form of parallelism is batch or statistical, where the statistical properties of a system are determined by a series of independent calculations, each of which can be run in parallel. Statistical parallelism has the prime advantage that the division of work leads to an ideal scaling behaviour, since each of the runs are entirely independent and require no intercommunication.

In magnetic simulations, the most common applications of statistical parallelism are sweeps of the parameter space for a particular system, or in determining thermodynamic averages. For the former, a given system is calculated for different values of key parameters, for example, anisotropy or exchange constants; for the latter, the same system is simulated, but each run is given a different seed for the random number generator. This leads to a different thermodynamic evolution, which can provide information about the statistical behaviour of the system. It should be noted that the correct seeding of the random number generator, where a number of uncorrelated random number sequences are generated, is quite complex \cite{Matsumoto:1998jt}. For magnetic simulations, the chaotic nature of the system, whereby a small change in the time evolution rapidly leads to a significantly different result, means that crude sequential number seeding is quite satisfactory.

\subsubsection*{Geometric Decomposition}
Although statistical parallelism is useful for some types of simulation, it has one significant limitation: it can only be applied to relatively small systems, as the entire problem must be solved on a single processor. For larger systems it is necessary to divide the system into smaller parts for parallel execution. The most efficient method for such parallelisation is generally geometric decomposition, where the space is divided into cells, and each processor is assigned a cell to simulate. If well implemented, geometric decomposition can be scaled to run on thousands of CPUs, and this is one of the key aims of our implementation.

The starting point for geometric decomposition is efficiently dividing the space to run on $N_{\mathrm{CPU}}$ CPUs. In order to achieve this, we have devised an algorithm which takes into account the physical system dimensions and which searches for a solution where
\begin{equation}\label{eqn:nxnynz}
n_x \cdot n_y \cdot n_z = N_{\mathrm{CPU}}
\end{equation}
while minimising the surface to volume ratio. If the dimensions of the overall system are given by $l_x$,$l_y$, and $l_z$, then the volume of each cell is:
\begin{equation}
V_{\mathrm{cell}} = \frac{l_x}{n_x} \cdot \frac{l_y}{n_y} \cdot \frac{l_z}{n_z}
\end{equation}
and the surface area of each cell is:
\begin{equation}
A_{\mathrm{cell}} = 2\left[\frac{l_x l_y}{n_x n_y} + \frac{l_y l_z}{n_y n_z} + \frac{l_x l_z}{n_x n_z} \right]\mathrm{.}
\end{equation}
The surface to volume ratio is then given by:
\begin{equation}
\frac{A_{\mathrm{cell}}}{V_{\mathrm{cell}}} = 2\left[\frac{n_z}{l_z} + \frac{n_y}{l_y} + \frac{n_x}{l_x} \right]\mathrm{.}
\end{equation}
It is clear that the minimum in the surface to volume ratio occurs for $n_{\alpha}/l_{\alpha}=1$ for all three dimensions, essentially showing that longer dimensions parallelise better with more CPUs.

\begin{figure}[!t]
\center
\includegraphics[width=8cm]{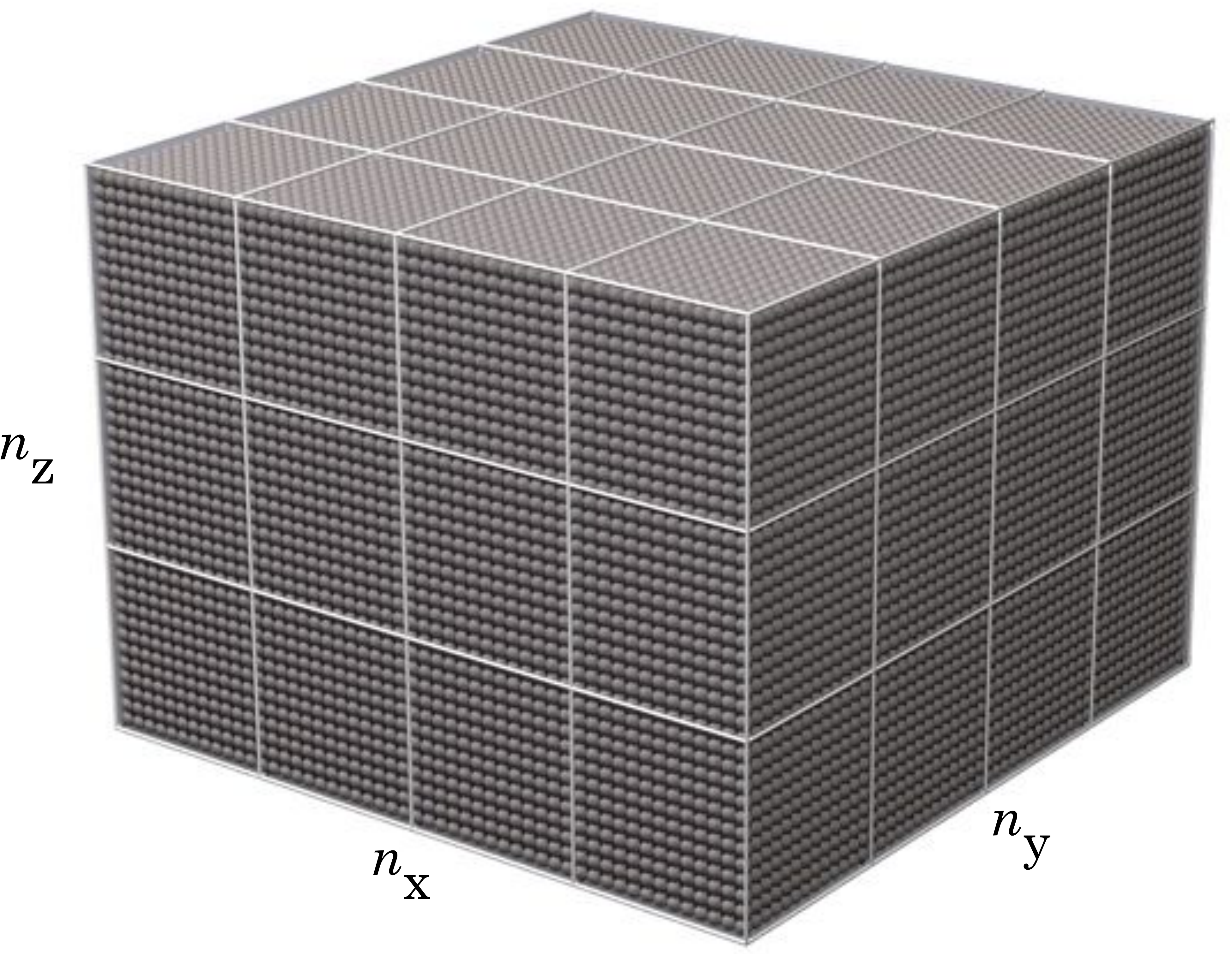}
\caption{Visualisation of the decomposition of a cubic system into 48 blocks of equal volume.}
\label{fig:GD}
\end{figure}

Given that the dimensions of the system are fixed, the only free variables are the number of CPUs in each dimension, $n_x$, $n_y$, and $n_z$. These are further constrained by equation \ref{eqn:nxnynz}. In order to find the optimal solution for a given number of CPUs, the starting point is $n_{\alpha} = \frac{l_x}{\sqrt{N_{\mathrm{CPU}}}}$. Exact solutions for $n_x$, $n_y$ and $n_z$ are then searched for and the one with the lowest surface to volume ratio is selected. This approach is very flexible and allows for efficient decomposition for any number of CPUs. The only problematic solution is for prime numbers of CPUs, where only one exact solution exists, though this is a rare occurrence for large numbers of CPUs. A visualisation of a cubic system decomposed into 48 blocks is shown in Fig.~\ref{fig:GD}.

\begin{figure}[!ht]
\center
\includegraphics[width=8cm]{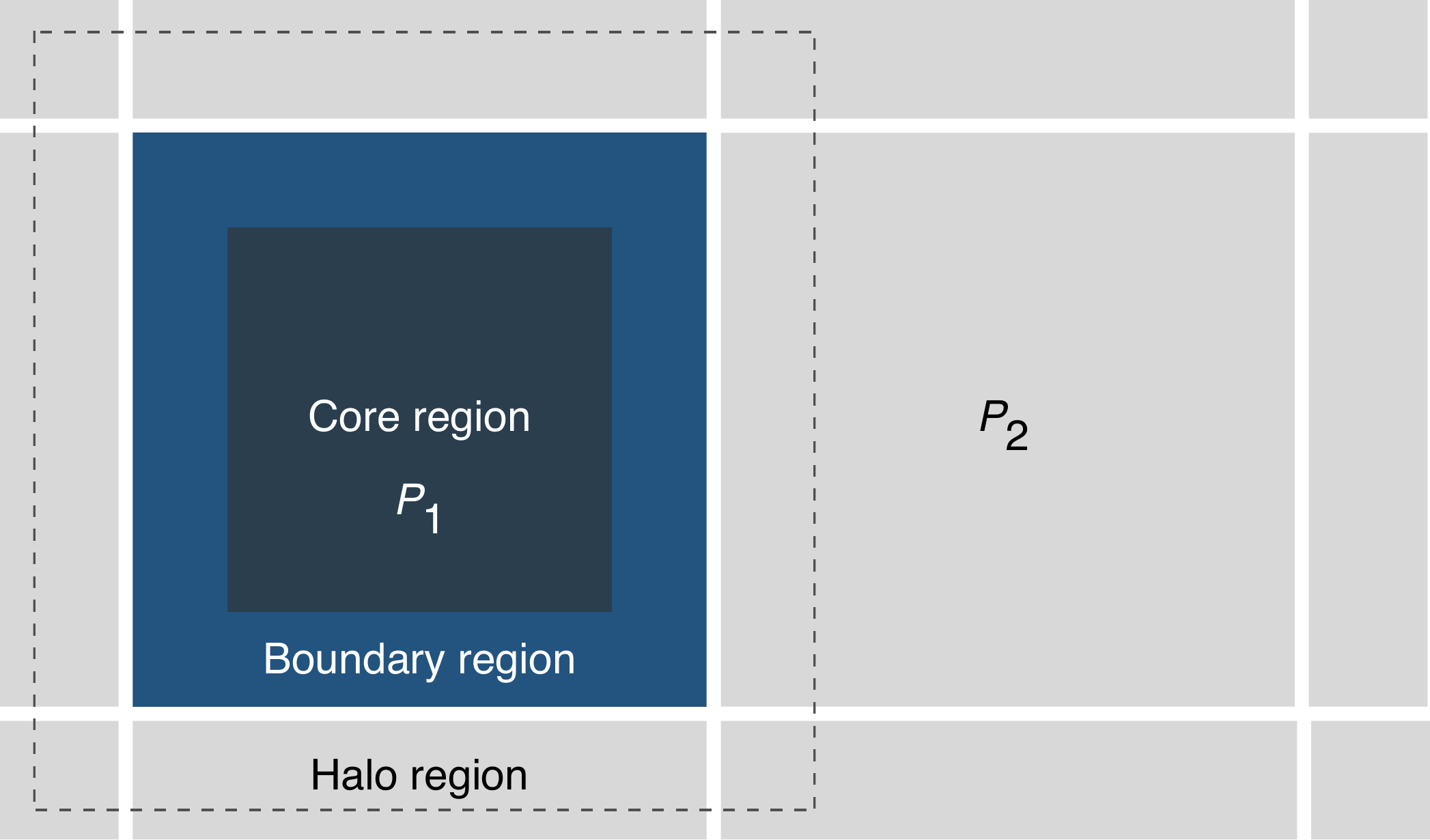}
\caption{Visualisation of the different categorisations of atoms on a processor, determined by their spatial location. The core and boundary regions exist on the local processor, denoted by the regions within the dashed line. The halo region contains atoms on remote processors which atoms on the local processor potentially interact with. (Colour online.)}
\label{fig:CBH}
\end{figure}

Having decomposed the system, each CPU is allocated a cell which defines its own spatial domain. In order to maintain maximum scalability, each CPU generates its own portion of the complete system, and all associated data. This has the advantage of minimising the memory footprint and also parallelising the system creation, which can become a significant bottleneck for very large numbers of processors. Once the local atoms have been generated it is necessary to know which atoms on remote processors (halo atoms) are potentially interacting with atoms on the local processor (boundary atoms), as well as which atoms are interacting locally only(core atoms). This essentially defines three distinct regions, as shown schematically in Fig.~\ref{fig:CBH}.

The maximum interaction range of the atoms is known globally, and so provided all atoms in this range are included, generation of the neighbour list is trivially the same as the serial case. In practice this is implemented by a global broadcast of each processor's domain, i.e., which regions of space are 'owned' by each processor. Each processor then looks at each atom in its boundary region, and then dispatches a copy of the atom to the appropriate neighbouring processors. This method has the advantage that it is quite general, and can be applied to any decomposition method, not necessarily cubes. At this point parallel periodic boundary conditions are easily implemented in the same manner, by copying atoms at the edge of the system to the desired processors. Once all boundary atoms have been sent, and all halo atoms have been received, the neighbour list is generated in the usual fashion with a linked-cell algorithm. After the actual neighbour list has been generated, it is likely that some of the copied halo atoms are in fact not needed, and so these atoms are deleted. Similarly some atoms in the in the boundary region may not interact with the halo, and these atoms are re-assigned to the core region. Following this book-keeping exercise, parallel simulation of the system can begin.

The method we have adopted for parallel simulation of the system makes use of latency-hiding, where requests for data from other processors are made prior to a locally compute-intensive period, after which the requested data should have arrived. Such latency-hiding is an important consideration when running the code on many processors. In practice atoms on each processor are ordered according to their interaction classification, ie: core; boundary; and finally halo atoms. The integration of the system proceeds as follows:
\begin{itemize}
\item A request is made for all halo data from other processors
\item The core region is then integrated
\item If the halo data has not arrived, then wait for it
\item Integrate the boundary region
\item Global synchronisation
\end{itemize}
The parallel integration is repeated the desired number of times during the simulation.

\subsubsection*{Replicated Data}
For continuous systems, geometric decomposition provides an efficient way of parallelisation of the calculation. However, for sparse systems geometric decomposition can be inefficient due to poor load balancing, where some processors have many more atoms than others. This means some processors spend a significant amount of time waiting for others to complete the integration step, leading to a reduction in scalability. In magnetism, such systems are typically granular, consisting of a small number of grains. One solution to sparse systems is to utilise a replicated data approach, where each processor has a complete copy of all data, similar to the statistical parallelism method. Each processor then simulates $(1/n_{\mathrm{cpu}})^{\mathrm{th}}$  of the total system, without any constraints on spatial locality. The atoms are classified in a similar way to the geometric decomposition approach, as core, boundary, and halo, and integrated in exactly the same way.

The principal disadvantages of the replicated data approach are the increased memory footprint (each processor must generate a complete copy of the system), and the tendency to have a high proportion of boundary spins. The latter can be mitigated by re-ordering spins in memory to ensure some degree of spatial locality. For granular systems this is fact trivial, since the assignment of each spin to a grain provides the necessary geometric information, and so the spins are ordered by a unique grain identification, which is also spatially correlated. In addition to its use in simulation of sparse systems, the replicated data approach is also the strategy adopted for the parallel calculation of the long-ranged demagnetisation fields. The method for the parallel code is identical to that described earlier, where the spins are allocated to macrocells which then interact with each other. Since the calculation of the demagnetisation field in each cell requires knowledge of all other cells, replicated data is the logical choice for the parallelisation. The demagnetisation field calculation proceeds as follows:
\begin{itemize}
\item The macrocell moments are initialised to zero on all processors
\item Each processor determines the contribution of its spins to each macrocell
\item The macrocell moments are summed globally, so that each processor has a complete copy of the macrocell magnetisations
\item Each processor calculates the demagnetisation field only for macrocells which contain local spins
\item The local demagnetisation field on each spin is determined from its macrocell demagnetisation field
\end{itemize}

This approach leads to excellent scaling, as for reasonable macrocell sizes the communication costs are minimal, and for $\gg$ 1 macrocell per processor the method scales linearly with $n_{\mathrm{cpu}}$. Fine macrocell discretisations ($\ll$ 27 unit cells/macrocell) can lead to significant memory and computation costs, but in general this is unnecessary for most atomistic scale calculations.

\begin{figure}[!t]
\center
\includegraphics[width=8cm, trim=10 0 10 0]{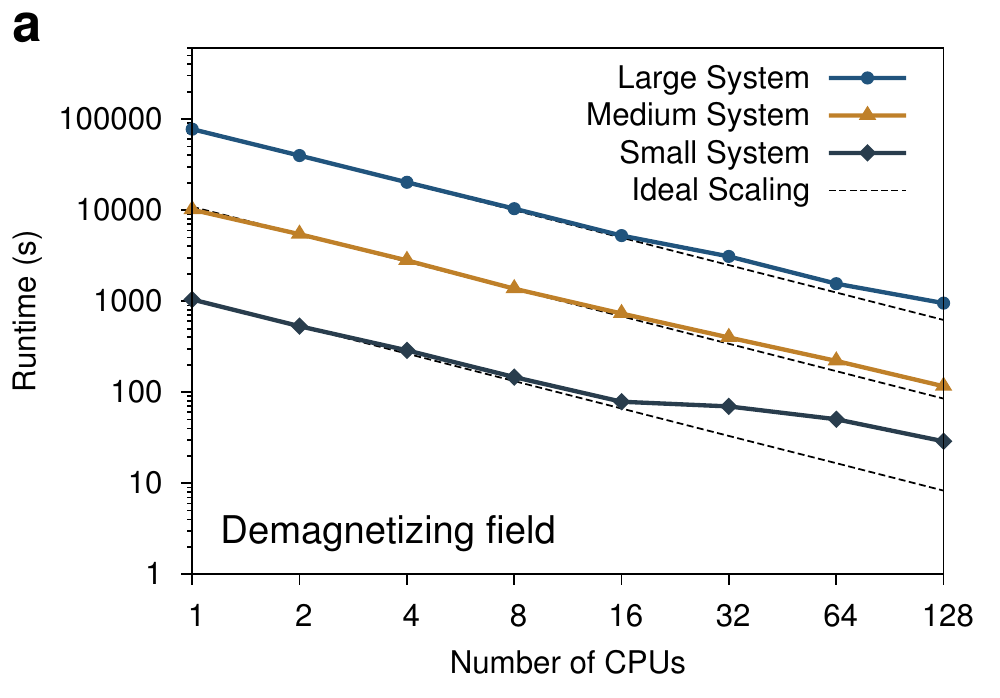}
\includegraphics[width=8cm, trim=10 0 10 0]{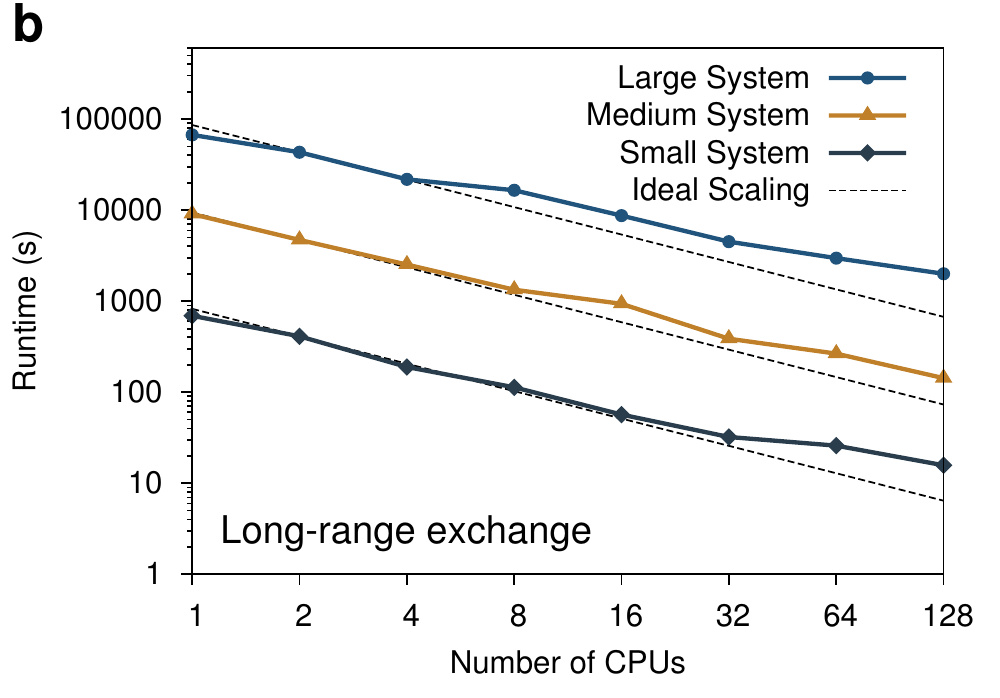}
\caption{Runtime scaling of \vampire for three different problem sizes including demagnetising fields (a) and a spin Hamiltonian including a long-ranged exchange interaction (b), normalised to the runtime for 2 cores for each problem size. The cluster has 8 processors per node, and an Infiniband interconnect for inter-node communications. (Colour online.)}
\label{fig:advanced-scaling}
\end{figure}

\subsubsection*{Additional scaling tests}
\vampire has already been shown to scale well for a generic system with nearest neighbour exchange interactions, but in order to verify the general usefulness of the parallelisation we have also considered an extended system including the demagnetisation field calculation, and a system using \textit{ab initio} parameters for FePt\cite{Mryasov:2005kj} which includes a long-ranged exchange interaction extending over several unit cells, as shown in Fig~\ref{fig:advanced-scaling}.

The simulations in Fig.~\ref{fig:advanced-scaling}(a) including the demagnetising fields use a $2 \times 2$ unit cell macrocell size, updated every 10 steps, for system sizes of approximately $10^4, 10^5 and 10^6$ spins respectively. The high spatial and temporal resolution are in some sense a worst case scenario as these are probably not needed for most problems, and so for these simulations, calculation of the demagnetising fields dominates the run time. Nevertheless, the scaling of the code remains very good, showing the effectiveness of the parallelisation of the demagnetising field calculation. For small system sizes the scaling breaks down as the number of macrocells approaches the number of processors. Here the scaling is limited by the time required to update all of the cell magnetisations and the time required to calculate the contributions of the reduced number of local macrocells. For all problem sizes scaling begins to reduce somewhat due to limitations in the network bandwidth. 

For the long-ranged exchange interactions shown in Fig.~\ref{fig:advanced-scaling}(b) the scaling is only good within a single node (up to 8 processors). Above that, the larger the system the worse the scaling. This arises due to the large amount of data which has to be shared between processors. For the long-ranged exchange interaction, each atom to be simulated must know the spin configurations of over 1,000 neighbouring atoms. In a parallel simulation these spin directions must be passed between processors twice per time step, which is a bandwidth intensive operation. Thus, the reduced scaling, particularly with arguer system size, is due to saturation of the network link. Due to the long range nature of the exchange interaction, memory use also becomes an issue in terms of storing the neighbour list. For the largest simulation size of around $1.6 \times 10^6$ atoms,33GB of RAM is required to store all the interactions. However, due to the parallel system generation this divides nicely between all of the processors, so the memory required per processor is quite reasonable for larger numbers of processors.

\bibliography{papers,/Users/rfle500/Documents/Work/Papers/thesis,local,/Users/rfle500/Documents/Work/Papers/Bibliography/library}

\begin{thebibliography}{133}%
\makeatletter
\providecommand \@ifxundefined [1]{%
 \@ifx{#1\undefined}
}%
\providecommand \@ifnum [1]{%
 \ifnum #1\expandafter \@firstoftwo
 \else \expandafter \@secondoftwo
 \fi
}%
\providecommand \@ifx [1]{%
 \ifx #1\expandafter \@firstoftwo
 \else \expandafter \@secondoftwo
 \fi
}%
\providecommand \natexlab [1]{#1}%
\providecommand \enquote  [1]{``#1''}%
\providecommand \bibnamefont  [1]{#1}%
\providecommand \bibfnamefont [1]{#1}%
\providecommand \citenamefont [1]{#1}%
\providecommand \href@noop [0]{\@secondoftwo}%
\providecommand \href [0]{\begingroup \@sanitize@url \@href}%
\providecommand \@href[1]{\@@startlink{#1}\@@href}%
\providecommand \@@href[1]{\endgroup#1\@@endlink}%
\providecommand \@sanitize@url [0]{\catcode `\\12\catcode `\$12\catcode
  `\&12\catcode `\#12\catcode `\^12\catcode `\_12\catcode `\%12\relax}%
\providecommand \@@startlink[1]{}%
\providecommand \@@endlink[0]{}%
\providecommand \url  [0]{\begingroup\@sanitize@url \@url }%
\providecommand \@url [1]{\endgroup\@href {#1}{\urlprefix }}%
\providecommand \urlprefix  [0]{URL }%
\providecommand \Eprint [0]{\href }%
\providecommand \doibase [0]{http://dx.doi.org/}%
\providecommand \selectlanguage [0]{\@gobble}%
\providecommand \bibinfo  [0]{\@secondoftwo}%
\providecommand \bibfield  [0]{\@secondoftwo}%
\providecommand \translation [1]{[#1]}%
\providecommand \BibitemOpen [0]{}%
\providecommand \bibitemStop [0]{}%
\providecommand \bibitemNoStop [0]{.\EOS\space}%
\providecommand \EOS [0]{\spacefactor3000\relax}%
\providecommand \BibitemShut  [1]{\csname bibitem#1\endcsname}%
\let\auto@bib@innerbib\@empty
\bibitem [{\citenamefont {Ising}(1925)}]{Ising:1925em}%
  \BibitemOpen
  \bibfield  {author} {\bibinfo {author} {\bibfnamefont {E.}~\bibnamefont
  {Ising}},\ }\href@noop {} {\bibfield  {journal} {\bibinfo  {journal}
  {Z.Phys.}\ }\textbf {\bibinfo {volume} {31}},\ \bibinfo {pages} {253}
  (\bibinfo {year} {1925})}\BibitemShut {NoStop}%
\bibitem [{\citenamefont {Watson}\ \emph {et~al.}(1969)\citenamefont {Watson},
  \citenamefont {Blume},\ and\ \citenamefont {Vineyard}}]{watson1969}%
  \BibitemOpen
  \bibfield  {author} {\bibinfo {author} {\bibfnamefont {R.~E.}\ \bibnamefont
  {Watson}}, \bibinfo {author} {\bibfnamefont {M.}~\bibnamefont {Blume}}, \
  and\ \bibinfo {author} {\bibfnamefont {G.~H.}\ \bibnamefont {Vineyard}},\
  }\href {\doibase 10.1103/PhysRev.181.811} {\bibfield  {journal} {\bibinfo
  {journal} {Phys. Rev.}\ }\textbf {\bibinfo {volume} {181}},\ \bibinfo {pages}
  {811} (\bibinfo {year} {1969})}\BibitemShut {NoStop}%
\bibitem [{\citenamefont {Binder}\ and\ \citenamefont
  {Rauch}(1969)}]{binder1969}%
  \BibitemOpen
  \bibfield  {author} {\bibinfo {author} {\bibfnamefont {K.}~\bibnamefont
  {Binder}}\ and\ \bibinfo {author} {\bibfnamefont {H.}~\bibnamefont {Rauch}},\
  }\href {\doibase 10.1007/BF01397564} {\bibfield  {journal} {\bibinfo
  {journal} {Zeitschrift f√ºr Physik}\ }\textbf {\bibinfo {volume} {219}},\
  \bibinfo {pages} {201} (\bibinfo {year} {1969})}\BibitemShut {NoStop}%
\bibitem [{\citenamefont {Kodama}\ and\ \citenamefont
  {Berkowitz}(1999)}]{Kodama:1999wk}%
  \BibitemOpen
  \bibfield  {author} {\bibinfo {author} {\bibfnamefont {R.}~\bibnamefont
  {Kodama}}\ and\ \bibinfo {author} {\bibfnamefont {A.}~\bibnamefont
  {Berkowitz}},\ }\href@noop {} {\bibfield  {journal} {\bibinfo  {journal}
  {Phys. Rev. B}\ }\textbf {\bibinfo {volume} {59}},\ \bibinfo {pages} {6321}
  (\bibinfo {year} {1999})}\BibitemShut {NoStop}%
\bibitem [{\citenamefont {Mitsumata}\ \emph {et~al.}(2003)\citenamefont
  {Mitsumata}, \citenamefont {Sakuma},\ and\ \citenamefont
  {Fukamichi}}]{Mitsumata:2003hd}%
  \BibitemOpen
  \bibfield  {author} {\bibinfo {author} {\bibfnamefont {C.}~\bibnamefont
  {Mitsumata}}, \bibinfo {author} {\bibfnamefont {A.}~\bibnamefont {Sakuma}}, \
  and\ \bibinfo {author} {\bibfnamefont {K.}~\bibnamefont {Fukamichi}},\ }\href
  {\doibase 10.1103/PhysRevB.68.014437} {\bibfield  {journal} {\bibinfo
  {journal} {Phys. Rev. B}\ }\textbf {\bibinfo {volume} {68}},\ \bibinfo
  {pages} {014437} (\bibinfo {year} {2003})}\BibitemShut {NoStop}%
\bibitem [{\citenamefont {Nowak}(2007)}]{Nowakspinmodels}%
  \BibitemOpen
  \bibfield  {author} {\bibinfo {author} {\bibfnamefont {U.}~\bibnamefont
  {Nowak}},\ }\enquote {\bibinfo {title} {Classical spin models},}\ in\ \href
  {\doibase 10.1002/9780470022184.hmm205} {\emph {\bibinfo {booktitle}
  {Handbook of Magnetism and Advanced Magnetic Materials}}}\ (\bibinfo
  {publisher} {John Wiley \& Sons, Ltd},\ \bibinfo {year} {2007})\BibitemShut
  {NoStop}%
\bibitem [{\citenamefont {Boerner}\ \emph {et~al.}(2005)\citenamefont
  {Boerner}, \citenamefont {Chubykalo-Fesenko}, \citenamefont {Chantrell},
  \citenamefont {Heinonen},\ and\ \citenamefont {Mryasov}}]{Boerner2005}%
  \BibitemOpen
  \bibfield  {author} {\bibinfo {author} {\bibfnamefont {E.}~\bibnamefont
  {Boerner}}, \bibinfo {author} {\bibfnamefont {O.}~\bibnamefont
  {Chubykalo-Fesenko}}, \bibinfo {author} {\bibfnamefont {R.}~\bibnamefont
  {Chantrell}}, \bibinfo {author} {\bibfnamefont {O.}~\bibnamefont {Heinonen}},
  \ and\ \bibinfo {author} {\bibfnamefont {O.}~\bibnamefont {Mryasov}},\
  }\href@noop {} {\bibfield  {journal} {\bibinfo  {journal} {IEEE Trans.
  Magn.}\ }\textbf {\bibinfo {volume} {41}},\ \bibinfo {pages} {936} (\bibinfo
  {year} {2005})}\BibitemShut {NoStop}%
\bibitem [{\citenamefont {Skubic}\ \emph {et~al.}(2008)\citenamefont {Skubic},
  \citenamefont {Hellsvik}, \citenamefont {Nordstr{\"o}m},\ and\ \citenamefont
  {Eriksson}}]{Skubic:2008gs}%
  \BibitemOpen
  \bibfield  {author} {\bibinfo {author} {\bibfnamefont {B.}~\bibnamefont
  {Skubic}}, \bibinfo {author} {\bibfnamefont {J.}~\bibnamefont {Hellsvik}},
  \bibinfo {author} {\bibfnamefont {L.}~\bibnamefont {Nordstr{\"o}m}}, \ and\
  \bibinfo {author} {\bibfnamefont {O.}~\bibnamefont {Eriksson}},\ }\href
  {\doibase 10.1088/0953-8984/20/31/315203} {\bibfield  {journal} {\bibinfo
  {journal} {J. Phys.: Condens. Matter}\ }\textbf {\bibinfo {volume} {20}},\
  \bibinfo {pages} {315203} (\bibinfo {year} {2008})}\BibitemShut {NoStop}%
\bibitem [{\citenamefont {Ostler}\ \emph {et~al.}(2011)\citenamefont {Ostler},
  \citenamefont {Evans}, \citenamefont {Chantrell}, \citenamefont {Atxitia},
  \citenamefont {Chubykalo-Fesenko}, \citenamefont {Radu}, \citenamefont
  {Abrudan}, \citenamefont {Radu}, \citenamefont {Tsukamoto}, \citenamefont
  {Itoh}, \citenamefont {Kirilyuk}, \citenamefont {Rasing},\ and\ \citenamefont
  {Kimel}}]{Ostler:2011jf}%
  \BibitemOpen
  \bibfield  {author} {\bibinfo {author} {\bibfnamefont {T.}~\bibnamefont
  {Ostler}}, \bibinfo {author} {\bibfnamefont {R.~F.~L.}\ \bibnamefont
  {Evans}}, \bibinfo {author} {\bibfnamefont {R.~W.}\ \bibnamefont
  {Chantrell}}, \bibinfo {author} {\bibfnamefont {U.}~\bibnamefont {Atxitia}},
  \bibinfo {author} {\bibfnamefont {O.}~\bibnamefont {Chubykalo-Fesenko}},
  \bibinfo {author} {\bibfnamefont {I.}~\bibnamefont {Radu}}, \bibinfo {author}
  {\bibfnamefont {R.}~\bibnamefont {Abrudan}}, \bibinfo {author} {\bibfnamefont
  {F.}~\bibnamefont {Radu}}, \bibinfo {author} {\bibfnamefont {A.}~\bibnamefont
  {Tsukamoto}}, \bibinfo {author} {\bibfnamefont {A.}~\bibnamefont {Itoh}},
  \bibinfo {author} {\bibfnamefont {A.}~\bibnamefont {Kirilyuk}}, \bibinfo
  {author} {\bibfnamefont {T.}~\bibnamefont {Rasing}}, \ and\ \bibinfo {author}
  {\bibfnamefont {A.}~\bibnamefont {Kimel}},\ }\href@noop {} {\bibfield
  {journal} {\bibinfo  {journal} {Phys. Rev. B}\ }\textbf {\bibinfo {volume}
  {84}} (\bibinfo {year} {2011})}\BibitemShut {NoStop}%
\bibitem [{\citenamefont {Ostler}\ \emph {et~al.}(2012)\citenamefont {Ostler},
  \citenamefont {Barker}, \citenamefont {Evans}, \citenamefont {Chantrell},
  \citenamefont {Atxitia}, \citenamefont {Chubykalo-Fesenko}, \citenamefont
  {El~Moussaoui}, \citenamefont {Le~Guyader}, \citenamefont {Mengotti},
  \citenamefont {Heyderman}, \citenamefont {Nolting}, \citenamefont
  {Tsukamoto}, \citenamefont {Itoh}, \citenamefont {Afanasiev}, \citenamefont
  {Ivanov}, \citenamefont {Kalashnikova}, \citenamefont {Vahaplar},
  \citenamefont {Mentink}, \citenamefont {Kirilyuk}, \citenamefont {Rasing},\
  and\ \citenamefont {Kimel}}]{Ostler:2012hx}%
  \BibitemOpen
  \bibfield  {author} {\bibinfo {author} {\bibfnamefont {T.~A.}\ \bibnamefont
  {Ostler}}, \bibinfo {author} {\bibfnamefont {J.}~\bibnamefont {Barker}},
  \bibinfo {author} {\bibfnamefont {R.~F.~L.}\ \bibnamefont {Evans}}, \bibinfo
  {author} {\bibfnamefont {R.~W.}\ \bibnamefont {Chantrell}}, \bibinfo {author}
  {\bibfnamefont {U.}~\bibnamefont {Atxitia}}, \bibinfo {author} {\bibfnamefont
  {O.}~\bibnamefont {Chubykalo-Fesenko}}, \bibinfo {author} {\bibfnamefont
  {S.}~\bibnamefont {El~Moussaoui}}, \bibinfo {author} {\bibfnamefont
  {L.}~\bibnamefont {Le~Guyader}}, \bibinfo {author} {\bibfnamefont
  {E.}~\bibnamefont {Mengotti}}, \bibinfo {author} {\bibfnamefont {L.~J.}\
  \bibnamefont {Heyderman}}, \bibinfo {author} {\bibfnamefont {F.}~\bibnamefont
  {Nolting}}, \bibinfo {author} {\bibfnamefont {A.}~\bibnamefont {Tsukamoto}},
  \bibinfo {author} {\bibfnamefont {A.}~\bibnamefont {Itoh}}, \bibinfo {author}
  {\bibfnamefont {D.}~\bibnamefont {Afanasiev}}, \bibinfo {author}
  {\bibfnamefont {B.~A.}\ \bibnamefont {Ivanov}}, \bibinfo {author}
  {\bibfnamefont {A.~M.}\ \bibnamefont {Kalashnikova}}, \bibinfo {author}
  {\bibfnamefont {K.}~\bibnamefont {Vahaplar}}, \bibinfo {author}
  {\bibfnamefont {J.}~\bibnamefont {Mentink}}, \bibinfo {author} {\bibfnamefont
  {A.}~\bibnamefont {Kirilyuk}}, \bibinfo {author} {\bibfnamefont
  {T.}~\bibnamefont {Rasing}}, \ and\ \bibinfo {author} {\bibfnamefont {A.~V.}\
  \bibnamefont {Kimel}},\ }\href {\doibase 10.1038/ncomms1666} {\bibfield
  {journal} {\bibinfo  {journal} {Nature Communications}\ }\textbf {\bibinfo
  {volume} {3}},\ \bibinfo {pages} {666} (\bibinfo {year} {2012})}\BibitemShut
  {NoStop}%
\bibitem [{\citenamefont {Radu}\ \emph {et~al.}(2011)\citenamefont {Radu},
  \citenamefont {Vahaplar}, \citenamefont {Stamm}, \citenamefont {Kachel},
  \citenamefont {Pontius}, \citenamefont {D{\"u}rr}, \citenamefont {Ostler},
  \citenamefont {Barker}, \citenamefont {Evans}, \citenamefont {Chantrell},
  \citenamefont {Tsukamoto}, \citenamefont {Itoh}, \citenamefont {Kirilyuk},
  \citenamefont {Rasing},\ and\ \citenamefont {Kimel}}]{Radu:2011krb}%
  \BibitemOpen
  \bibfield  {author} {\bibinfo {author} {\bibfnamefont {I.}~\bibnamefont
  {Radu}}, \bibinfo {author} {\bibfnamefont {K.}~\bibnamefont {Vahaplar}},
  \bibinfo {author} {\bibfnamefont {C.}~\bibnamefont {Stamm}}, \bibinfo
  {author} {\bibfnamefont {T.}~\bibnamefont {Kachel}}, \bibinfo {author}
  {\bibfnamefont {N.}~\bibnamefont {Pontius}}, \bibinfo {author} {\bibfnamefont
  {H.~A.}\ \bibnamefont {D{\"u}rr}}, \bibinfo {author} {\bibfnamefont {T.~A.}\
  \bibnamefont {Ostler}}, \bibinfo {author} {\bibfnamefont {J.}~\bibnamefont
  {Barker}}, \bibinfo {author} {\bibfnamefont {R.~F.~L.}\ \bibnamefont
  {Evans}}, \bibinfo {author} {\bibfnamefont {R.~W.}\ \bibnamefont
  {Chantrell}}, \bibinfo {author} {\bibfnamefont {A.}~\bibnamefont
  {Tsukamoto}}, \bibinfo {author} {\bibfnamefont {A.}~\bibnamefont {Itoh}},
  \bibinfo {author} {\bibfnamefont {A.}~\bibnamefont {Kirilyuk}}, \bibinfo
  {author} {\bibfnamefont {T.}~\bibnamefont {Rasing}}, \ and\ \bibinfo {author}
  {\bibfnamefont {A.~V.}\ \bibnamefont {Kimel}},\ }\href {\doibase
  10.1038/nature09901} {\bibfield  {journal} {\bibinfo  {journal} {Nature}\
  }\textbf {\bibinfo {volume} {472}},\ \bibinfo {pages} {205} (\bibinfo {year}
  {2011})}\BibitemShut {NoStop}%
\bibitem [{\citenamefont {Iglesias}\ \emph {et~al.}(2005)\citenamefont
  {Iglesias}, \citenamefont {Batlle},\ and\ \citenamefont
  {Labarta}}]{Iglesias:2005cl}%
  \BibitemOpen
  \bibfield  {author} {\bibinfo {author} {\bibfnamefont {{\`O}.}~\bibnamefont
  {Iglesias}}, \bibinfo {author} {\bibfnamefont {X.}~\bibnamefont {Batlle}}, \
  and\ \bibinfo {author} {\bibfnamefont {A.}~\bibnamefont {Labarta}},\ }\href
  {\doibase 10.1103/PhysRevB.72.212401} {\bibfield  {journal} {\bibinfo
  {journal} {Phys. Rev. B}\ }\textbf {\bibinfo {volume} {72}},\ \bibinfo
  {pages} {212401} (\bibinfo {year} {2005})}\BibitemShut {NoStop}%
\bibitem [{\citenamefont {Evans}\ \emph {et~al.}(2011)\citenamefont {Evans},
  \citenamefont {Bate}, \citenamefont {Chantrell}, \citenamefont {Yanes},\ and\
  \citenamefont {Chubykalo-Fesenko}}]{Evans:2011tb}%
  \BibitemOpen
  \bibfield  {author} {\bibinfo {author} {\bibfnamefont {R.~F.~L.}\
  \bibnamefont {Evans}}, \bibinfo {author} {\bibfnamefont {D.}~\bibnamefont
  {Bate}}, \bibinfo {author} {\bibfnamefont {R.~W.}\ \bibnamefont {Chantrell}},
  \bibinfo {author} {\bibfnamefont {R.}~\bibnamefont {Yanes}}, \ and\ \bibinfo
  {author} {\bibfnamefont {O.}~\bibnamefont {Chubykalo-Fesenko}},\ }\href@noop
  {} {\bibfield  {journal} {\bibinfo  {journal} {Phys. Rev. B}\ }\textbf
  {\bibinfo {volume} {84}},\ \bibinfo {pages} {092404} (\bibinfo {year}
  {2011})}\BibitemShut {NoStop}%
\bibitem [{\citenamefont {Evans}\ \emph {et~al.}(2009)\citenamefont {Evans},
  \citenamefont {Yanes}, \citenamefont {Mryasov}, \citenamefont {Chantrell},\
  and\ \citenamefont {Chubykalo-Fesenko}}]{Evans:2009kf}%
  \BibitemOpen
  \bibfield  {author} {\bibinfo {author} {\bibfnamefont {R.~F.~L.}\
  \bibnamefont {Evans}}, \bibinfo {author} {\bibfnamefont {R.}~\bibnamefont
  {Yanes}}, \bibinfo {author} {\bibfnamefont {O.}~\bibnamefont {Mryasov}},
  \bibinfo {author} {\bibfnamefont {R.~W.}\ \bibnamefont {Chantrell}}, \ and\
  \bibinfo {author} {\bibfnamefont {O.}~\bibnamefont {Chubykalo-Fesenko}},\
  }\href {\doibase 10.1209/0295-5075/88/57004} {\bibfield  {journal} {\bibinfo
  {journal} {EPL (Europhysics Letters)}\ }\textbf {\bibinfo {volume} {88}},\
  \bibinfo {pages} {57004} (\bibinfo {year} {2009})}\BibitemShut {NoStop}%
\bibitem [{\citenamefont {Ali}\ \emph {et~al.}(2003)\citenamefont {Ali},
  \citenamefont {Marrows}, \citenamefont {Al-Jawad}, \citenamefont {Hickey},
  \citenamefont {Misra}, \citenamefont {Nowak},\ and\ \citenamefont
  {Usadel}}]{Ali:2003vi}%
  \BibitemOpen
  \bibfield  {author} {\bibinfo {author} {\bibfnamefont {M.}~\bibnamefont
  {Ali}}, \bibinfo {author} {\bibfnamefont {C.~H.}\ \bibnamefont {Marrows}},
  \bibinfo {author} {\bibfnamefont {M.}~\bibnamefont {Al-Jawad}}, \bibinfo
  {author} {\bibfnamefont {B.~J.}\ \bibnamefont {Hickey}}, \bibinfo {author}
  {\bibfnamefont {A.}~\bibnamefont {Misra}}, \bibinfo {author} {\bibfnamefont
  {U.}~\bibnamefont {Nowak}}, \ and\ \bibinfo {author} {\bibfnamefont {K.-D.}\
  \bibnamefont {Usadel}},\ }\href
  {http://prb.aps.org/abstract/PRB/v68/i21/e214420} {\bibfield  {journal}
  {\bibinfo  {journal} {Phys. Rev. B}\ }\textbf {\bibinfo {volume} {68}},\
  \bibinfo {pages} {214420} (\bibinfo {year} {2003})}\BibitemShut {NoStop}%
\bibitem [{\citenamefont {Garanin}\ and\ \citenamefont
  {Kachkachi}(2003)}]{Garanin:2003ho}%
  \BibitemOpen
  \bibfield  {author} {\bibinfo {author} {\bibfnamefont {D.}~\bibnamefont
  {Garanin}}\ and\ \bibinfo {author} {\bibfnamefont {H.}~\bibnamefont
  {Kachkachi}},\ }\href {\doibase 10.1103/PhysRevLett.90.065504} {\bibfield
  {journal} {\bibinfo  {journal} {Phys. Rev. Lett.}\ }\textbf {\bibinfo
  {volume} {90}},\ \bibinfo {pages} {065504} (\bibinfo {year}
  {2003})}\BibitemShut {NoStop}%
\bibitem [{\citenamefont {Yanes}\ \emph {et~al.}(2007)\citenamefont {Yanes},
  \citenamefont {Chubykalo-Fesenko}, \citenamefont {Kachkachi}, \citenamefont
  {Garanin}, \citenamefont {Evans},\ and\ \citenamefont
  {Chantrell}}]{Yanes:2007gu}%
  \BibitemOpen
  \bibfield  {author} {\bibinfo {author} {\bibfnamefont {R.}~\bibnamefont
  {Yanes}}, \bibinfo {author} {\bibfnamefont {O.}~\bibnamefont
  {Chubykalo-Fesenko}}, \bibinfo {author} {\bibfnamefont {H.}~\bibnamefont
  {Kachkachi}}, \bibinfo {author} {\bibfnamefont {D.}~\bibnamefont {Garanin}},
  \bibinfo {author} {\bibfnamefont {R.}~\bibnamefont {Evans}}, \ and\ \bibinfo
  {author} {\bibfnamefont {R.}~\bibnamefont {Chantrell}},\ }\href {\doibase
  10.1103/PhysRevB.76.064416} {\bibfield  {journal} {\bibinfo  {journal} {Phys.
  Rev. B}\ }\textbf {\bibinfo {volume} {76}},\ \bibinfo {pages} {064416}
  (\bibinfo {year} {2007})}\BibitemShut {NoStop}%
\bibitem [{\citenamefont {Ma}\ \emph {et~al.}(2012{\natexlab{a}})\citenamefont
  {Ma}, \citenamefont {Dudarev},\ and\ \citenamefont {Woo}}]{Ma:2012el}%
  \BibitemOpen
  \bibfield  {author} {\bibinfo {author} {\bibfnamefont {P.-W.}\ \bibnamefont
  {Ma}}, \bibinfo {author} {\bibfnamefont {S.~L.}\ \bibnamefont {Dudarev}}, \
  and\ \bibinfo {author} {\bibfnamefont {C.~H.}\ \bibnamefont {Woo}},\ }\href
  {\doibase 10.1063/1.3673859} {\bibfield  {journal} {\bibinfo  {journal} {J.
  Appl. Phys.}\ }\textbf {\bibinfo {volume} {111}},\ \bibinfo {pages} {07D114}
  (\bibinfo {year} {2012}{\natexlab{a}})}\BibitemShut {NoStop}%
\bibitem [{\citenamefont {Evans}\ \emph {et~al.}(2006)\citenamefont {Evans},
  \citenamefont {Nowak}, \citenamefont {Dorfbauer}, \citenamefont {Shrefl},
  \citenamefont {Mryasov}, \citenamefont {Chantrell},\ and\ \citenamefont
  {Grochola}}]{Evans:2006cf}%
  \BibitemOpen
  \bibfield  {author} {\bibinfo {author} {\bibfnamefont {R.}~\bibnamefont
  {Evans}}, \bibinfo {author} {\bibfnamefont {U.}~\bibnamefont {Nowak}},
  \bibinfo {author} {\bibfnamefont {F.}~\bibnamefont {Dorfbauer}}, \bibinfo
  {author} {\bibfnamefont {T.}~\bibnamefont {Shrefl}}, \bibinfo {author}
  {\bibfnamefont {O.}~\bibnamefont {Mryasov}}, \bibinfo {author} {\bibfnamefont
  {R.~W.}\ \bibnamefont {Chantrell}}, \ and\ \bibinfo {author} {\bibfnamefont
  {G.}~\bibnamefont {Grochola}},\ }\href {\doibase 10.1063/1.2167636}
  {\bibfield  {journal} {\bibinfo  {journal} {J. Appl. Phys.}\ }\textbf
  {\bibinfo {volume} {99}},\ \bibinfo {pages} {08G703} (\bibinfo {year}
  {2006})}\BibitemShut {NoStop}%
\bibitem [{\citenamefont {Evans}\ \emph {et~al.}(2007)\citenamefont {Evans},
  \citenamefont {Dorfbauer}, \citenamefont {Myrasov}, \citenamefont
  {Chubykalo-Fesenko}, \citenamefont {Schrefl},\ and\ \citenamefont
  {Chantrell}}]{Evans:2007tw}%
  \BibitemOpen
  \bibfield  {author} {\bibinfo {author} {\bibfnamefont {R.}~\bibnamefont
  {Evans}}, \bibinfo {author} {\bibfnamefont {F.}~\bibnamefont {Dorfbauer}},
  \bibinfo {author} {\bibfnamefont {O.}~\bibnamefont {Myrasov}}, \bibinfo
  {author} {\bibfnamefont {O.}~\bibnamefont {Chubykalo-Fesenko}}, \bibinfo
  {author} {\bibfnamefont {T.}~\bibnamefont {Schrefl}}, \ and\ \bibinfo
  {author} {\bibfnamefont {R.}~\bibnamefont {Chantrell}},\ }\href@noop {}
  {\bibfield  {journal} {\bibinfo  {journal} {IEEE Tran. Magn.}\ }\textbf
  {\bibinfo {volume} {43}},\ \bibinfo {pages} {3106} (\bibinfo {year}
  {2007})}\BibitemShut {NoStop}%
\bibitem [{\citenamefont {Ho}\ \emph {et~al.}(2011)\citenamefont {Ho},
  \citenamefont {Evans}, \citenamefont {Chantrell}, \citenamefont {Han},
  \citenamefont {Chow},\ and\ \citenamefont {Chen}}]{Ho:eb}%
  \BibitemOpen
  \bibfield  {author} {\bibinfo {author} {\bibfnamefont {P.}~\bibnamefont
  {Ho}}, \bibinfo {author} {\bibfnamefont {R.~F.~L.}\ \bibnamefont {Evans}},
  \bibinfo {author} {\bibfnamefont {R.~W.}\ \bibnamefont {Chantrell}}, \bibinfo
  {author} {\bibfnamefont {G.}~\bibnamefont {Han}}, \bibinfo {author}
  {\bibfnamefont {G.-M.}\ \bibnamefont {Chow}}, \ and\ \bibinfo {author}
  {\bibfnamefont {J.}~\bibnamefont {Chen}},\ }\href {\doibase
  10.1109/TMAG.2011.2147765} {\bibfield  {journal} {\bibinfo  {journal} {IEEE
  Trans. Magn}\ }\textbf {\bibinfo {volume} {47}},\ \bibinfo {pages} {2646}
  (\bibinfo {year} {2011})}\BibitemShut {NoStop}%
\bibitem [{\citenamefont {Chureemart}\ \emph {et~al.}(2011)\citenamefont
  {Chureemart}, \citenamefont {Evans},\ and\ \citenamefont
  {Chantrell}}]{Chureemart:2011cr}%
  \BibitemOpen
  \bibfield  {author} {\bibinfo {author} {\bibfnamefont {P.}~\bibnamefont
  {Chureemart}}, \bibinfo {author} {\bibfnamefont {R.~F.~L.}\ \bibnamefont
  {Evans}}, \ and\ \bibinfo {author} {\bibfnamefont {R.~W.}\ \bibnamefont
  {Chantrell}},\ }\href {http://link.aps.org/doi/10.1103/PhysRevB.83.184416}
  {\bibfield  {journal} {\bibinfo  {journal} {Phys. Rev. B}\ }\textbf {\bibinfo
  {volume} {83}} (\bibinfo {year} {2011})}\BibitemShut {NoStop}%
\bibitem [{\citenamefont {Asselin}\ \emph {et~al.}(2010)\citenamefont
  {Asselin}, \citenamefont {Evans}, \citenamefont {Barker}, \citenamefont
  {Chantrell}, \citenamefont {Yanes}, \citenamefont {Chubykalo-Fesenko},
  \citenamefont {Hinzke},\ and\ \citenamefont {Nowak}}]{Asselin:2010iv}%
  \BibitemOpen
  \bibfield  {author} {\bibinfo {author} {\bibfnamefont {P.}~\bibnamefont
  {Asselin}}, \bibinfo {author} {\bibfnamefont {R.~F.~L.}\ \bibnamefont
  {Evans}}, \bibinfo {author} {\bibfnamefont {J.}~\bibnamefont {Barker}},
  \bibinfo {author} {\bibfnamefont {R.~W.}\ \bibnamefont {Chantrell}}, \bibinfo
  {author} {\bibfnamefont {R.}~\bibnamefont {Yanes}}, \bibinfo {author}
  {\bibfnamefont {O.}~\bibnamefont {Chubykalo-Fesenko}}, \bibinfo {author}
  {\bibfnamefont {D.}~\bibnamefont {Hinzke}}, \ and\ \bibinfo {author}
  {\bibfnamefont {U.}~\bibnamefont {Nowak}},\ }\href@noop {} {\bibfield
  {journal} {\bibinfo  {journal} {Phys. Rev. B}\ }\textbf {\bibinfo {volume}
  {82}},\ \bibinfo {pages} {054415} (\bibinfo {year} {2010})}\BibitemShut
  {NoStop}%
\bibitem [{\citenamefont {Barker}\ \emph {et~al.}(2010)\citenamefont {Barker},
  \citenamefont {Evans}, \citenamefont {Chantrell}, \citenamefont {Hinzke},\
  and\ \citenamefont {Nowak}}]{Barker:2010hfa}%
  \BibitemOpen
  \bibfield  {author} {\bibinfo {author} {\bibfnamefont {J.}~\bibnamefont
  {Barker}}, \bibinfo {author} {\bibfnamefont {R.~F.~L.}\ \bibnamefont
  {Evans}}, \bibinfo {author} {\bibfnamefont {R.~W.}\ \bibnamefont
  {Chantrell}}, \bibinfo {author} {\bibfnamefont {D.}~\bibnamefont {Hinzke}}, \
  and\ \bibinfo {author} {\bibfnamefont {U.}~\bibnamefont {Nowak}},\ }\href
  {\doibase 10.1063/1.3515928} {\bibfield  {journal} {\bibinfo  {journal}
  {Appl. Phys. Lett.}\ }\textbf {\bibinfo {volume} {97}},\ \bibinfo {pages}
  {192504} (\bibinfo {year} {2010})}\BibitemShut {NoStop}%
\bibitem [{\citenamefont {Yanes}\ \emph {et~al.}(2010)\citenamefont {Yanes},
  \citenamefont {Chubykalo-Fesenko}, \citenamefont {Evans},\ and\ \citenamefont
  {Chantrell}}]{Yanes:2010we}%
  \BibitemOpen
  \bibfield  {author} {\bibinfo {author} {\bibfnamefont {R.}~\bibnamefont
  {Yanes}}, \bibinfo {author} {\bibfnamefont {O.}~\bibnamefont
  {Chubykalo-Fesenko}}, \bibinfo {author} {\bibfnamefont {R.~F.~L.}\
  \bibnamefont {Evans}}, \ and\ \bibinfo {author} {\bibfnamefont {R.~W.}\
  \bibnamefont {Chantrell}},\ }\href@noop {} {\bibfield  {journal} {\bibinfo
  {journal} {J. Phys. D: Appl. Phys.}\ }\textbf {\bibinfo {volume} {43}},\
  \bibinfo {pages} {474009} (\bibinfo {year} {2010})}\BibitemShut {NoStop}%
\bibitem [{\citenamefont {Bauer}\ \emph {et~al.}(2011)\citenamefont {Bauer},
  \citenamefont {Mavropoulos}, \citenamefont {Lounis},\ and\ \citenamefont
  {Bl\"{u}gel}}]{Bauer2011}%
  \BibitemOpen
  \bibfield  {author} {\bibinfo {author} {\bibfnamefont {D.~S.~G.}\
  \bibnamefont {Bauer}}, \bibinfo {author} {\bibfnamefont {P.}~\bibnamefont
  {Mavropoulos}}, \bibinfo {author} {\bibfnamefont {S.}~\bibnamefont {Lounis}},
  \ and\ \bibinfo {author} {\bibfnamefont {S.}~\bibnamefont {Bl\"{u}gel}},\
  }\href {http://stacks.iop.org/0953-8984/23/i=39/a=394204} {\bibfield
  {journal} {\bibinfo  {journal} {Journal of Physics: Condensed Matter}\
  }\textbf {\bibinfo {volume} {23}},\ \bibinfo {pages} {394204} (\bibinfo
  {year} {2011})}\BibitemShut {NoStop}%
\bibitem [{\citenamefont {Fal}\ \emph {et~al.}(2013)\citenamefont {Fal},
  \citenamefont {Plumer}, \citenamefont {Whitehead}, \citenamefont {Mercer},
  \citenamefont {van Ek},\ and\ \citenamefont {Srinivasan}}]{Fal:2013ba}%
  \BibitemOpen
  \bibfield  {author} {\bibinfo {author} {\bibfnamefont {T.~J.}\ \bibnamefont
  {Fal}}, \bibinfo {author} {\bibfnamefont {M.~L.}\ \bibnamefont {Plumer}},
  \bibinfo {author} {\bibfnamefont {J.~P.}\ \bibnamefont {Whitehead}}, \bibinfo
  {author} {\bibfnamefont {J.~I.}\ \bibnamefont {Mercer}}, \bibinfo {author}
  {\bibfnamefont {J.}~\bibnamefont {van Ek}}, \ and\ \bibinfo {author}
  {\bibfnamefont {K.}~\bibnamefont {Srinivasan}},\ }\href {\doibase
  10.1063/1.4807501} {\bibfield  {journal} {\bibinfo  {journal} {Appl. Phys.
  Lett.}\ }\textbf {\bibinfo {volume} {102}},\ \bibinfo {pages} {202404}
  (\bibinfo {year} {2013})}\BibitemShut {NoStop}%
\bibitem [{\citenamefont {Victora}\ and\ \citenamefont
  {Huang}(2013)}]{Victora:2013ki}%
  \BibitemOpen
  \bibfield  {author} {\bibinfo {author} {\bibfnamefont {R.~H.}\ \bibnamefont
  {Victora}}\ and\ \bibinfo {author} {\bibfnamefont {P.-W.}\ \bibnamefont
  {Huang}},\ }\href {\doibase 10.1109/TMAG.2012.2219300} {\bibfield  {journal}
  {\bibinfo  {journal} {IEEE Trans. Magn}\ }\textbf {\bibinfo {volume} {49}},\
  \bibinfo {pages} {751} (\bibinfo {year} {2013})}\BibitemShut {NoStop}%
\bibitem [{\citenamefont {Mryasov}\ \emph {et~al.}(2005)\citenamefont
  {Mryasov}, \citenamefont {Nowak}, \citenamefont {Guslienko},\ and\
  \citenamefont {Chantrell}}]{Mryasov:2005kj}%
  \BibitemOpen
  \bibfield  {author} {\bibinfo {author} {\bibfnamefont {O.}~\bibnamefont
  {Mryasov}}, \bibinfo {author} {\bibfnamefont {U.}~\bibnamefont {Nowak}},
  \bibinfo {author} {\bibfnamefont {K.}~\bibnamefont {Guslienko}}, \ and\
  \bibinfo {author} {\bibfnamefont {R.~W.}\ \bibnamefont {Chantrell}},\ }\href
  {\doibase 10.1209/epl/i2004-10404-2} {\bibfield  {journal} {\bibinfo
  {journal} {EPL (Europhysics Letters)}\ }\textbf {\bibinfo {volume} {69}},\
  \bibinfo {pages} {805} (\bibinfo {year} {2005})}\BibitemShut {NoStop}%
\bibitem [{\citenamefont {Szunyogh}\ \emph {et~al.}(2011)\citenamefont
  {Szunyogh}, \citenamefont {Udvardi}, \citenamefont {Jackson}, \citenamefont
  {Nowak},\ and\ \citenamefont {Chantrell}}]{Szunyogh:2011db}%
  \BibitemOpen
  \bibfield  {author} {\bibinfo {author} {\bibfnamefont {L.}~\bibnamefont
  {Szunyogh}}, \bibinfo {author} {\bibfnamefont {L.}~\bibnamefont {Udvardi}},
  \bibinfo {author} {\bibfnamefont {J.}~\bibnamefont {Jackson}}, \bibinfo
  {author} {\bibfnamefont {U.}~\bibnamefont {Nowak}}, \ and\ \bibinfo {author}
  {\bibfnamefont {R.~W.}\ \bibnamefont {Chantrell}},\ }\href
  {http://link.aps.org/doi/10.1103/PhysRevB.83.024401} {\bibfield  {journal}
  {\bibinfo  {journal} {Phys. Rev. B}\ }\textbf {\bibinfo {volume} {83}}
  (\bibinfo {year} {2011})}\BibitemShut {NoStop}%
\bibitem [{\citenamefont {Szunyogh}\ \emph {et~al.}(2009)\citenamefont
  {Szunyogh}, \citenamefont {Lazarovits}, \citenamefont {Udvardi},
  \citenamefont {Jackson},\ and\ \citenamefont {Nowak}}]{Szunyogh:2009kf}%
  \BibitemOpen
  \bibfield  {author} {\bibinfo {author} {\bibfnamefont {L.}~\bibnamefont
  {Szunyogh}}, \bibinfo {author} {\bibfnamefont {B.}~\bibnamefont
  {Lazarovits}}, \bibinfo {author} {\bibfnamefont {L.}~\bibnamefont {Udvardi}},
  \bibinfo {author} {\bibfnamefont {J.}~\bibnamefont {Jackson}}, \ and\
  \bibinfo {author} {\bibfnamefont {U.}~\bibnamefont {Nowak}},\ }\href
  {http://link.aps.org/doi/10.1103/PhysRevB.79.020403} {\bibfield  {journal}
  {\bibinfo  {journal} {Phys. Rev. B}\ }\textbf {\bibinfo {volume} {79}}
  (\bibinfo {year} {2009})}\BibitemShut {NoStop}%
\bibitem [{\citenamefont {Sandratskii}\ and\ \citenamefont
  {Mavropoulos}(2011)}]{FeRh2011}%
  \BibitemOpen
  \bibfield  {author} {\bibinfo {author} {\bibfnamefont {L.~M.}\ \bibnamefont
  {Sandratskii}}\ and\ \bibinfo {author} {\bibfnamefont {P.}~\bibnamefont
  {Mavropoulos}},\ }\href {\doibase 10.1103/PhysRevB.83.174408} {\bibfield
  {journal} {\bibinfo  {journal} {Phys. Rev. B}\ }\textbf {\bibinfo {volume}
  {83}},\ \bibinfo {pages} {174408} (\bibinfo {year} {2011})}\BibitemShut
  {NoStop}%
\bibitem [{\citenamefont {Kazantseva}\ \emph {et~al.}(2008)\citenamefont
  {Kazantseva}, \citenamefont {Hinzke}, \citenamefont {Nowak}, \citenamefont
  {Chantrell}, \citenamefont {Atxitia},\ and\ \citenamefont
  {Chubykalo-Fesenko}}]{Kazantseva:2008ja}%
  \BibitemOpen
  \bibfield  {author} {\bibinfo {author} {\bibfnamefont {N.}~\bibnamefont
  {Kazantseva}}, \bibinfo {author} {\bibfnamefont {D.}~\bibnamefont {Hinzke}},
  \bibinfo {author} {\bibfnamefont {U.}~\bibnamefont {Nowak}}, \bibinfo
  {author} {\bibfnamefont {R.}~\bibnamefont {Chantrell}}, \bibinfo {author}
  {\bibfnamefont {U.}~\bibnamefont {Atxitia}}, \ and\ \bibinfo {author}
  {\bibfnamefont {O.}~\bibnamefont {Chubykalo-Fesenko}},\ }\href {\doibase
  10.1103/PhysRevB.77.184428} {\bibfield  {journal} {\bibinfo  {journal} {Phys.
  Rev. B}\ }\textbf {\bibinfo {volume} {77}},\ \bibinfo {pages} {184428}
  (\bibinfo {year} {2008})}\BibitemShut {NoStop}%
\bibitem [{\citenamefont {Atxitia}\ \emph {et~al.}(2010)\citenamefont
  {Atxitia}, \citenamefont {Hinzke}, \citenamefont {Chubykalo-Fesenko},
  \citenamefont {Nowak}, \citenamefont {Kachkachi}, \citenamefont {Mryasov},
  \citenamefont {Evans},\ and\ \citenamefont {Chantrell}}]{Atxitia:2010fu}%
  \BibitemOpen
  \bibfield  {author} {\bibinfo {author} {\bibfnamefont {U.}~\bibnamefont
  {Atxitia}}, \bibinfo {author} {\bibfnamefont {D.}~\bibnamefont {Hinzke}},
  \bibinfo {author} {\bibfnamefont {O.}~\bibnamefont {Chubykalo-Fesenko}},
  \bibinfo {author} {\bibfnamefont {U.}~\bibnamefont {Nowak}}, \bibinfo
  {author} {\bibfnamefont {H.}~\bibnamefont {Kachkachi}}, \bibinfo {author}
  {\bibfnamefont {O.}~\bibnamefont {Mryasov}}, \bibinfo {author} {\bibfnamefont
  {R.~F.~L.}\ \bibnamefont {Evans}}, \ and\ \bibinfo {author} {\bibfnamefont
  {R.~W.}\ \bibnamefont {Chantrell}},\ }\href {\doibase
  10.1103/PhysRevB.82.134440} {\bibfield  {journal} {\bibinfo  {journal} {Phys.
  Rev. B}\ }\textbf {\bibinfo {volume} {82}},\ \bibinfo {pages} {134440}
  (\bibinfo {year} {2010})}\BibitemShut {NoStop}%
\bibitem [{\citenamefont {Jourdan}\ \emph {et~al.}(2008)\citenamefont
  {Jourdan}, \citenamefont {Marty},\ and\ \citenamefont {Lan{\c
  c}on}}]{Jourdan:2008ij}%
  \BibitemOpen
  \bibfield  {author} {\bibinfo {author} {\bibfnamefont {T.}~\bibnamefont
  {Jourdan}}, \bibinfo {author} {\bibfnamefont {A.}~\bibnamefont {Marty}}, \
  and\ \bibinfo {author} {\bibfnamefont {F.}~\bibnamefont {Lan{\c c}on}},\
  }\href {http://link.aps.org/doi/10.1103/PhysRevB.77.224428} {\bibfield
  {journal} {\bibinfo  {journal} {Phys. Rev. B}\ }\textbf {\bibinfo {volume}
  {77}},\ \bibinfo {pages} {224428} (\bibinfo {year} {2008})}\BibitemShut
  {NoStop}%
\bibitem [{\citenamefont {Garcia~Sanchez}\ \emph {et~al.}(2006)\citenamefont
  {Garcia~Sanchez}, \citenamefont {Chubykalo-Fesenko}, \citenamefont
  {Mryasov},\ and\ \citenamefont {Chantrell}}]{GarciaSanchez:2006tt}%
  \BibitemOpen
  \bibfield  {author} {\bibinfo {author} {\bibfnamefont {F.}~\bibnamefont
  {Garcia~Sanchez}}, \bibinfo {author} {\bibfnamefont {O.}~\bibnamefont
  {Chubykalo-Fesenko}}, \bibinfo {author} {\bibfnamefont {O.}~\bibnamefont
  {Mryasov}}, \ and\ \bibinfo {author} {\bibfnamefont {R.~W.}\ \bibnamefont
  {Chantrell}},\ }\href@noop {} {\bibfield  {journal} {\bibinfo  {journal}
  {Physica B: Condensed Matter}\ }\textbf {\bibinfo {volume} {372}},\ \bibinfo
  {pages} {328} (\bibinfo {year} {2006})}\BibitemShut {NoStop}%
\bibitem [{\citenamefont {http:/math.nist.gov/oommf}()}]{oommf}%
  \BibitemOpen
  \bibfield  {author} {\bibinfo {author} {\bibnamefont
  {http:/math.nist.gov/oommf}},\ }\href@noop {} {}\BibitemShut {NoStop}%
\bibitem [{\citenamefont {Scholz}\ \emph {et~al.}(2003)\citenamefont {Scholz},
  \citenamefont {Fidler}, \citenamefont {Schrefl}, \citenamefont {Suess},
  \citenamefont {Dittrich}, \citenamefont {Forster},\ and\ \citenamefont
  {Tsiantos}}]{Scholz:2003ty}%
  \BibitemOpen
  \bibfield  {author} {\bibinfo {author} {\bibfnamefont {W.}~\bibnamefont
  {Scholz}}, \bibinfo {author} {\bibfnamefont {J.}~\bibnamefont {Fidler}},
  \bibinfo {author} {\bibfnamefont {T.}~\bibnamefont {Schrefl}}, \bibinfo
  {author} {\bibfnamefont {D.}~\bibnamefont {Suess}}, \bibinfo {author}
  {\bibfnamefont {R.}~\bibnamefont {Dittrich}}, \bibinfo {author}
  {\bibfnamefont {H.}~\bibnamefont {Forster}}, \ and\ \bibinfo {author}
  {\bibfnamefont {V.}~\bibnamefont {Tsiantos}},\ }\href@noop {} {\bibfield
  {journal} {\bibinfo  {journal} {Computational Materials Science}\ }\textbf
  {\bibinfo {volume} {28}},\ \bibinfo {pages} {366} (\bibinfo {year}
  {2003})}\BibitemShut {NoStop}%
\bibitem [{\citenamefont {Fischbacher}\ \emph {et~al.}(2007)\citenamefont
  {Fischbacher}, \citenamefont {Franchin}, \citenamefont {Bordignon},\ and\
  \citenamefont {Fangohr}}]{Fischbacher:ii}%
  \BibitemOpen
  \bibfield  {author} {\bibinfo {author} {\bibfnamefont {T.}~\bibnamefont
  {Fischbacher}}, \bibinfo {author} {\bibfnamefont {M.}~\bibnamefont
  {Franchin}}, \bibinfo {author} {\bibfnamefont {G.}~\bibnamefont {Bordignon}},
  \ and\ \bibinfo {author} {\bibfnamefont {H.}~\bibnamefont {Fangohr}},\ }\href
  {\doibase 10.1109/TMAG.2007.893843} {\bibfield  {journal} {\bibinfo
  {journal} {IEEE Trans. Magn}\ }\textbf {\bibinfo {volume} {43}},\ \bibinfo
  {pages} {2896} (\bibinfo {year} {2007})}\BibitemShut {NoStop}%
\bibitem [{\citenamefont {Mistral}\ \emph {et~al.}(2008)\citenamefont
  {Mistral}, \citenamefont {van Kampen}, \citenamefont {Hrkac}, \citenamefont
  {Kim}, \citenamefont {Devolder}, \citenamefont {Crozat}, \citenamefont
  {Chappert}, \citenamefont {Lagae},\ and\ \citenamefont
  {Schrefl}}]{Mistral:2008js}%
  \BibitemOpen
  \bibfield  {author} {\bibinfo {author} {\bibfnamefont {Q.}~\bibnamefont
  {Mistral}}, \bibinfo {author} {\bibfnamefont {M.}~\bibnamefont {van Kampen}},
  \bibinfo {author} {\bibfnamefont {G.}~\bibnamefont {Hrkac}}, \bibinfo
  {author} {\bibfnamefont {J.-V.}\ \bibnamefont {Kim}}, \bibinfo {author}
  {\bibfnamefont {T.}~\bibnamefont {Devolder}}, \bibinfo {author}
  {\bibfnamefont {P.}~\bibnamefont {Crozat}}, \bibinfo {author} {\bibfnamefont
  {C.}~\bibnamefont {Chappert}}, \bibinfo {author} {\bibfnamefont
  {L.}~\bibnamefont {Lagae}}, \ and\ \bibinfo {author} {\bibfnamefont
  {T.}~\bibnamefont {Schrefl}},\ }\href {\doibase
  10.1103/PhysRevLett.100.257201} {\bibfield  {journal} {\bibinfo  {journal}
  {Phys. Rev. Lett.}\ }\textbf {\bibinfo {volume} {100}},\ \bibinfo {pages}
  {257201} (\bibinfo {year} {2008})}\BibitemShut {NoStop}%
\bibitem [{\citenamefont {Dobin}\ and\ \citenamefont
  {Richter}(2006)}]{Dobin:2006hv}%
  \BibitemOpen
  \bibfield  {author} {\bibinfo {author} {\bibfnamefont {A.~Y.}\ \bibnamefont
  {Dobin}}\ and\ \bibinfo {author} {\bibfnamefont {H.~J.}\ \bibnamefont
  {Richter}},\ }\href {\doibase 10.1063/1.2335590} {\bibfield  {journal}
  {\bibinfo  {journal} {Appl. Phys. Lett.}\ }\textbf {\bibinfo {volume} {89}},\
  \bibinfo {pages} {062512} (\bibinfo {year} {2006})}\BibitemShut {NoStop}%
\bibitem [{\citenamefont {Schrefl}\ \emph {et~al.}(1997)\citenamefont
  {Schrefl}, \citenamefont {Fidler}, \citenamefont {Kirk},\ and\ \citenamefont
  {Chapman}}]{Schrefl:1997tl}%
  \BibitemOpen
  \bibfield  {author} {\bibinfo {author} {\bibfnamefont {T.}~\bibnamefont
  {Schrefl}}, \bibinfo {author} {\bibfnamefont {J.}~\bibnamefont {Fidler}},
  \bibinfo {author} {\bibfnamefont {K.~J.}\ \bibnamefont {Kirk}}, \ and\
  \bibinfo {author} {\bibfnamefont {J.~N.}\ \bibnamefont {Chapman}},\ }\href
  {http://www.sciencedirect.com/science/article/pii/S030488539700156X}
  {\bibfield  {journal} {\bibinfo  {journal} {J. Magn. Magn. Mater}\ }\textbf
  {\bibinfo {volume} {175}},\ \bibinfo {pages} {193} (\bibinfo {year}
  {1997})}\BibitemShut {NoStop}%
\bibitem [{\citenamefont {Kryder}\ \emph {et~al.}(2008)\citenamefont {Kryder},
  \citenamefont {Gage}, \citenamefont {McDaniel}, \citenamefont {Challener},
  \citenamefont {Rottmayer}, \citenamefont {Ju}, \citenamefont {Hsia},\ and\
  \citenamefont {Erden}}]{Kryder:2008kt}%
  \BibitemOpen
  \bibfield  {author} {\bibinfo {author} {\bibfnamefont {M.~H.}\ \bibnamefont
  {Kryder}}, \bibinfo {author} {\bibfnamefont {E.~C.}\ \bibnamefont {Gage}},
  \bibinfo {author} {\bibfnamefont {T.~W.}\ \bibnamefont {McDaniel}}, \bibinfo
  {author} {\bibfnamefont {W.~A.}\ \bibnamefont {Challener}}, \bibinfo {author}
  {\bibfnamefont {R.~E.}\ \bibnamefont {Rottmayer}}, \bibinfo {author}
  {\bibfnamefont {G.}~\bibnamefont {Ju}}, \bibinfo {author} {\bibfnamefont
  {Y.-T.}\ \bibnamefont {Hsia}}, \ and\ \bibinfo {author} {\bibfnamefont
  {M.~F.}\ \bibnamefont {Erden}},\ }\href {\doibase 10.1109/JPROC.2008.2004315}
  {\bibfield  {journal} {\bibinfo  {journal} {Proceedings of the IEEE}\
  }\textbf {\bibinfo {volume} {96}},\ \bibinfo {pages} {1810} (\bibinfo {year}
  {2008})}\BibitemShut {NoStop}%
\bibitem [{\citenamefont {Beaurepaire}\ \emph {et~al.}(1996)\citenamefont
  {Beaurepaire}, \citenamefont {Merle}, \citenamefont {Daunois},\ and\
  \citenamefont {Bigot}}]{Beaurepaire:1996es}%
  \BibitemOpen
  \bibfield  {author} {\bibinfo {author} {\bibfnamefont {E.}~\bibnamefont
  {Beaurepaire}}, \bibinfo {author} {\bibfnamefont {J.~C.}\ \bibnamefont
  {Merle}}, \bibinfo {author} {\bibfnamefont {A.}~\bibnamefont {Daunois}}, \
  and\ \bibinfo {author} {\bibfnamefont {J.~Y.}\ \bibnamefont {Bigot}},\ }\href
  {\doibase 10.1103/PhysRevLett.76.4250} {\bibfield  {journal} {\bibinfo
  {journal} {Phys. Rev. Lett.}\ }\textbf {\bibinfo {volume} {76}},\ \bibinfo
  {pages} {4250} (\bibinfo {year} {1996})}\BibitemShut {NoStop}%
\bibitem [{\citenamefont {Stanciu}\ \emph {et~al.}(2007)\citenamefont
  {Stanciu}, \citenamefont {Tsukamoto}, \citenamefont {Kimel}, \citenamefont
  {Hansteen}, \citenamefont {Kirilyuk}, \citenamefont {Itoh},\ and\
  \citenamefont {Rasing}}]{Stanciu:2007co}%
  \BibitemOpen
  \bibfield  {author} {\bibinfo {author} {\bibfnamefont {C.}~\bibnamefont
  {Stanciu}}, \bibinfo {author} {\bibfnamefont {A.}~\bibnamefont {Tsukamoto}},
  \bibinfo {author} {\bibfnamefont {A.}~\bibnamefont {Kimel}}, \bibinfo
  {author} {\bibfnamefont {F.}~\bibnamefont {Hansteen}}, \bibinfo {author}
  {\bibfnamefont {A.}~\bibnamefont {Kirilyuk}}, \bibinfo {author}
  {\bibfnamefont {A.}~\bibnamefont {Itoh}}, \ and\ \bibinfo {author}
  {\bibfnamefont {T.}~\bibnamefont {Rasing}},\ }\href {\doibase
  10.1103/PhysRevLett.99.217204} {\bibfield  {journal} {\bibinfo  {journal}
  {Phys. Rev. Lett.}\ }\textbf {\bibinfo {volume} {99}},\ \bibinfo {pages}
  {217204} (\bibinfo {year} {2007})}\BibitemShut {NoStop}%
\bibitem [{\citenamefont {O'Grady}\ \emph {et~al.}(2010)\citenamefont
  {O'Grady}, \citenamefont {Fernandez-Outon},\ and\ \citenamefont
  {Vallejo-Fernandez}}]{OGrady:2010gn}%
  \BibitemOpen
  \bibfield  {author} {\bibinfo {author} {\bibfnamefont {K.}~\bibnamefont
  {O'Grady}}, \bibinfo {author} {\bibfnamefont {L.~E.}\ \bibnamefont
  {Fernandez-Outon}}, \ and\ \bibinfo {author} {\bibfnamefont {G.}~\bibnamefont
  {Vallejo-Fernandez}},\ }\href {\doibase 10.1016/j.jmmm.2009.12.011}
  {\bibfield  {journal} {\bibinfo  {journal} {J. Magn. Magn. Mater.}\ }\textbf
  {\bibinfo {volume} {322}},\ \bibinfo {pages} {883} (\bibinfo {year}
  {2010})}\BibitemShut {NoStop}%
\bibitem [{\citenamefont {Ikeda}\ \emph {et~al.}(2010)\citenamefont {Ikeda},
  \citenamefont {Miura}, \citenamefont {Yamamoto}, \citenamefont {Mizunuma},
  \citenamefont {Gan}, \citenamefont {Endo}, \citenamefont {Kanai},
  \citenamefont {Hayakawa}, \citenamefont {Matsukura},\ and\ \citenamefont
  {Ohno}}]{Ikeda:2010iz}%
  \BibitemOpen
  \bibfield  {author} {\bibinfo {author} {\bibfnamefont {S.}~\bibnamefont
  {Ikeda}}, \bibinfo {author} {\bibfnamefont {K.}~\bibnamefont {Miura}},
  \bibinfo {author} {\bibfnamefont {H.}~\bibnamefont {Yamamoto}}, \bibinfo
  {author} {\bibfnamefont {K.}~\bibnamefont {Mizunuma}}, \bibinfo {author}
  {\bibfnamefont {H.~D.}\ \bibnamefont {Gan}}, \bibinfo {author} {\bibfnamefont
  {M.}~\bibnamefont {Endo}}, \bibinfo {author} {\bibfnamefont {S.}~\bibnamefont
  {Kanai}}, \bibinfo {author} {\bibfnamefont {J.}~\bibnamefont {Hayakawa}},
  \bibinfo {author} {\bibfnamefont {F.}~\bibnamefont {Matsukura}}, \ and\
  \bibinfo {author} {\bibfnamefont {H.}~\bibnamefont {Ohno}},\ }\href {\doibase
  10.1038/nmat2804} {\bibfield  {journal} {\bibinfo  {journal} {Nature
  Materials}\ }\textbf {\bibinfo {volume} {9}},\ \bibinfo {pages} {721}
  (\bibinfo {year} {2010})}\BibitemShut {NoStop}%
\bibitem [{\citenamefont {Jamet}\ \emph {et~al.}(2001)\citenamefont {Jamet},
  \citenamefont {Wernsdorfer}, \citenamefont {Thirion}, \citenamefont {Mailly},
  \citenamefont {Dupuis}, \citenamefont {M{\'e}linon},\ and\ \citenamefont
  {P{\'e}rez}}]{Jamet:2001em}%
  \BibitemOpen
  \bibfield  {author} {\bibinfo {author} {\bibfnamefont {M.}~\bibnamefont
  {Jamet}}, \bibinfo {author} {\bibfnamefont {W.}~\bibnamefont {Wernsdorfer}},
  \bibinfo {author} {\bibfnamefont {C.}~\bibnamefont {Thirion}}, \bibinfo
  {author} {\bibfnamefont {D.}~\bibnamefont {Mailly}}, \bibinfo {author}
  {\bibfnamefont {V.}~\bibnamefont {Dupuis}}, \bibinfo {author} {\bibfnamefont
  {P.}~\bibnamefont {M{\'e}linon}}, \ and\ \bibinfo {author} {\bibfnamefont
  {A.}~\bibnamefont {P{\'e}rez}},\ }\href {\doibase
  10.1103/PhysRevLett.86.4676} {\bibfield  {journal} {\bibinfo  {journal}
  {Phys. Rev. Lett.}\ }\textbf {\bibinfo {volume} {86}},\ \bibinfo {pages}
  {4676} (\bibinfo {year} {2001})}\BibitemShut {NoStop}%
\bibitem [{\citenamefont {Klemmer}\ \emph {et~al.}(2002)\citenamefont
  {Klemmer}, \citenamefont {Shukla}, \citenamefont {Liu}, \citenamefont {Wu},
  \citenamefont {Svedberg}, \citenamefont {Mryasov}, \citenamefont {Chantrell},
  \citenamefont {Weller}, \citenamefont {Tanase},\ and\ \citenamefont
  {Laughlin}}]{Klemmer:2002wq}%
  \BibitemOpen
  \bibfield  {author} {\bibinfo {author} {\bibfnamefont {T.}~\bibnamefont
  {Klemmer}}, \bibinfo {author} {\bibfnamefont {N.}~\bibnamefont {Shukla}},
  \bibinfo {author} {\bibfnamefont {C.}~\bibnamefont {Liu}}, \bibinfo {author}
  {\bibfnamefont {X.}~\bibnamefont {Wu}}, \bibinfo {author} {\bibfnamefont
  {E.}~\bibnamefont {Svedberg}}, \bibinfo {author} {\bibfnamefont
  {O.}~\bibnamefont {Mryasov}}, \bibinfo {author} {\bibfnamefont
  {R.}~\bibnamefont {Chantrell}}, \bibinfo {author} {\bibfnamefont
  {D.}~\bibnamefont {Weller}}, \bibinfo {author} {\bibfnamefont
  {M.}~\bibnamefont {Tanase}}, \ and\ \bibinfo {author} {\bibfnamefont
  {D.}~\bibnamefont {Laughlin}},\ }\href@noop {} {\bibfield  {journal}
  {\bibinfo  {journal} {Appl. Phys. Lett.}\ }\textbf {\bibinfo {volume} {81}},\
  \bibinfo {pages} {2220} (\bibinfo {year} {2002})}\BibitemShut {NoStop}%
\bibitem [{\citenamefont {Garanin}(1997)}]{Garanin:1997uf}%
  \BibitemOpen
  \bibfield  {author} {\bibinfo {author} {\bibfnamefont {D.}~\bibnamefont
  {Garanin}},\ }\href@noop {} {\bibfield  {journal} {\bibinfo  {journal} {Phys.
  Rev. B}\ }\textbf {\bibinfo {volume} {55}},\ \bibinfo {pages} {3050}
  (\bibinfo {year} {1997})}\BibitemShut {NoStop}%
\bibitem [{\citenamefont {Evans}\ \emph {et~al.}(2012)\citenamefont {Evans},
  \citenamefont {Hinzke}, \citenamefont {Atxitia}, \citenamefont {Nowak},
  \citenamefont {Chantrell},\ and\ \citenamefont
  {Chubykalo-Fesenko}}]{Evans:2012ex}%
  \BibitemOpen
  \bibfield  {author} {\bibinfo {author} {\bibfnamefont {R.~F.~L.}\
  \bibnamefont {Evans}}, \bibinfo {author} {\bibfnamefont {D.}~\bibnamefont
  {Hinzke}}, \bibinfo {author} {\bibfnamefont {U.}~\bibnamefont {Atxitia}},
  \bibinfo {author} {\bibfnamefont {U.}~\bibnamefont {Nowak}}, \bibinfo
  {author} {\bibfnamefont {R.~W.}\ \bibnamefont {Chantrell}}, \ and\ \bibinfo
  {author} {\bibfnamefont {O.}~\bibnamefont {Chubykalo-Fesenko}},\ }\href
  {\doibase 10.1103/PhysRevB.85.014433} {\bibfield  {journal} {\bibinfo
  {journal} {Phys. Rev. B}\ }\textbf {\bibinfo {volume} {85}},\ \bibinfo
  {pages} {014433} (\bibinfo {year} {2012})}\BibitemShut {NoStop}%
\bibitem [{\citenamefont {Atxitia}\ \emph {et~al.}(2012)\citenamefont
  {Atxitia}, \citenamefont {Nieves},\ and\ \citenamefont
  {Chubykalo-Fesenko}}]{Atxitia:2012bk}%
  \BibitemOpen
  \bibfield  {author} {\bibinfo {author} {\bibfnamefont {U.}~\bibnamefont
  {Atxitia}}, \bibinfo {author} {\bibfnamefont {P.}~\bibnamefont {Nieves}}, \
  and\ \bibinfo {author} {\bibfnamefont {O.}~\bibnamefont
  {Chubykalo-Fesenko}},\ }\href {\doibase 10.1103/PhysRevB.86.104414}
  {\bibfield  {journal} {\bibinfo  {journal} {Phys. Rev. B}\ }\textbf {\bibinfo
  {volume} {86}},\ \bibinfo {pages} {104414} (\bibinfo {year}
  {2012})}\BibitemShut {NoStop}%
\bibitem [{\citenamefont {McDaniel}(2012)}]{McDaniel:2012ep}%
  \BibitemOpen
  \bibfield  {author} {\bibinfo {author} {\bibfnamefont {T.~W.}\ \bibnamefont
  {McDaniel}},\ }\href {\doibase 10.1063/1.4764336} {\bibfield  {journal}
  {\bibinfo  {journal} {J. Appl. Phys.}\ }\textbf {\bibinfo {volume} {112}},\
  \bibinfo {pages} {093920} (\bibinfo {year} {2012})}\BibitemShut {NoStop}%
\bibitem [{\citenamefont {Atxitia}\ and\ \citenamefont
  {Chubykalo-Fesenko}(2011)}]{Atxitia:2011ds}%
  \BibitemOpen
  \bibfield  {author} {\bibinfo {author} {\bibfnamefont {U.}~\bibnamefont
  {Atxitia}}\ and\ \bibinfo {author} {\bibfnamefont {O.}~\bibnamefont
  {Chubykalo-Fesenko}},\ }\href {\doibase 10.1038/nmat2593} {\bibfield
  {journal} {\bibinfo  {journal} {Phys. Rev. B}\ }\textbf {\bibinfo {volume}
  {84}},\ \bibinfo {pages} {144414} (\bibinfo {year} {2011})}\BibitemShut
  {NoStop}%
\bibitem [{\citenamefont {Sultan}\ \emph {et~al.}(2012)\citenamefont {Sultan},
  \citenamefont {Atxitia}, \citenamefont {Melnikov}, \citenamefont
  {Chubykalo-Fesenko},\ and\ \citenamefont {Bovensiepen}}]{Sultan:2012bh}%
  \BibitemOpen
  \bibfield  {author} {\bibinfo {author} {\bibfnamefont {M.}~\bibnamefont
  {Sultan}}, \bibinfo {author} {\bibfnamefont {U.}~\bibnamefont {Atxitia}},
  \bibinfo {author} {\bibfnamefont {A.}~\bibnamefont {Melnikov}}, \bibinfo
  {author} {\bibfnamefont {O.}~\bibnamefont {Chubykalo-Fesenko}}, \ and\
  \bibinfo {author} {\bibfnamefont {U.}~\bibnamefont {Bovensiepen}},\ }\href
  {\doibase 10.1103/PhysRevB.85.184407} {\bibfield  {journal} {\bibinfo
  {journal} {Phys. Rev. B}\ }\textbf {\bibinfo {volume} {85}},\ \bibinfo
  {pages} {184407} (\bibinfo {year} {2012})}\BibitemShut {NoStop}%
\bibitem [{\citenamefont {Atxitia}\ \emph {et~al.}(2013)\citenamefont
  {Atxitia}, \citenamefont {Ostler}, \citenamefont {Barker}, \citenamefont
  {Evans}, \citenamefont {Chantrell},\ and\ \citenamefont
  {Chubykalo-Fesenko}}]{Atxitia:2013ji}%
  \BibitemOpen
  \bibfield  {author} {\bibinfo {author} {\bibfnamefont {U.}~\bibnamefont
  {Atxitia}}, \bibinfo {author} {\bibfnamefont {T.}~\bibnamefont {Ostler}},
  \bibinfo {author} {\bibfnamefont {J.}~\bibnamefont {Barker}}, \bibinfo
  {author} {\bibfnamefont {R.~F.~L.}\ \bibnamefont {Evans}}, \bibinfo {author}
  {\bibfnamefont {R.~W.}\ \bibnamefont {Chantrell}}, \ and\ \bibinfo {author}
  {\bibfnamefont {O.}~\bibnamefont {Chubykalo-Fesenko}},\ }\href {\doibase
  10.1103/PhysRevB.87.224417} {\bibfield  {journal} {\bibinfo  {journal} {Phys.
  Rev. B}\ }\textbf {\bibinfo {volume} {87}},\ \bibinfo {pages} {224417}
  (\bibinfo {year} {2013})}\BibitemShut {NoStop}%
\bibitem [{\citenamefont {Hovorka}\ \emph {et~al.}(2012)\citenamefont
  {Hovorka}, \citenamefont {Devos}, \citenamefont {Coopman}, \citenamefont
  {Fan}, \citenamefont {Aas}, \citenamefont {Evans}, \citenamefont {Chen},
  \citenamefont {Ju},\ and\ \citenamefont {Chantrell}}]{Hovorka:2012kx}%
  \BibitemOpen
  \bibfield  {author} {\bibinfo {author} {\bibfnamefont {O.}~\bibnamefont
  {Hovorka}}, \bibinfo {author} {\bibfnamefont {S.}~\bibnamefont {Devos}},
  \bibinfo {author} {\bibfnamefont {Q.}~\bibnamefont {Coopman}}, \bibinfo
  {author} {\bibfnamefont {W.~J.}\ \bibnamefont {Fan}}, \bibinfo {author}
  {\bibfnamefont {C.~J.}\ \bibnamefont {Aas}}, \bibinfo {author} {\bibfnamefont
  {R.~F.~L.}\ \bibnamefont {Evans}}, \bibinfo {author} {\bibfnamefont
  {X.}~\bibnamefont {Chen}}, \bibinfo {author} {\bibfnamefont {G.}~\bibnamefont
  {Ju}}, \ and\ \bibinfo {author} {\bibfnamefont {R.~W.}\ \bibnamefont
  {Chantrell}},\ }\href {\doibase 10.1063/1.4740075} {\bibfield  {journal}
  {\bibinfo  {journal} {Appl. Phys. Lett.}\ }\textbf {\bibinfo {volume}
  {101}},\ \bibinfo {pages} {052406} (\bibinfo {year} {2012})}\BibitemShut
  {NoStop}%
\bibitem [{Note1()}]{Note1}%
  \BibitemOpen
  \bibinfo {note} {Details available from vampire.york.ac.uk}\BibitemShut
  {NoStop}%
\bibitem [{\citenamefont {Dorfbauer}\ \emph {et~al.}(2006)\citenamefont
  {Dorfbauer}, \citenamefont {Schrefl}, \citenamefont {Kirschner},
  \citenamefont {Hrkac}, \citenamefont {Suess}, \citenamefont {Ertl},\ and\
  \citenamefont {Fidler}}]{Dorfbauer:2006uq}%
  \BibitemOpen
  \bibfield  {author} {\bibinfo {author} {\bibfnamefont {F.}~\bibnamefont
  {Dorfbauer}}, \bibinfo {author} {\bibfnamefont {T.}~\bibnamefont {Schrefl}},
  \bibinfo {author} {\bibfnamefont {M.}~\bibnamefont {Kirschner}}, \bibinfo
  {author} {\bibfnamefont {G.}~\bibnamefont {Hrkac}}, \bibinfo {author}
  {\bibfnamefont {D.}~\bibnamefont {Suess}}, \bibinfo {author} {\bibfnamefont
  {O.}~\bibnamefont {Ertl}}, \ and\ \bibinfo {author} {\bibfnamefont
  {J.}~\bibnamefont {Fidler}},\ }\href@noop {} {\bibfield  {journal} {\bibinfo
  {journal} {J. Appl. Phys.}\ }\textbf {\bibinfo {volume} {99}},\ \bibinfo
  {pages} {08G706} (\bibinfo {year} {2006})}\BibitemShut {NoStop}%
\bibitem [{\citenamefont {Jiles}(1991)}]{Jiles:1991wd}%
  \BibitemOpen
  \bibfield  {author} {\bibinfo {author} {\bibfnamefont {D.}~\bibnamefont
  {Jiles}},\ }\href@noop {} {\emph {\bibinfo {title} {{Introduction to
  magnetism and magnetic materials}}}}\ (\bibinfo  {publisher} {Chapman {\&}
  Hall, London, UK},\ \bibinfo {year} {1991})\BibitemShut {NoStop}%
\bibitem [{\citenamefont {Schwarz}\ \emph {et~al.}(1984)\citenamefont
  {Schwarz}, \citenamefont {Mohn}, \citenamefont {Blaha},\ and\ \citenamefont
  {Kubler}}]{Schwarz:1984wn}%
  \BibitemOpen
  \bibfield  {author} {\bibinfo {author} {\bibfnamefont {K.}~\bibnamefont
  {Schwarz}}, \bibinfo {author} {\bibfnamefont {P.}~\bibnamefont {Mohn}},
  \bibinfo {author} {\bibfnamefont {P.}~\bibnamefont {Blaha}}, \ and\ \bibinfo
  {author} {\bibfnamefont {J.}~\bibnamefont {Kubler}},\ }\href
  {http://iopscience.iop.org/0305-4608/14/11/021} {\bibfield  {journal}
  {\bibinfo  {journal} {Journal of Physics F: Metal Physics}\ }\textbf
  {\bibinfo {volume} {14}},\ \bibinfo {pages} {2659} (\bibinfo {year}
  {1984})}\BibitemShut {NoStop}%
\bibitem [{\citenamefont {Liechtenstein}\ \emph {et~al.}(1987)\citenamefont
  {Liechtenstein}, \citenamefont {Katsnelson}, \citenamefont {Antropov},\ and\
  \citenamefont {Gubanov}}]{rtm1}%
  \BibitemOpen
  \bibfield  {author} {\bibinfo {author} {\bibfnamefont {A.}~\bibnamefont
  {Liechtenstein}}, \bibinfo {author} {\bibfnamefont {M.}~\bibnamefont
  {Katsnelson}}, \bibinfo {author} {\bibfnamefont {V.}~\bibnamefont
  {Antropov}}, \ and\ \bibinfo {author} {\bibfnamefont {V.}~\bibnamefont
  {Gubanov}},\ }\href@noop {} {\bibfield  {journal} {\bibinfo  {journal} {J.
  Magn. Magn. Mater.}\ }\textbf {\bibinfo {volume} {67}},\ \bibinfo {pages}
  {65} (\bibinfo {year} {1987})}\BibitemShut {NoStop}%
\bibitem [{\citenamefont {Udvardi}\ \emph {et~al.}(2003)\citenamefont
  {Udvardi}, \citenamefont {Szunyogh}, \citenamefont {Palotas},\ and\
  \citenamefont {Weinberger}}]{rtm2}%
  \BibitemOpen
  \bibfield  {author} {\bibinfo {author} {\bibfnamefont {L.}~\bibnamefont
  {Udvardi}}, \bibinfo {author} {\bibfnamefont {L.}~\bibnamefont {Szunyogh}},
  \bibinfo {author} {\bibfnamefont {K.}~\bibnamefont {Palotas}}, \ and\
  \bibinfo {author} {\bibfnamefont {P.}~\bibnamefont {Weinberger}},\
  }\href@noop {} {\bibfield  {journal} {\bibinfo  {journal} {Phys. Rev. B}\
  }\textbf {\bibinfo {volume} {68}},\ \bibinfo {pages} {104436} (\bibinfo
  {year} {2003})}\BibitemShut {NoStop}%
\bibitem [{\citenamefont {Antal}\ \emph {et~al.}(2008)\citenamefont {Antal},
  \citenamefont {Lazarovits}, \citenamefont {Udvardi}, \citenamefont
  {Szunyogh}, \citenamefont {Ujfalussy},\ and\ \citenamefont
  {Weinberger}}]{rtm3}%
  \BibitemOpen
  \bibfield  {author} {\bibinfo {author} {\bibfnamefont {A.}~\bibnamefont
  {Antal}}, \bibinfo {author} {\bibfnamefont {B.}~\bibnamefont {Lazarovits}},
  \bibinfo {author} {\bibfnamefont {L.}~\bibnamefont {Udvardi}}, \bibinfo
  {author} {\bibfnamefont {L.}~\bibnamefont {Szunyogh}}, \bibinfo {author}
  {\bibfnamefont {B.}~\bibnamefont {Ujfalussy}}, \ and\ \bibinfo {author}
  {\bibfnamefont {P.}~\bibnamefont {Weinberger}},\ }\href@noop {} {\bibfield
  {journal} {\bibinfo  {journal} {Phys. Rev. B}\ }\textbf {\bibinfo {volume}
  {77}},\ \bibinfo {pages} {174429} (\bibinfo {year} {2008})}\BibitemShut
  {NoStop}%
\bibitem [{\citenamefont {Ebert}\ and\ \citenamefont {Mankovsky}(2009)}]{rtm4}%
  \BibitemOpen
  \bibfield  {author} {\bibinfo {author} {\bibfnamefont {H.}~\bibnamefont
  {Ebert}}\ and\ \bibinfo {author} {\bibfnamefont {S.}~\bibnamefont
  {Mankovsky}},\ }\href@noop {} {\bibfield  {journal} {\bibinfo  {journal}
  {Phys. Rev. B}\ }\textbf {\bibinfo {volume} {79}},\ \bibinfo {pages} {045209}
  (\bibinfo {year} {2009})}\BibitemShut {NoStop}%
\bibitem [{\citenamefont {Drautz}\ and\ \citenamefont {F\"ahnle}(2004)}]{sce1}%
  \BibitemOpen
  \bibfield  {author} {\bibinfo {author} {\bibfnamefont {R.}~\bibnamefont
  {Drautz}}\ and\ \bibinfo {author} {\bibfnamefont {M.}~\bibnamefont
  {F\"ahnle}},\ }\href@noop {} {\bibfield  {journal} {\bibinfo  {journal}
  {Phys. Rev. B}\ }\textbf {\bibinfo {volume} {69}},\ \bibinfo {pages} {104404}
  (\bibinfo {year} {2004})}\BibitemShut {NoStop}%
\bibitem [{\citenamefont {Drautz}\ and\ \citenamefont
  {F\"ahnle}(2005)}]{sce1b}%
  \BibitemOpen
  \bibfield  {author} {\bibinfo {author} {\bibfnamefont {R.}~\bibnamefont
  {Drautz}}\ and\ \bibinfo {author} {\bibfnamefont {M.}~\bibnamefont
  {F\"ahnle}},\ }\href@noop {} {\bibfield  {journal} {\bibinfo  {journal}
  {Phys. Rev. B}\ }\textbf {\bibinfo {volume} {72}},\ \bibinfo {pages} {212405}
  (\bibinfo {year} {2005})}\BibitemShut {NoStop}%
\bibitem [{\citenamefont {De\'ak}\ \emph {et~al.}(2011)\citenamefont {De\'ak},
  \citenamefont {Szunyogh},\ and\ \citenamefont {Ujfalussy}}]{sce3}%
  \BibitemOpen
  \bibfield  {author} {\bibinfo {author} {\bibfnamefont {A.}~\bibnamefont
  {De\'ak}}, \bibinfo {author} {\bibfnamefont {L.}~\bibnamefont {Szunyogh}}, \
  and\ \bibinfo {author} {\bibfnamefont {B.}~\bibnamefont {Ujfalussy}},\
  }\href@noop {} {\bibfield  {journal} {\bibinfo  {journal} {Phys. Rev. B}\
  }\textbf {\bibinfo {volume} {84}},\ \bibinfo {pages} {224413} (\bibinfo
  {year} {2011})}\BibitemShut {NoStop}%
\bibitem [{\citenamefont {Skomski}(2012)}]{Skomski:2012uh}%
  \BibitemOpen
  \bibfield  {author} {\bibinfo {author} {\bibfnamefont {R.}~\bibnamefont
  {Skomski}},\ }\href
  {http://books.google.co.uk/books?id=1VNxuQAACAAJ&dq=simple+models+of+magnetism&hl=&cd=1&source=gbs_api}
  {\emph {\bibinfo {title} {{Simple Models of Magnetism}}}}\ (\bibinfo
  {publisher} {OUP Oxford},\ \bibinfo {year} {2012})\BibitemShut {NoStop}%
\bibitem [{\citenamefont {Bruno}(1993)}]{Bruno:1993wc}%
  \BibitemOpen
  \bibfield  {author} {\bibinfo {author} {\bibfnamefont {P.}~\bibnamefont
  {Bruno}},\ }\href@noop {} {\emph {\bibinfo {title} {{Physical Origins and
  Theoretical Models of Magnetic Anisotropy}}}}\ (\bibinfo  {publisher}
  {Forschungszentrums Julich},\ \bibinfo {year} {1993})\BibitemShut {NoStop}%
\bibitem [{\citenamefont {Hohenberg}\ and\ \citenamefont {Kohn}(1964)}]{KS1}%
  \BibitemOpen
  \bibfield  {author} {\bibinfo {author} {\bibfnamefont {P.}~\bibnamefont
  {Hohenberg}}\ and\ \bibinfo {author} {\bibfnamefont {W.}~\bibnamefont
  {Kohn}},\ }\href@noop {} {\bibfield  {journal} {\bibinfo  {journal} {Phys.
  Rev.}\ }\textbf {\bibinfo {volume} {136}},\ \bibinfo {pages} {B864} (\bibinfo
  {year} {1964})}\BibitemShut {NoStop}%
\bibitem [{\citenamefont {Kohn}\ and\ \citenamefont {Sham}(1965)}]{KS2}%
  \BibitemOpen
  \bibfield  {author} {\bibinfo {author} {\bibfnamefont {W.}~\bibnamefont
  {Kohn}}\ and\ \bibinfo {author} {\bibfnamefont {L.~J.}\ \bibnamefont
  {Sham}},\ }\href@noop {} {\bibfield  {journal} {\bibinfo  {journal} {Phys.
  Rev.}\ }\textbf {\bibinfo {volume} {140}},\ \bibinfo {pages} {A1133}
  (\bibinfo {year} {1965})}\BibitemShut {NoStop}%
\bibitem [{\citenamefont {{von Barth}}\ and\ \citenamefont
  {Hedin}(1972)}]{SP1}%
  \BibitemOpen
  \bibfield  {author} {\bibinfo {author} {\bibfnamefont {U.}~\bibnamefont {{von
  Barth}}}\ and\ \bibinfo {author} {\bibfnamefont {L.}~\bibnamefont {Hedin}},\
  }\href@noop {} {\bibfield  {journal} {\bibinfo  {journal} {J. Phys. C: Solid
  State Phys.}\ }\textbf {\bibinfo {volume} {5}},\ \bibinfo {pages} {1629}
  (\bibinfo {year} {1972})}\BibitemShut {NoStop}%
\bibitem [{\citenamefont {Kresse}\ and\ \citenamefont
  {Furthm\"uller}(1996)}]{vasp}%
  \BibitemOpen
  \bibfield  {author} {\bibinfo {author} {\bibfnamefont {G.}~\bibnamefont
  {Kresse}}\ and\ \bibinfo {author} {\bibfnamefont {J.}~\bibnamefont
  {Furthm\"uller}},\ }\href {\doibase 10.1103/PhysRevB.54.11169} {\bibfield
  {journal} {\bibinfo  {journal} {Phys. Rev. B}\ }\textbf {\bibinfo {volume}
  {54}},\ \bibinfo {pages} {11169} (\bibinfo {year} {1996})}\BibitemShut
  {NoStop}%
\bibitem [{\citenamefont {Segall}\ \emph {et~al.}(2002)\citenamefont {Segall},
  \citenamefont {Lindan}, \citenamefont {Probert}, \citenamefont {Pickard},
  \citenamefont {Hasnip}, \citenamefont {Clark},\ and\ \citenamefont
  {Payne}}]{castep1}%
  \BibitemOpen
  \bibfield  {author} {\bibinfo {author} {\bibfnamefont {M.~D.}\ \bibnamefont
  {Segall}}, \bibinfo {author} {\bibfnamefont {P.~J.~D.}\ \bibnamefont
  {Lindan}}, \bibinfo {author} {\bibfnamefont {M.~J.}\ \bibnamefont {Probert}},
  \bibinfo {author} {\bibfnamefont {C.~J.}\ \bibnamefont {Pickard}}, \bibinfo
  {author} {\bibfnamefont {P.~J.}\ \bibnamefont {Hasnip}}, \bibinfo {author}
  {\bibfnamefont {S.~J.}\ \bibnamefont {Clark}}, \ and\ \bibinfo {author}
  {\bibfnamefont {M.~C.}\ \bibnamefont {Payne}},\ }\href
  {http://stacks.iop.org/0953-8984/14/i=11/a=301} {\bibfield  {journal}
  {\bibinfo  {journal} {J. Phys.: Condens. Matter}\ }\textbf {\bibinfo {volume}
  {14}},\ \bibinfo {pages} {2717} (\bibinfo {year} {2002})}\BibitemShut
  {NoStop}%
\bibitem [{\citenamefont {Payne}\ \emph {et~al.}(1992)\citenamefont {Payne},
  \citenamefont {Teter}, \citenamefont {Allan}, \citenamefont {Arias},\ and\
  \citenamefont {Joannopoulos}}]{castep2}%
  \BibitemOpen
  \bibfield  {author} {\bibinfo {author} {\bibfnamefont {M.~C.}\ \bibnamefont
  {Payne}}, \bibinfo {author} {\bibfnamefont {M.~P.}\ \bibnamefont {Teter}},
  \bibinfo {author} {\bibfnamefont {D.~C.}\ \bibnamefont {Allan}}, \bibinfo
  {author} {\bibfnamefont {T.~A.}\ \bibnamefont {Arias}}, \ and\ \bibinfo
  {author} {\bibfnamefont {J.~D.}\ \bibnamefont {Joannopoulos}},\ }\href
  {\doibase 10.1103/RevModPhys.64.1045} {\bibfield  {journal} {\bibinfo
  {journal} {Rev. Mod. Phys.}\ }\textbf {\bibinfo {volume} {64}},\ \bibinfo
  {pages} {1045} (\bibinfo {year} {1992})}\BibitemShut {NoStop}%
\bibitem [{\citenamefont {Soler}\ \emph {et~al.}(2002)\citenamefont {Soler},
  \citenamefont {Artacho}, \citenamefont {Gale}, \citenamefont {Garc\'ia},
  \citenamefont {Junquera}, \citenamefont {Ordej\'on},\ and\ \citenamefont
  {S\'anchez-Portal}}]{siesta}%
  \BibitemOpen
  \bibfield  {author} {\bibinfo {author} {\bibfnamefont {J.~M.}\ \bibnamefont
  {Soler}}, \bibinfo {author} {\bibfnamefont {E.}~\bibnamefont {Artacho}},
  \bibinfo {author} {\bibfnamefont {J.~D.}\ \bibnamefont {Gale}}, \bibinfo
  {author} {\bibfnamefont {A.}~\bibnamefont {Garc\'ia}}, \bibinfo {author}
  {\bibfnamefont {J.}~\bibnamefont {Junquera}}, \bibinfo {author}
  {\bibfnamefont {P.}~\bibnamefont {Ordej\'on}}, \ and\ \bibinfo {author}
  {\bibfnamefont {D.}~\bibnamefont {S\'anchez-Portal}},\ }\href
  {http://stacks.iop.org/0953-8984/14/i=11/a=302} {\bibfield  {journal}
  {\bibinfo  {journal} {J. Phys.: Condens. Matter}\ }\textbf {\bibinfo {volume}
  {14}},\ \bibinfo {pages} {2745} (\bibinfo {year} {2002})}\BibitemShut
  {NoStop}%
\bibitem [{\citenamefont {Zabloudil}\ \emph {et~al.}(2005)\citenamefont
  {Zabloudil}, \citenamefont {Hammerling}, \citenamefont {Szunyogh},\ and\
  \citenamefont {Weinberger}}]{skkr}%
  \BibitemOpen
  \bibfield  {author} {\bibinfo {author} {\bibfnamefont {J.}~\bibnamefont
  {Zabloudil}}, \bibinfo {author} {\bibfnamefont {R.}~\bibnamefont
  {Hammerling}}, \bibinfo {author} {\bibfnamefont {L.}~\bibnamefont
  {Szunyogh}}, \ and\ \bibinfo {author} {\bibfnamefont {P.}~\bibnamefont
  {Weinberger}},\ }\href@noop {} {\emph {\bibinfo {title} {Electron Scattering
  in Solid Matter}}}\ (\bibinfo  {publisher} {Springer, Berlin},\ \bibinfo
  {year} {2005})\BibitemShut {NoStop}%
\bibitem [{\citenamefont {Ebert}\ \emph {et~al.}(2011)\citenamefont {Ebert},
  \citenamefont {K\"odderitzsch},\ and\ \citenamefont {Min\'ar}}]{skkr2}%
  \BibitemOpen
  \bibfield  {author} {\bibinfo {author} {\bibfnamefont {H.}~\bibnamefont
  {Ebert}}, \bibinfo {author} {\bibfnamefont {D.}~\bibnamefont
  {K\"odderitzsch}}, \ and\ \bibinfo {author} {\bibfnamefont {J.}~\bibnamefont
  {Min\'ar}},\ }\href@noop {} {\bibfield  {journal} {\bibinfo  {journal}
  {Reports on Progress in Physics}\ }\textbf {\bibinfo {volume} {74}},\
  \bibinfo {pages} {096501} (\bibinfo {year} {2011})}\BibitemShut {NoStop}%
\bibitem [{\citenamefont {Katsnelson}\ and\ \citenamefont
  {Lichtenstein}(2000)}]{magneticforce2000}%
  \BibitemOpen
  \bibfield  {author} {\bibinfo {author} {\bibfnamefont {M.~I.}\ \bibnamefont
  {Katsnelson}}\ and\ \bibinfo {author} {\bibfnamefont {A.~I.}\ \bibnamefont
  {Lichtenstein}},\ }\href {\doibase 10.1103/PhysRevB.61.8906} {\bibfield
  {journal} {\bibinfo  {journal} {Phys. Rev. B}\ }\textbf {\bibinfo {volume}
  {61}},\ \bibinfo {pages} {8906} (\bibinfo {year} {2000})}\BibitemShut
  {NoStop}%
\bibitem [{\citenamefont {Pajda}\ \emph {et~al.}(2001)\citenamefont {Pajda},
  \citenamefont {Kudrnovsk{\'y}}, \citenamefont {Turek}, \citenamefont
  {Drchal},\ and\ \citenamefont {Bruno}}]{Pajda:2001ix}%
  \BibitemOpen
  \bibfield  {author} {\bibinfo {author} {\bibfnamefont {M.}~\bibnamefont
  {Pajda}}, \bibinfo {author} {\bibfnamefont {J.}~\bibnamefont
  {Kudrnovsk{\'y}}}, \bibinfo {author} {\bibfnamefont {I.}~\bibnamefont
  {Turek}}, \bibinfo {author} {\bibfnamefont {V.}~\bibnamefont {Drchal}}, \
  and\ \bibinfo {author} {\bibfnamefont {P.}~\bibnamefont {Bruno}},\ }\href
  {\doibase 10.1103/PhysRevB.64.174402} {\bibfield  {journal} {\bibinfo
  {journal} {Phys. Rev. B}\ }\textbf {\bibinfo {volume} {64}},\ \bibinfo
  {pages} {174402} (\bibinfo {year} {2001})}\BibitemShut {NoStop}%
\bibitem [{\citenamefont {Sandratskii}(1986)}]{SS1}%
  \BibitemOpen
  \bibfield  {author} {\bibinfo {author} {\bibfnamefont {L.~M.}\ \bibnamefont
  {Sandratskii}},\ }\href@noop {} {\bibfield  {journal} {\bibinfo  {journal}
  {Phys. Status Solidi B}\ }\textbf {\bibinfo {volume} {136}},\ \bibinfo
  {pages} {167} (\bibinfo {year} {1986})}\BibitemShut {NoStop}%
\bibitem [{\citenamefont {Halilov}\ \emph {et~al.}(1998)\citenamefont
  {Halilov}, \citenamefont {Eschrig}, \citenamefont {Perlov},\ and\
  \citenamefont {Oppeneer}}]{SS2}%
  \BibitemOpen
  \bibfield  {author} {\bibinfo {author} {\bibfnamefont {S.~V.}\ \bibnamefont
  {Halilov}}, \bibinfo {author} {\bibfnamefont {H.}~\bibnamefont {Eschrig}},
  \bibinfo {author} {\bibfnamefont {A.~Y.}\ \bibnamefont {Perlov}}, \ and\
  \bibinfo {author} {\bibfnamefont {P.~M.}\ \bibnamefont {Oppeneer}},\
  }\href@noop {} {\bibfield  {journal} {\bibinfo  {journal} {Phys. Rev. B}\
  }\textbf {\bibinfo {volume} {58}},\ \bibinfo {pages} {293} (\bibinfo {year}
  {1998})}\BibitemShut {NoStop}%
\bibitem [{\citenamefont {Le\ifmmode \check{z}\else
  \v{z}\fi{}ai\ifmmode~\acute{c}\else \'{c}\fi{}}\ \emph
  {et~al.}(2013)\citenamefont {Le\ifmmode \check{z}\else
  \v{z}\fi{}ai\ifmmode~\acute{c}\else \'{c}\fi{}}, \citenamefont {Mavropoulos},
  \citenamefont {Bihlmayer},\ and\ \citenamefont
  {Bl\"ugel}}]{inducedmoments2013}%
  \BibitemOpen
  \bibfield  {author} {\bibinfo {author} {\bibfnamefont {M.}~\bibnamefont
  {Le\ifmmode \check{z}\else \v{z}\fi{}ai\ifmmode~\acute{c}\else \'{c}\fi{}}},
  \bibinfo {author} {\bibfnamefont {P.}~\bibnamefont {Mavropoulos}}, \bibinfo
  {author} {\bibfnamefont {G.}~\bibnamefont {Bihlmayer}}, \ and\ \bibinfo
  {author} {\bibfnamefont {S.}~\bibnamefont {Bl\"ugel}},\ }\href {\doibase
  10.1103/PhysRevB.88.134403} {\bibfield  {journal} {\bibinfo  {journal} {Phys.
  Rev. B}\ }\textbf {\bibinfo {volume} {88}},\ \bibinfo {pages} {134403}
  (\bibinfo {year} {2013})}\BibitemShut {NoStop}%
\bibitem [{\citenamefont {Uhl}\ and\ \citenamefont
  {Siberchicot}(1999)}]{Uhl:1999bz}%
  \BibitemOpen
  \bibfield  {author} {\bibinfo {author} {\bibfnamefont {M.}~\bibnamefont
  {Uhl}}\ and\ \bibinfo {author} {\bibfnamefont {B.}~\bibnamefont
  {Siberchicot}},\ }\href {\doibase 10.1088/0953-8984/7/22/006} {\bibfield
  {journal} {\bibinfo  {journal} {J. Phys.: Condens. Matter}\ }\textbf
  {\bibinfo {volume} {7}},\ \bibinfo {pages} {4227} (\bibinfo {year}
  {1999})}\BibitemShut {NoStop}%
\bibitem [{\citenamefont {Skubic}\ \emph {et~al.}(2009)\citenamefont {Skubic},
  \citenamefont {Peil}, \citenamefont {Hellsvik}, \citenamefont {Nordblad},
  \citenamefont {Nordstr{\"o}m},\ and\ \citenamefont
  {Eriksson}}]{Skubic:2009bv}%
  \BibitemOpen
  \bibfield  {author} {\bibinfo {author} {\bibfnamefont {B.}~\bibnamefont
  {Skubic}}, \bibinfo {author} {\bibfnamefont {O.}~\bibnamefont {Peil}},
  \bibinfo {author} {\bibfnamefont {J.}~\bibnamefont {Hellsvik}}, \bibinfo
  {author} {\bibfnamefont {P.}~\bibnamefont {Nordblad}}, \bibinfo {author}
  {\bibfnamefont {L.}~\bibnamefont {Nordstr{\"o}m}}, \ and\ \bibinfo {author}
  {\bibfnamefont {O.}~\bibnamefont {Eriksson}},\ }\href {\doibase
  10.1103/PhysRevB.79.024411} {\bibfield  {journal} {\bibinfo  {journal} {Phys.
  Rev. B}\ }\textbf {\bibinfo {volume} {79}},\ \bibinfo {pages} {024411}
  (\bibinfo {year} {2009})}\BibitemShut {NoStop}%
\bibitem [{\citenamefont {{Aas}}\ \emph {et~al.}(2013)\citenamefont {{Aas}},
  \citenamefont {{Hasnip}}, \citenamefont {{Cuadrado}}, \citenamefont
  {{Plotnikova}}, \citenamefont {{Szunyogh}}, \citenamefont {{Udvardi}},\ and\
  \citenamefont {{Chantrell}}}]{AasFePtFe2013}%
  \BibitemOpen
  \bibfield  {author} {\bibinfo {author} {\bibfnamefont {C.~J.}\ \bibnamefont
  {{Aas}}}, \bibinfo {author} {\bibfnamefont {P.}~\bibnamefont {{Hasnip}}},
  \bibinfo {author} {\bibfnamefont {R.}~\bibnamefont {{Cuadrado}}}, \bibinfo
  {author} {\bibfnamefont {E.}~\bibnamefont {{Plotnikova}}}, \bibinfo {author}
  {\bibfnamefont {L.}~\bibnamefont {{Szunyogh}}}, \bibinfo {author}
  {\bibfnamefont {L.}~\bibnamefont {{Udvardi}}}, \ and\ \bibinfo {author}
  {\bibfnamefont {R.~W.}\ \bibnamefont {{Chantrell}}},\ }\href@noop {}
  {\bibfield  {journal} {\bibinfo  {journal} {Phys. Rev. B}\ }\textbf {\bibinfo
  {volume} {88}},\ \bibinfo {pages} {174409} (\bibinfo {year}
  {2013})}\BibitemShut {NoStop}%
\bibitem [{\citenamefont {Sato}\ \emph {et~al.}(2010)\citenamefont {Sato},
  \citenamefont {Bergqvist}, \citenamefont {Kudrnovský}, \citenamefont
  {Dederichs}, \citenamefont {Eriksson}, \citenamefont {Turek}, \citenamefont
  {Sanyal}, \citenamefont {Bouzerar}, \citenamefont {Katayama-Yoshida},
  \citenamefont {Dinh}, \citenamefont {Fukushima}, \citenamefont {Kizaki},\
  and\ \citenamefont {R}}]{disorder1}%
  \BibitemOpen
  \bibfield  {author} {\bibinfo {author} {\bibfnamefont {K.}~\bibnamefont
  {Sato}}, \bibinfo {author} {\bibfnamefont {L.}~\bibnamefont {Bergqvist}},
  \bibinfo {author} {\bibfnamefont {J.}~\bibnamefont {Kudrnovský}}, \bibinfo
  {author} {\bibfnamefont {P.~H.}\ \bibnamefont {Dederichs}}, \bibinfo {author}
  {\bibfnamefont {O.}~\bibnamefont {Eriksson}}, \bibinfo {author}
  {\bibfnamefont {I.}~\bibnamefont {Turek}}, \bibinfo {author} {\bibfnamefont
  {B.}~\bibnamefont {Sanyal}}, \bibinfo {author} {\bibfnamefont
  {G.}~\bibnamefont {Bouzerar}}, \bibinfo {author} {\bibfnamefont
  {H.}~\bibnamefont {Katayama-Yoshida}}, \bibinfo {author} {\bibfnamefont
  {V.~A.}\ \bibnamefont {Dinh}}, \bibinfo {author} {\bibfnamefont
  {T.}~\bibnamefont {Fukushima}}, \bibinfo {author} {\bibfnamefont
  {H.}~\bibnamefont {Kizaki}}, \ and\ \bibinfo {author} {\bibfnamefont
  {Z.}~\bibnamefont {R}},\ }\href@noop {} {\bibfield  {journal} {\bibinfo
  {journal} {Rev. Mod. Phys.}\ }\textbf {\bibinfo {volume} {82}},\ \bibinfo
  {pages} {1633} (\bibinfo {year} {2010})}\BibitemShut {NoStop}%
\bibitem [{\citenamefont {Le\v{z}ai\'{c}}\ \emph {et~al.}(2011)\citenamefont
  {Le\v{z}ai\'{c}}, \citenamefont {Mavropoulos}, \citenamefont {Bl\"{u}gel},\
  and\ \citenamefont {Ebert}}]{disorder2}%
  \BibitemOpen
  \bibfield  {author} {\bibinfo {author} {\bibfnamefont {M.}~\bibnamefont
  {Le\v{z}ai\'{c}}}, \bibinfo {author} {\bibfnamefont {P.}~\bibnamefont
  {Mavropoulos}}, \bibinfo {author} {\bibfnamefont {S.}~\bibnamefont
  {Bl\"{u}gel}}, \ and\ \bibinfo {author} {\bibfnamefont {H.}~\bibnamefont
  {Ebert}},\ }\href@noop {} {\bibfield  {journal} {\bibinfo  {journal} {Phys.
  Rev. B}\ }\textbf {\bibinfo {volume} {83}},\ \bibinfo {pages} {094434}
  (\bibinfo {year} {2011})}\BibitemShut {NoStop}%
\bibitem [{\citenamefont {Danan}(1968)}]{Danan:1968jw}%
  \BibitemOpen
  \bibfield  {author} {\bibinfo {author} {\bibfnamefont {H.}~\bibnamefont
  {Danan}},\ }\href {\doibase 10.1063/1.2163571} {\bibfield  {journal}
  {\bibinfo  {journal} {J. Appl. Phys.}\ }\textbf {\bibinfo {volume} {39}},\
  \bibinfo {pages} {669} (\bibinfo {year} {1968})}\BibitemShut {NoStop}%
\bibitem [{\citenamefont {Garanin}(1996)}]{Garanin:351283}%
  \BibitemOpen
  \bibfield  {author} {\bibinfo {author} {\bibfnamefont {D.~A.}\ \bibnamefont
  {Garanin}},\ }\href
  {http://www.google.com/search?client=safari&rls=10_7_4&q=Self+Consistent+Gaussian+Approximation+for+Classical+Spin+Systems+Thermodynamics&ie=UTF-8&oe=UTF-8}
  {\bibfield  {journal} {\bibinfo  {journal} {Phys. Rev. B}\ }\textbf {\bibinfo
  {volume} {53}},\ \bibinfo {pages} {11593} (\bibinfo {year}
  {1996})}\BibitemShut {NoStop}%
\bibitem [{\citenamefont {Wysin}(2000)}]{Wysin:2000uf}%
  \BibitemOpen
  \bibfield  {author} {\bibinfo {author} {\bibfnamefont {G.~M.}\ \bibnamefont
  {Wysin}},\ }\href {http://prb.aps.org/abstract/PRB/v62/i5/p3251_1} {\bibfield
   {journal} {\bibinfo  {journal} {Phys. Rev. B}\ }\textbf {\bibinfo {volume}
  {62}},\ \bibinfo {pages} {3251} (\bibinfo {year} {2000})}\BibitemShut
  {NoStop}%
\bibitem [{\citenamefont {Colarieti-Tosti}\ \emph {et~al.}(2003)\citenamefont
  {Colarieti-Tosti}, \citenamefont {Simak}, \citenamefont {Ahuja},
  \citenamefont {Nordstr{\"o}m}, \citenamefont {Eriksson}, \citenamefont
  {{\AA}berg}, \citenamefont {Edvardsson},\ and\ \citenamefont
  {Brooks}}]{ColarietiTosti:2003iw}%
  \BibitemOpen
  \bibfield  {author} {\bibinfo {author} {\bibfnamefont {M.}~\bibnamefont
  {Colarieti-Tosti}}, \bibinfo {author} {\bibfnamefont {S.}~\bibnamefont
  {Simak}}, \bibinfo {author} {\bibfnamefont {R.}~\bibnamefont {Ahuja}},
  \bibinfo {author} {\bibfnamefont {L.}~\bibnamefont {Nordstr{\"o}m}}, \bibinfo
  {author} {\bibfnamefont {O.}~\bibnamefont {Eriksson}}, \bibinfo {author}
  {\bibfnamefont {D.}~\bibnamefont {{\AA}berg}}, \bibinfo {author}
  {\bibfnamefont {S.}~\bibnamefont {Edvardsson}}, \ and\ \bibinfo {author}
  {\bibfnamefont {M.}~\bibnamefont {Brooks}},\ }\href {\doibase
  10.1103/PhysRevLett.91.157201} {\bibfield  {journal} {\bibinfo  {journal}
  {Phys. Rev. Lett.}\ }\textbf {\bibinfo {volume} {91}},\ \bibinfo {pages}
  {157201} (\bibinfo {year} {2003})}\BibitemShut {NoStop}%
\bibitem [{\citenamefont {Radu}\ \emph {et~al.}(2013)\citenamefont {Radu},
  \citenamefont {Stamm}, \citenamefont {Eschenlohr}, \citenamefont {Vahaplar},
  \citenamefont {Kachel}, \citenamefont {Pontius}, \citenamefont {Mitzner},
  \citenamefont {Holldack}, \citenamefont {Föhlish}, \citenamefont {Radu},
  \citenamefont {Evans}, \citenamefont {Ostler}, \citenamefont {Mentink},
  \citenamefont {Chantrell}, \citenamefont {Tsukamoto}, \citenamefont {Itoh},
  \citenamefont {Kirilyuk}, \citenamefont {Kimel},\ and\ \citenamefont
  {Rasing}}]{raduPRX2013}%
  \BibitemOpen
  \bibfield  {author} {\bibinfo {author} {\bibfnamefont {I.}~\bibnamefont
  {Radu}}, \bibinfo {author} {\bibfnamefont {C.}~\bibnamefont {Stamm}},
  \bibinfo {author} {\bibfnamefont {A.}~\bibnamefont {Eschenlohr}}, \bibinfo
  {author} {\bibfnamefont {K.}~\bibnamefont {Vahaplar}}, \bibinfo {author}
  {\bibfnamefont {T.}~\bibnamefont {Kachel}}, \bibinfo {author} {\bibfnamefont
  {N.}~\bibnamefont {Pontius}}, \bibinfo {author} {\bibfnamefont
  {R.}~\bibnamefont {Mitzner}}, \bibinfo {author} {\bibfnamefont
  {K.}~\bibnamefont {Holldack}}, \bibinfo {author} {\bibfnamefont
  {A.}~\bibnamefont {Föhlish}}, \bibinfo {author} {\bibfnamefont
  {F.}~\bibnamefont {Radu}}, \bibinfo {author} {\bibfnamefont {R.~F.~L.}\
  \bibnamefont {Evans}}, \bibinfo {author} {\bibfnamefont {T.~A.}\ \bibnamefont
  {Ostler}}, \bibinfo {author} {\bibfnamefont {J.}~\bibnamefont {Mentink}},
  \bibinfo {author} {\bibfnamefont {R.~W.}\ \bibnamefont {Chantrell}}, \bibinfo
  {author} {\bibfnamefont {A.}~\bibnamefont {Tsukamoto}}, \bibinfo {author}
  {\bibfnamefont {A.}~\bibnamefont {Itoh}}, \bibinfo {author} {\bibfnamefont
  {A.}~\bibnamefont {Kirilyuk}}, \bibinfo {author} {\bibfnamefont {A.~V.}\
  \bibnamefont {Kimel}}, \ and\ \bibinfo {author} {\bibfnamefont
  {T.}~\bibnamefont {Rasing}},\ }\href@noop {} {\bibfield  {journal} {\bibinfo
  {journal} {Submitted}\ } (\bibinfo {year} {2013})}\BibitemShut {NoStop}%
\bibitem [{\citenamefont {Fan}\ \emph {et~al.}(2011)\citenamefont {Fan},
  \citenamefont {Evans}, \citenamefont {Hancock},\ and\ \citenamefont
  {Chantrell}}]{Fan:2011es}%
  \BibitemOpen
  \bibfield  {author} {\bibinfo {author} {\bibfnamefont {W.~J.}\ \bibnamefont
  {Fan}}, \bibinfo {author} {\bibfnamefont {R.~F.~L.}\ \bibnamefont {Evans}},
  \bibinfo {author} {\bibfnamefont {Y.}~\bibnamefont {Hancock}}, \ and\
  \bibinfo {author} {\bibfnamefont {R.~W.}\ \bibnamefont {Chantrell}},\ }\href
  {\doibase 10.1063/1.3561446} {\bibfield  {journal} {\bibinfo  {journal} {J.
  Appl. Phys.}\ }\textbf {\bibinfo {volume} {109}},\  (\bibinfo {year}
  {2011})}\BibitemShut {NoStop}%
\bibitem [{\citenamefont {Schr\"oder}\ \emph
  {et~al.}(2005{\natexlab{a}})\citenamefont {Schr\"oder}, \citenamefont
  {Schmidt}, \citenamefont {Schnack},\ and\ \citenamefont
  {Luban}}]{SchroderPRL2005}%
  \BibitemOpen
  \bibfield  {author} {\bibinfo {author} {\bibfnamefont {C.}~\bibnamefont
  {Schr\"oder}}, \bibinfo {author} {\bibfnamefont {H.-J.}\ \bibnamefont
  {Schmidt}}, \bibinfo {author} {\bibfnamefont {J.}~\bibnamefont {Schnack}}, \
  and\ \bibinfo {author} {\bibfnamefont {M.}~\bibnamefont {Luban}},\ }\href
  {\doibase 10.1103/PhysRevLett.94.207203} {\bibfield  {journal} {\bibinfo
  {journal} {Phys. Rev. Lett.}\ }\textbf {\bibinfo {volume} {94}},\ \bibinfo
  {pages} {207203} (\bibinfo {year} {2005}{\natexlab{a}})}\BibitemShut
  {NoStop}%
\bibitem [{\citenamefont {Schr\"oder}\ \emph
  {et~al.}(2005{\natexlab{b}})\citenamefont {Schr\"oder}, \citenamefont
  {Nojiri}, \citenamefont {Schnack}, \citenamefont {Hage}, \citenamefont
  {Luban},\ and\ \citenamefont {K\"ogerler}}]{SchroderPRL2005-2}%
  \BibitemOpen
  \bibfield  {author} {\bibinfo {author} {\bibfnamefont {C.}~\bibnamefont
  {Schr\"oder}}, \bibinfo {author} {\bibfnamefont {H.}~\bibnamefont {Nojiri}},
  \bibinfo {author} {\bibfnamefont {J.}~\bibnamefont {Schnack}}, \bibinfo
  {author} {\bibfnamefont {P.}~\bibnamefont {Hage}}, \bibinfo {author}
  {\bibfnamefont {M.}~\bibnamefont {Luban}}, \ and\ \bibinfo {author}
  {\bibfnamefont {P.}~\bibnamefont {K\"ogerler}},\ }\href {\doibase
  10.1103/PhysRevLett.94.017205} {\bibfield  {journal} {\bibinfo  {journal}
  {Phys. Rev. Lett.}\ }\textbf {\bibinfo {volume} {94}},\ \bibinfo {pages}
  {017205} (\bibinfo {year} {2005}{\natexlab{b}})}\BibitemShut {NoStop}%
\bibitem [{\citenamefont {Dorfbauer}\ \emph {et~al.}(2007)\citenamefont
  {Dorfbauer}, \citenamefont {Evans}, \citenamefont {Kirschner}, \citenamefont
  {Chubykalo-Fesenko}, \citenamefont {Chantrell},\ and\ \citenamefont
  {Schrefl}}]{Dorfbauer:2007vp}%
  \BibitemOpen
  \bibfield  {author} {\bibinfo {author} {\bibfnamefont {F.}~\bibnamefont
  {Dorfbauer}}, \bibinfo {author} {\bibfnamefont {R.}~\bibnamefont {Evans}},
  \bibinfo {author} {\bibfnamefont {M.}~\bibnamefont {Kirschner}}, \bibinfo
  {author} {\bibfnamefont {O.}~\bibnamefont {Chubykalo-Fesenko}}, \bibinfo
  {author} {\bibfnamefont {T.}~\bibnamefont {Chantrell}}, \ and\ \bibinfo
  {author} {\bibfnamefont {T.}~\bibnamefont {Schrefl}},\ }\href
  {http://linkinghub.elsevier.com/retrieve/pii/S0304885307004337} {\bibfield
  {journal} {\bibinfo  {journal} {J. Magn. Magn. Mater.}\ }\textbf {\bibinfo
  {volume} {316}},\ \bibinfo {pages} {791} (\bibinfo {year}
  {2007})}\BibitemShut {NoStop}%
\bibitem [{\citenamefont {Landau}\ and\ \citenamefont
  {Lifshitz}(1935)}]{LandauLifshitz1935}%
  \BibitemOpen
  \bibfield  {author} {\bibinfo {author} {\bibfnamefont {L.~D.}\ \bibnamefont
  {Landau}}\ and\ \bibinfo {author} {\bibfnamefont {E.~M.}\ \bibnamefont
  {Lifshitz}},\ }\href@noop {} {\bibfield  {journal} {\bibinfo  {journal}
  {Phys. Z. Sowietunion}\ }\textbf {\bibinfo {volume} {8}},\ \bibinfo {pages}
  {153} (\bibinfo {year} {1935})}\BibitemShut {NoStop}%
\bibitem [{\citenamefont {Gilbert}(1955)}]{Gilbert1955}%
  \BibitemOpen
  \bibfield  {author} {\bibinfo {author} {\bibfnamefont {T.}~\bibnamefont
  {Gilbert}},\ }\href@noop {} {\bibfield  {journal} {\bibinfo  {journal}
  {Physical Review}\ }\textbf {\bibinfo {volume} {100}},\ \bibinfo {pages}
  {1243} (\bibinfo {year} {1955})}\BibitemShut {NoStop}%
\bibitem [{\citenamefont {Karakurt}\ \emph {et~al.}(2007)\citenamefont
  {Karakurt}, \citenamefont {Chantrell},\ and\ \citenamefont
  {Nowak}}]{Karakurt:2007cz}%
  \BibitemOpen
  \bibfield  {author} {\bibinfo {author} {\bibfnamefont {S.}~\bibnamefont
  {Karakurt}}, \bibinfo {author} {\bibfnamefont {R.~W.}\ \bibnamefont
  {Chantrell}}, \ and\ \bibinfo {author} {\bibfnamefont {U.}~\bibnamefont
  {Nowak}},\ }\href {\doibase 10.1016/j.jmmm.2007.02.118} {\bibfield  {journal}
  {\bibinfo  {journal} {J. Magn. Magn. Mater.}\ }\textbf {\bibinfo {volume}
  {316}},\ \bibinfo {pages} {e280} (\bibinfo {year} {2007})}\BibitemShut
  {NoStop}%
\bibitem [{\citenamefont {F{\"a}hnle}\ \emph {et~al.}(2010)\citenamefont
  {F{\"a}hnle}, \citenamefont {Seib},\ and\ \citenamefont
  {Illg}}]{Fahnle:2010ev}%
  \BibitemOpen
  \bibfield  {author} {\bibinfo {author} {\bibfnamefont {M.}~\bibnamefont
  {F{\"a}hnle}}, \bibinfo {author} {\bibfnamefont {J.}~\bibnamefont {Seib}}, \
  and\ \bibinfo {author} {\bibfnamefont {C.}~\bibnamefont {Illg}},\ }\href
  {\doibase 10.1103/PhysRevB.82.144405} {\bibfield  {journal} {\bibinfo
  {journal} {Phys. Rev. B}\ }\textbf {\bibinfo {volume} {82}},\ \bibinfo
  {pages} {144405} (\bibinfo {year} {2010})}\BibitemShut {NoStop}%
\bibitem [{\citenamefont {Dobin}\ and\ \citenamefont
  {Victora}(2004)}]{Dobin:2004vq}%
  \BibitemOpen
  \bibfield  {author} {\bibinfo {author} {\bibfnamefont {A.~Y.}\ \bibnamefont
  {Dobin}}\ and\ \bibinfo {author} {\bibfnamefont {R.~H.}\ \bibnamefont
  {Victora}},\ }\href
  {http://www.freewebs.com/alexdobin/Papers/PhysRevLett_92_257204.pdf}
  {\bibfield  {journal} {\bibinfo  {journal} {Phys. Rev. Lett.}\ }\textbf
  {\bibinfo {volume} {92}},\ \bibinfo {pages} {257204} (\bibinfo {year}
  {2004})}\BibitemShut {NoStop}%
\bibitem [{\citenamefont {Ellis}\ \emph {et~al.}(2012)\citenamefont {Ellis},
  \citenamefont {Ostler},\ and\ \citenamefont {Chantrell}}]{Ellis:2012bb}%
  \BibitemOpen
  \bibfield  {author} {\bibinfo {author} {\bibfnamefont {M.}~\bibnamefont
  {Ellis}}, \bibinfo {author} {\bibfnamefont {T.~A.}\ \bibnamefont {Ostler}}, \
  and\ \bibinfo {author} {\bibfnamefont {R.~W.}\ \bibnamefont {Chantrell}},\
  }\href {\doibase 10.1103/PhysRevLett.102.117201} {\bibfield  {journal}
  {\bibinfo  {journal} {Phys. Rev. B}\ }\textbf {\bibinfo {volume} {86}},\
  \bibinfo {pages} {174418} (\bibinfo {year} {2012})}\BibitemShut {NoStop}%
\bibitem [{\citenamefont {Jr.}(1979)}]{WFBrown1979}%
  \BibitemOpen
  \bibfield  {author} {\bibinfo {author} {\bibfnamefont {W.~F.~B.}\
  \bibnamefont {Jr.}},\ }\href@noop {} {\bibfield  {journal} {\bibinfo
  {journal} {IEEE Trans. Mag.}\ }\textbf {\bibinfo {volume} {15}},\ \bibinfo
  {pages} {1196} (\bibinfo {year} {1979})}\BibitemShut {NoStop}%
\bibitem [{\citenamefont {Garc{\'\i}a-Palacios}\ and\ \citenamefont
  {L{\'a}zaro}(1998)}]{GarciaPalacios:1998wz}%
  \BibitemOpen
  \bibfield  {author} {\bibinfo {author} {\bibfnamefont {J.~L.}\ \bibnamefont
  {Garc{\'\i}a-Palacios}}\ and\ \bibinfo {author} {\bibfnamefont {F.~J.}\
  \bibnamefont {L{\'a}zaro}},\ }\href
  {http://prb.aps.org/abstract/PRB/v58/i22/p14937_1} {\bibfield  {journal}
  {\bibinfo  {journal} {Phys. Rev. B}\ }\textbf {\bibinfo {volume} {58}},\
  \bibinfo {pages} {14937} (\bibinfo {year} {1998})}\BibitemShut {NoStop}%
\bibitem [{\citenamefont {Lyberatos}\ \emph {et~al.}(1999)\citenamefont
  {Lyberatos}, \citenamefont {Berkov},\ and\ \citenamefont
  {Chantrell}}]{Lyberatos:1999da}%
  \BibitemOpen
  \bibfield  {author} {\bibinfo {author} {\bibfnamefont {A.}~\bibnamefont
  {Lyberatos}}, \bibinfo {author} {\bibfnamefont {D.~V.}\ \bibnamefont
  {Berkov}}, \ and\ \bibinfo {author} {\bibfnamefont {R.~W.}\ \bibnamefont
  {Chantrell}},\ }\href {\doibase 10.1088/0953-8984/5/47/016} {\bibfield
  {journal} {\bibinfo  {journal} {J. Phys.: Condens. Matter}\ }\textbf
  {\bibinfo {volume} {5}},\ \bibinfo {pages} {8911} (\bibinfo {year}
  {1999})}\BibitemShut {NoStop}%
\bibitem [{\citenamefont {Nowak}\ \emph {et~al.}(2005)\citenamefont {Nowak},
  \citenamefont {Mryasov}, \citenamefont {Wieser}, \citenamefont {Guslienko},\
  and\ \citenamefont {Chantrell}}]{Nowak:2005eg}%
  \BibitemOpen
  \bibfield  {author} {\bibinfo {author} {\bibfnamefont {U.}~\bibnamefont
  {Nowak}}, \bibinfo {author} {\bibfnamefont {O.}~\bibnamefont {Mryasov}},
  \bibinfo {author} {\bibfnamefont {R.}~\bibnamefont {Wieser}}, \bibinfo
  {author} {\bibfnamefont {K.}~\bibnamefont {Guslienko}}, \ and\ \bibinfo
  {author} {\bibfnamefont {R.}~\bibnamefont {Chantrell}},\ }\href
  {http://link.aps.org/doi/10.1103/PhysRevB.72.172410} {\bibfield  {journal}
  {\bibinfo  {journal} {Phys. Rev. B}\ }\textbf {\bibinfo {volume} {72}}
  (\bibinfo {year} {2005})}\BibitemShut {NoStop}%
\bibitem [{\citenamefont {Matsumoto}\ and\ \citenamefont
  {Nishimura}(1998)}]{Matsumoto:1998jt}%
  \BibitemOpen
  \bibfield  {author} {\bibinfo {author} {\bibfnamefont {M.}~\bibnamefont
  {Matsumoto}}\ and\ \bibinfo {author} {\bibfnamefont {T.}~\bibnamefont
  {Nishimura}},\ }\href {\doibase 10.1145/272991.272995} {\bibfield  {journal}
  {\bibinfo  {journal} {ACM Transactions on Modeling and Computer Simulation}\
  }\textbf {\bibinfo {volume} {8}},\ \bibinfo {pages} {3} (\bibinfo {year}
  {1998})}\BibitemShut {NoStop}%
\bibitem [{\citenamefont {Marsaglia}\ and\ \citenamefont
  {Tsang}(2000)}]{Marsaglia:2000wj}%
  \BibitemOpen
  \bibfield  {author} {\bibinfo {author} {\bibfnamefont {G.}~\bibnamefont
  {Marsaglia}}\ and\ \bibinfo {author} {\bibfnamefont {W.-W.}\ \bibnamefont
  {Tsang}},\ }\href
  {http://scholar.google.com/scholar?q=related:zJrGfCvZC_8J:scholar.google.com/&hl=en&num=20&as_sdt=0,5}
  {\bibfield  {journal} {\bibinfo  {journal} {Journal of Statistical Software}\
  }\textbf {\bibinfo {volume} {5}} (\bibinfo {year} {2000})}\BibitemShut
  {NoStop}%
\bibitem [{\citenamefont {Atxitia}\ \emph {et~al.}(2009)\citenamefont
  {Atxitia}, \citenamefont {Chubykalo-Fesenko}, \citenamefont {Chantrell},
  \citenamefont {Nowak},\ and\ \citenamefont {Rebei}}]{Atxitia:2009io}%
  \BibitemOpen
  \bibfield  {author} {\bibinfo {author} {\bibfnamefont {U.}~\bibnamefont
  {Atxitia}}, \bibinfo {author} {\bibfnamefont {O.}~\bibnamefont
  {Chubykalo-Fesenko}}, \bibinfo {author} {\bibfnamefont {R.}~\bibnamefont
  {Chantrell}}, \bibinfo {author} {\bibfnamefont {U.}~\bibnamefont {Nowak}}, \
  and\ \bibinfo {author} {\bibfnamefont {A.}~\bibnamefont {Rebei}},\ }\href
  {\doibase 10.1103/PhysRevLett.102.057203} {\bibfield  {journal} {\bibinfo
  {journal} {Phys. Rev. Lett.}\ }\textbf {\bibinfo {volume} {102}},\ \bibinfo
  {pages} {057203} (\bibinfo {year} {2009})}\BibitemShut {NoStop}%
\bibitem [{\citenamefont {Schellekens}\ and\ \citenamefont
  {Koopmans}(2013)}]{Schellekens:2013ci}%
  \BibitemOpen
  \bibfield  {author} {\bibinfo {author} {\bibfnamefont {A.~J.}\ \bibnamefont
  {Schellekens}}\ and\ \bibinfo {author} {\bibfnamefont {B.}~\bibnamefont
  {Koopmans}},\ }\href {\doibase 10.1103/PhysRevLett.110.217204} {\bibfield
  {journal} {\bibinfo  {journal} {Phys. Rev. Lett.}\ }\textbf {\bibinfo
  {volume} {110}},\ \bibinfo {pages} {217204} (\bibinfo {year}
  {2013})}\BibitemShut {NoStop}%
\bibitem [{\citenamefont {Grossmann}\ and\ \citenamefont
  {Rancourt}(1996)}]{Grossmann:1996vi}%
  \BibitemOpen
  \bibfield  {author} {\bibinfo {author} {\bibfnamefont {B.}~\bibnamefont
  {Grossmann}}\ and\ \bibinfo {author} {\bibfnamefont {D.}~\bibnamefont
  {Rancourt}},\ }\href@noop {} {\bibfield  {journal} {\bibinfo  {journal}
  {Phys. Rev. B}\ }\textbf {\bibinfo {volume} {54}},\ \bibinfo {pages} {12294}
  (\bibinfo {year} {1996})}\BibitemShut {NoStop}%
\bibitem [{\citenamefont {Ma}\ \emph {et~al.}(2012{\natexlab{b}})\citenamefont
  {Ma}, \citenamefont {Dudarev},\ and\ \citenamefont {Woo}}]{Ma:2012hp}%
  \BibitemOpen
  \bibfield  {author} {\bibinfo {author} {\bibfnamefont {P.-W.}\ \bibnamefont
  {Ma}}, \bibinfo {author} {\bibfnamefont {S.~L.}\ \bibnamefont {Dudarev}}, \
  and\ \bibinfo {author} {\bibfnamefont {C.~H.}\ \bibnamefont {Woo}},\ }\href
  {\doibase 10.1103/PhysRevB.85.184301} {\bibfield  {journal} {\bibinfo
  {journal} {Phys. Rev. B}\ }\textbf {\bibinfo {volume} {85}},\ \bibinfo
  {pages} {184301} (\bibinfo {year} {2012}{\natexlab{b}})}\BibitemShut
  {NoStop}%
\bibitem [{\citenamefont {Battiato}\ \emph {et~al.}(2010)\citenamefont
  {Battiato}, \citenamefont {Carva},\ and\ \citenamefont
  {Oppeneer}}]{Battiato:2010br}%
  \BibitemOpen
  \bibfield  {author} {\bibinfo {author} {\bibfnamefont {M.}~\bibnamefont
  {Battiato}}, \bibinfo {author} {\bibfnamefont {K.}~\bibnamefont {Carva}}, \
  and\ \bibinfo {author} {\bibfnamefont {P.~M.}\ \bibnamefont {Oppeneer}},\
  }\href {\doibase 10.1103/PhysRevLett.105.027203} {\bibfield  {journal}
  {\bibinfo  {journal} {Phys. Rev. Lett.}\ }\textbf {\bibinfo {volume} {105}},\
  \bibinfo {pages} {027203} (\bibinfo {year} {2010})}\BibitemShut {NoStop}%
\bibitem [{\citenamefont {Schellekens}\ \emph {et~al.}(2013)\citenamefont
  {Schellekens}, \citenamefont {Verhoeven}, \citenamefont {Vader},\ and\
  \citenamefont {Koopmans}}]{Schellekens:2013em}%
  \BibitemOpen
  \bibfield  {author} {\bibinfo {author} {\bibfnamefont {A.~J.}\ \bibnamefont
  {Schellekens}}, \bibinfo {author} {\bibfnamefont {W.}~\bibnamefont
  {Verhoeven}}, \bibinfo {author} {\bibfnamefont {T.~N.}\ \bibnamefont
  {Vader}}, \ and\ \bibinfo {author} {\bibfnamefont {B.}~\bibnamefont
  {Koopmans}},\ }\href {\doibase 10.1063/1.4812658} {\bibfield  {journal}
  {\bibinfo  {journal} {Appl. Phys. Lett.}\ }\textbf {\bibinfo {volume}
  {102}},\ \bibinfo {pages} {252408} (\bibinfo {year} {2013})}\BibitemShut
  {NoStop}%
\bibitem [{\citenamefont {Berkov}\ and\ \citenamefont
  {Gorn}(2002)}]{Berkov2002}%
  \BibitemOpen
  \bibfield  {author} {\bibinfo {author} {\bibfnamefont {D.~V.}\ \bibnamefont
  {Berkov}}\ and\ \bibinfo {author} {\bibfnamefont {N.~L.}\ \bibnamefont
  {Gorn}},\ }\href {http://stacks.iop.org/0953-8984/14/i=13/a=101} {\bibfield
  {journal} {\bibinfo  {journal} {J. Phys.: Condens. Matter}\ }\textbf
  {\bibinfo {volume} {14}},\ \bibinfo {pages} {L281} (\bibinfo {year}
  {2002})}\BibitemShut {NoStop}%
\bibitem [{\citenamefont {d'Aquino}\ \emph {et~al.}(2005)\citenamefont
  {d'Aquino}, \citenamefont {Serpico},\ and\ \citenamefont
  {Miano}}]{dAquino:2005fx}%
  \BibitemOpen
  \bibfield  {author} {\bibinfo {author} {\bibfnamefont {M.}~\bibnamefont
  {d'Aquino}}, \bibinfo {author} {\bibfnamefont {C.}~\bibnamefont {Serpico}}, \
  and\ \bibinfo {author} {\bibfnamefont {G.}~\bibnamefont {Miano}},\ }\href
  {\doibase 10.1016/j.jcp.2005.04.001} {\bibfield  {journal} {\bibinfo
  {journal} {J. Comput. Phys.}\ }\textbf {\bibinfo {volume} {209}},\ \bibinfo
  {pages} {730} (\bibinfo {year} {2005})}\BibitemShut {NoStop}%
\bibitem [{\citenamefont {Mentink}\ \emph {et~al.}(2010)\citenamefont
  {Mentink}, \citenamefont {Tretyakov}, \citenamefont {Fasolino}, \citenamefont
  {Katsnelson},\ and\ \citenamefont {Rasing}}]{Mentink:2010vk}%
  \BibitemOpen
  \bibfield  {author} {\bibinfo {author} {\bibfnamefont {J.}~\bibnamefont
  {Mentink}}, \bibinfo {author} {\bibfnamefont {M.}~\bibnamefont {Tretyakov}},
  \bibinfo {author} {\bibfnamefont {A.}~\bibnamefont {Fasolino}}, \bibinfo
  {author} {\bibfnamefont {M.}~\bibnamefont {Katsnelson}}, \ and\ \bibinfo
  {author} {\bibfnamefont {T.}~\bibnamefont {Rasing}},\ }\href@noop {}
  {\bibfield  {journal} {\bibinfo  {journal} {J. Phys.: Condens. Matter}\
  }\textbf {\bibinfo {volume} {22}},\ \bibinfo {pages} {176001} (\bibinfo
  {year} {2010})}\BibitemShut {NoStop}%
\bibitem [{\citenamefont {d'Aquino}\ \emph {et~al.}(2006)\citenamefont
  {d'Aquino}, \citenamefont {Serpico}, \citenamefont {Coppola}, \citenamefont
  {Mayergoyz},\ and\ \citenamefont {Bertotti}}]{dAquino:2006kj}%
  \BibitemOpen
  \bibfield  {author} {\bibinfo {author} {\bibfnamefont {M.}~\bibnamefont
  {d'Aquino}}, \bibinfo {author} {\bibfnamefont {C.}~\bibnamefont {Serpico}},
  \bibinfo {author} {\bibfnamefont {G.}~\bibnamefont {Coppola}}, \bibinfo
  {author} {\bibfnamefont {I.~D.}\ \bibnamefont {Mayergoyz}}, \ and\ \bibinfo
  {author} {\bibfnamefont {G.}~\bibnamefont {Bertotti}},\ }\href {\doibase
  10.1063/1.2169472} {\bibfield  {journal} {\bibinfo  {journal} {J. Appl.
  Phys.}\ }\textbf {\bibinfo {volume} {99}},\ \bibinfo {pages} {08B905}
  (\bibinfo {year} {2006})}\BibitemShut {NoStop}%
\bibitem [{\citenamefont {Kloeden}\ and\ \citenamefont
  {Platen}(1995)}]{HeunzeroK}%
  \BibitemOpen
  \bibfield  {author} {\bibinfo {author} {\bibfnamefont {P.~E.}\ \bibnamefont
  {Kloeden}}\ and\ \bibinfo {author} {\bibfnamefont {E.}~\bibnamefont
  {Platen}},\ }\href@noop {} {\emph {\bibinfo {title} {{Numerical Solution of
  Stochastic Differential Equations}}}}\ (\bibinfo  {publisher} {Springer,
  Heidelberg, Berlin},\ \bibinfo {year} {1995})\BibitemShut {NoStop}%
\bibitem [{\citenamefont {Hannay}(2001)}]{Hannay}%
  \BibitemOpen
  \bibfield  {author} {\bibinfo {author} {\bibfnamefont {J.~D.}\ \bibnamefont
  {Hannay}},\ }\emph {\bibinfo {title} {{Computational Simulations of Thermally
  Activated Magnetisation Dynamics at High Frequencies}}},\ \href@noop {}
  {\bibinfo {type} {{PhD Thesis}}},\ \bibinfo  {school} {School of Informatics,
  University of Wales, Bangor, Gwynedd, UK} (\bibinfo {year}
  {2001})\BibitemShut {NoStop}%
\bibitem [{\citenamefont {Metropolis}\ \emph {et~al.}(1953)\citenamefont
  {Metropolis}, \citenamefont {Rosenbluth}, \citenamefont {Rosenbluth},
  \citenamefont {Teller},\ and\ \citenamefont {Teller}}]{Metropolis:1953in}%
  \BibitemOpen
  \bibfield  {author} {\bibinfo {author} {\bibfnamefont {N.}~\bibnamefont
  {Metropolis}}, \bibinfo {author} {\bibfnamefont {A.~W.}\ \bibnamefont
  {Rosenbluth}}, \bibinfo {author} {\bibfnamefont {M.~N.}\ \bibnamefont
  {Rosenbluth}}, \bibinfo {author} {\bibfnamefont {A.~H.}\ \bibnamefont
  {Teller}}, \ and\ \bibinfo {author} {\bibfnamefont {E.}~\bibnamefont
  {Teller}},\ }\href {\doibase 10.1063/1.1699114} {\bibfield  {journal}
  {\bibinfo  {journal} {J. Chem. Phys.}\ }\textbf {\bibinfo {volume} {21}},\
  \bibinfo {pages} {1087} (\bibinfo {year} {1953})}\BibitemShut {NoStop}%
\bibitem [{\citenamefont {Hinzke}\ and\ \citenamefont
  {Nowak}(1999)}]{Hinzke:1999ud}%
  \BibitemOpen
  \bibfield  {author} {\bibinfo {author} {\bibfnamefont {D.}~\bibnamefont
  {Hinzke}}\ and\ \bibinfo {author} {\bibfnamefont {U.}~\bibnamefont {Nowak}},\
  }\href
  {http://158.197.33.91/~horvath/download/articles/Monte_Carlo_switching_Heisenberg_model_small_ferromagnetic_particles_Hinzke_Nowak_1999.pdf}
  {\bibfield  {journal} {\bibinfo  {journal} {Comput. Phys. Commun.}\ }\textbf
  {\bibinfo {volume} {121}},\ \bibinfo {pages} {334} (\bibinfo {year}
  {1999})}\BibitemShut {NoStop}%
\bibitem [{\citenamefont
  {http://www.ctcms.nist.gov/$\sim$rdm/mumag.org}()}]{mmagstandardproblems}%
  \BibitemOpen
  \bibfield  {author} {\bibinfo {author} {\bibnamefont
  {http://www.ctcms.nist.gov/$\sim$rdm/mumag.org}},\ }\href@noop {}
  {}\BibitemShut {NoStop}%
\bibitem [{\citenamefont {Stoner}\ and\ \citenamefont
  {Wohlfarth}(1948)}]{Stoner:1948ur}%
  \BibitemOpen
  \bibfield  {author} {\bibinfo {author} {\bibfnamefont {E.~C.}\ \bibnamefont
  {Stoner}}\ and\ \bibinfo {author} {\bibfnamefont {E.~P.}\ \bibnamefont
  {Wohlfarth}},\ }\href {http://www.jstor.org/stable/10.2307/91421} {\bibfield
  {journal} {\bibinfo  {journal} {Philos. T. Roy. Soc. A}\ }\textbf {\bibinfo
  {volume} {240}},\ \bibinfo {pages} {599} (\bibinfo {year}
  {1948})}\BibitemShut {NoStop}%
\bibitem [{\citenamefont {Binder}(1969)}]{BinderPLA1969}%
  \BibitemOpen
  \bibfield  {author} {\bibinfo {author} {\bibfnamefont {K.}~\bibnamefont
  {Binder}},\ }\href {\doibase http://dx.doi.org/10.1016/0375-9601(69)90989-X}
  {\bibfield  {journal} {\bibinfo  {journal} {Physics Letters A}\ }\textbf
  {\bibinfo {volume} {30}},\ \bibinfo {pages} {273 } (\bibinfo {year}
  {1969})}\BibitemShut {NoStop}%
\bibitem [{\citenamefont {Yuan}\ and\ \citenamefont
  {Bertram}(1992)}]{Yuan:1992ue}%
  \BibitemOpen
  \bibfield  {author} {\bibinfo {author} {\bibfnamefont {S.~W.}\ \bibnamefont
  {Yuan}}\ and\ \bibinfo {author} {\bibfnamefont {H.~N.}\ \bibnamefont
  {Bertram}},\ }\href
  {http://ieeexplore.ieee.org/xpls/abs_all.jsp?arnumber=179394} {\bibfield
  {journal} {\bibinfo  {journal} {IEEE Trans. Magn}\ }\textbf {\bibinfo
  {volume} {28}},\ \bibinfo {pages} {2031} (\bibinfo {year}
  {1992})}\BibitemShut {NoStop}%
\bibitem [{\citenamefont {Berkov}\ \emph {et~al.}(1993)\citenamefont {Berkov},
  \citenamefont {Ramst{\"o}cck},\ and\ \citenamefont {Hubert}}]{Berkov:1993th}%
  \BibitemOpen
  \bibfield  {author} {\bibinfo {author} {\bibfnamefont {D.~V.}\ \bibnamefont
  {Berkov}}, \bibinfo {author} {\bibfnamefont {K.}~\bibnamefont
  {Ramst{\"o}cck}}, \ and\ \bibinfo {author} {\bibfnamefont {A.}~\bibnamefont
  {Hubert}},\ }\href
  {http://onlinelibrary.wiley.com/doi/10.1002/pssa.2211370118/abstract}
  {\bibfield  {journal} {\bibinfo  {journal} {physica status solidi (a)}\
  }\textbf {\bibinfo {volume} {137}},\ \bibinfo {pages} {207} (\bibinfo {year}
  {1993})}\BibitemShut {NoStop}%
\bibitem [{\citenamefont {Kurzak}(2005)}]{Kurzak:2005ur}%
  \BibitemOpen
  \bibfield  {author} {\bibinfo {author} {\bibfnamefont {J.}~\bibnamefont
  {Kurzak}},\ }\href {http://portal.acm.org/citation.cfm?id=1088526} {\bibfield
   {journal} {\bibinfo  {journal} {J. Parallel. Distr. Com.}\ }\textbf
  {\bibinfo {volume} {65}},\ \bibinfo {pages} {870} (\bibinfo {year}
  {2005})}\BibitemShut {NoStop}%
\bibitem [{\citenamefont {Aharoni}(1998)}]{Aharoni:1998ku}%
  \BibitemOpen
  \bibfield  {author} {\bibinfo {author} {\bibfnamefont {A.}~\bibnamefont
  {Aharoni}},\ }\href {\doibase 10.1063/1.367113} {\bibfield  {journal}
  {\bibinfo  {journal} {J. Appl. Phys.}\ }\textbf {\bibinfo {volume} {83}},\
  \bibinfo {pages} {3432} (\bibinfo {year} {1998})}\BibitemShut {NoStop}%
\bibitem [{\citenamefont {Osborn}(1945)}]{Osborn:1945to}%
  \BibitemOpen
  \bibfield  {author} {\bibinfo {author} {\bibfnamefont {J.~A.}\ \bibnamefont
  {Osborn}},\ }\href
  {http://www.cmap.polytechnique.fr/~alouges/coursm2/Osborn.pdf} {\bibfield
  {journal} {\bibinfo  {journal} {Physical Review}\ }\textbf {\bibinfo {volume}
  {67}},\ \bibinfo {pages} {351} (\bibinfo {year} {1945})}\BibitemShut
  {NoStop}%
\bibitem [{\citenamefont {Hinzke}\ and\ \citenamefont
  {Nowak}(2011)}]{Hinzke:2011gp}%
  \BibitemOpen
  \bibfield  {author} {\bibinfo {author} {\bibfnamefont {D.}~\bibnamefont
  {Hinzke}}\ and\ \bibinfo {author} {\bibfnamefont {U.}~\bibnamefont {Nowak}},\
  }\href {\doibase 10.1103/PhysRevLett.107.027205} {\bibfield  {journal}
  {\bibinfo  {journal} {Phys. Rev. Lett.}\ }\textbf {\bibinfo {volume} {107}},\
  \bibinfo {pages} {027205} (\bibinfo {year} {2011})}\BibitemShut {NoStop}%
\end{thebibliography}%

\end{document}